\documentclass[11pt]{article}
\usepackage[letterpaper, margin=1in]{geometry}
\usepackage{mathtools, amssymb, amsthm, physics, tikz, ifthen, multirow, adjustbox, rotating, array, mleftright, soul, bbm, makecell, nicefrac, booktabs, subcaption}
\usepackage[textsize=footnotesize]{todonotes}
\usepackage[usestackEOL]{stackengine}

\usepackage{amsmath}
\usepackage{authblk}
\usepackage{lineno}
\usepackage{hyperref}
\hypersetup{
    colorlinks,
    linkcolor={red!80!black},
    citecolor={blue},
    urlcolor={blue!80!black}
}
\usepackage[capitalize,noabbrev]{cleveref}

\makeatletter
\renewcommand\AB@affilsepx{\\ \protect\Affilfont}
\makeatother

\newcommand{\eps}{\epsilon}

\newcommand{\plainfrac}[2]{#1/#2}

\newcommand*\cno[1][0.8ex]{\tikz\draw (0,0) circle (#1);} 
\newcommand*\cboth[1][0.8ex]{%
  \begin{tikzpicture}
  \draw[fill] (0,0) -- (180:#1) arc (180:360:#1) -- cycle;
  \draw (0,0) circle (#1);  
  \end{tikzpicture}}
\newcommand*\cyes[1][0.8ex]{\tikz\fill (0,0) circle (#1);} 

\def\header{\vspace{1mm} \noindent}

\newcommand{\p}[1]{\mleft( #1 \mright)}
\newcommand{\np}[1]{( #1 )}

\newcommand{\curly}[1]{\mleft\{ #1 \mright\}}
\newcommand{\ncurly}[1]{\{ #1 \}}
\renewcommand{\square}[1]{\mleft[ #1 \mright]}

\renewcommand{\P}[1]{\mathbb{P} \square{ #1 } }
\newcommand{\E}[1]{\mathbb{E} \square{ #1 } }

\newcommand{\Ind}[1]{\mathbbm{1} \curly{ #1 } }

\newcommand{\DegIn}{\textnormal{\texttt{DEG-IN}}}
\newcommand{\DegOut}{\textnormal{\texttt{DEG-OUT}}}
\newcommand{\In}{\textnormal{\texttt{IN}}}
\newcommand{\Out}{\textnormal{\texttt{OUT}}}
\newcommand{\Jump}{\textnormal{\texttt{JUMP}}}
\newcommand{\InOrd}{\textnormal{\texttt{IN-SORTED}}}
\newcommand{\Adj}{\textnormal{\texttt{ADJ}}}

\newcommand{\pw}{\textnormal{\texttt{PowerIteration}}}
\newcommand{\bpush}{\textnormal{\texttt{ApproxContributions}}}
\newcommand{\BiPPR}{\textnormal{\texttt{BiPPR}}}
\newcommand{\FastPPR}{\textnormal{\texttt{FastPPR}}}

\newcommand{\Push}{\textnormal{\texttt{PushBack}}}

\newcommand{\RBS}{\textnormal{\texttt{RBS}}}

\newcommand{\ranPushthres}{\textnormal{\texttt{RandPushThreshold}}}
\newcommand{\spppr}{\textnormal{\texttt{SinglePairPPR}}}

\theoremstyle{plain}
\newtheorem{theorem}{Theorem}[section]
\newtheorem{lemma}[theorem]{Lemma}
\newtheorem{claim}[theorem]{Claim}

\newtheorem{result}{Result}

\theoremstyle{definition}
\newtheorem{definition}[theorem]{Definition}

\theoremstyle{remark}

\usepackage{algorithm}
\usepackage{algorithmicx}
\usepackage{algpseudocode}
\algtext*{EndWhile}
\algtext*{EndIf}
\algtext*{EndFor}
\algrenewcommand\algorithmicrequire{\textbf{Input:}}
\algrenewcommand\algorithmicensure{\textbf{Output:}}

\newcommand{\rh}{\hat{r}}
\newcommand{\rhh}{\hat{r}'}
\newcommand{\erh}{\hat{R}}

\newcommand{\er}{R}
\newcommand{\dr}{\chi}
\newcommand{\rmax}{\theta}
\newcommand{\rmaxx}{\theta'}
\newcommand{\gap}{\gamma}
\newcommand{\ph}{\hat{p}}
\newcommand{\nopush}{\mathbbm{1}}
\newcommand{\nr}{n_r}
\newcommand{\ns}{n_s}
\newcommand{\tpi}{\tilde{\pi}}
\newcommand{\pth}{\tau}
\newcommand{\Vp}{V_{\pth}}
\newcommand{\din}{d_\mathrm{in}}
\newcommand{\dout}{d_\mathrm{out}}

\newcommand{\Nin}{\mathcal{N}_\mathrm{in}}
\newcommand{\Nout}{\mathcal{N}_\mathrm{out}}
\newcommand{\pih}{\hat{\pi}}

\newcommand{\pit}{\tilde{\pi}}
\newcommand{\pitt}{\hat{\pi}}
\newcommand{\rela}{\varepsilon}
\newcommand{\pf}{p_f}
\newcommand{\vpi}{\pi}
\newcommand{\epi}{\hat{\pi}}

\newcommand{\residue}{r}
\newcommand{\reserve}{p}
\newcommand{\pushthreshold}{r_{\mathrm{max}}}
\newcommand{\rannum}{rand}

\newcommand{\Deltain}{\Delta_{\mathrm{in}}}
\newcommand{\Deltaout}{\Delta_{\mathrm{out}}}
\def\smallexpo{\gamma}

\title{Estimating Random-Walk Probabilities in Directed Graphs \footnote{This work was supported by the VILLUM Foundation grant 54451.}}

\author[1]{Christian Bertram}
\author[1]{Mads Vestergaard Jensen}
\author[2]{Mikkel Thorup}
\author[3]{
\and \vspace{-1em} Hanzhi Wang}
\author[1]{Shuyi Yan}

\vspace{0.3em}
\affil[1,2]{BARC, University of Copenhagen \vspace{0.2em}}
\affil[3]{The University of Melbourne \vspace{1.3em}}
\affil[1]{\texttt{\{chbe, mvje, shya\}@di.ku.dk}\vspace{0.2em}}
\affil[2,3]{\texttt{\{mikkel2thorup, hanzhi.hzwang\}@gmail.com}}

\date{}

\begin{document}
% \linenumbers

\pagenumbering{roman}
\setcounter{page}{1}

\begin{titlepage}
    \maketitle

    \begin{abstract}
        We study discounted random walks in directed graphs. In each step, the walk either terminates with a constant probability $\alpha$, or proceeds to a random out-neighbor. Our goal is to estimate the probability $\pi(s, t)$ that a discounted random walk starting from $s$ terminates at $t$. This probability is also known as the Personalized PageRank (PPR) score, which measures the relevance of $t$ to $s$, for instance, when $s$ and $t$ are web pages on the Internet. We aim to estimate $\pi(s, t)$ within a constant relative error with constant probability.
        
        A variety of algorithms have been developed for several problem variants, such as single-pair, single-source, single-target, and single-node estimation, under both worst-case and average-case settings, and for different combinations of allowed graph queries. However, in many important cases, there remain polynomial gaps between known upper and lower bounds.
        
        In this paper, we establish tight bounds for all problem variants and query combinations, closing all existing gaps in both the worst-case and average-case settings. We provide tight (up to logarithmic factors) lower bounds, showing that for all but one query combination, existing algorithms are already optimal.
        
        For the remaining case, we design a novel algorithm that matches our new lower bound up to logarithmic factors, thereby nearly achieving optimality. This is the first algorithm to exploit this specific query combination. It uses a new randomized bidirectional framework that combines randomized backward propagation with selective Monte Carlo estimation.
    \end{abstract}
\end{titlepage}
\newpage
\setcounter{page}{2}
\tableofcontents
\newpage

\pagenumbering{arabic}
\setcounter{page}{1}

\section{Introduction}\label{sec:intro}

Random walks in directed graphs are a fundamental algorithmic tool in modern network analysis. One important type is the discounted random walk, where at each step, the walk either terminates with some probability $\alpha \in (0,1)$ or moves to a random out-neighbor. Following previous works~\cite{RBS, BPP23, stoc/WangWW024, Mingji_ICDT}, we assume $\alpha$ to be a constant.
The stationary distribution of such a random walk is unique, fast-mixing, and always guaranteed to exist.
Estimating the probability $\vpi(s,t)$ that a discounted random walk starting from node $s$ ends at node $t$ has been a subject of extensive research for over a decade~\cite{BackwardPush_InternetMath, reversePageRank_Bar-YossefM08, FORA, MaeharaAIK14, FastPPR, BiPPR, stoc/WangWW024, walkingMassively, BPP18, BPP23, dynamicPageRank_ICALP, MC, Mingji_ICDT, GuoLST17_parallelPPR, heavyhitters_WangT18, edgepush, RBS, SpeedPPR, almostLinearTime_stoc/CohenKPPRSV17, randomwalk_streaming, fasterAlgorithms_focs/CohenKPPSV16} and has been widely applied in diverse areas such as web search~\cite{pagerank98, scalingwebsearch_JehWidom03, GleichP07_PPR_Websearch, rankwebFrontier_www/EironMT04}, recommender systems~\cite{WTF_www/GuptaGLSWZ13, video_recommendation}, 
spam filtering~\cite{spam_trustrank},
among others~\cite{TKDE_Survey_YangWWWW24, BeyondWeb_Survey_Gleich15}.
A notable example is Google's celebrated PageRank algorithm~\cite{pagerank98, page1999pagerank}, which ranks a web-page $t$ based on the average value of $\pi(s,t)$ over all web-pages $s$.
The probability $\vpi(s,t)$ is also referred to as the Personalized PageRank (PPR) score of $t$ with respect to $s$, indicating the relative importance of $t$ to $s$. 
We note that these walks have also been studied in the special case of symmetric (undirected) graphs~\cite{undirectedBiPPR_LofgrenBG15, Nibble, Fountoulakis_Variational, simpleSDD, SDDlinearsystem, BackMC}, but in this paper we focus on the general case of directed graphs.

Specifically, we study the computational complexity of estimating $\vpi(s,t)$ in this paper.
We are given a directed graph $G=(V,E)$ comprising $n$ nodes and $m$ edges, together with a source node $s \in V$, a target node $t \in V$, and an approximation threshold parameter $\delta \in (0,1)$.
Our goal is, roughly speaking, to estimate $\pi(s,t)$ with constant relative error with constant probability unless $\vpi(s,t)<\delta$. 
A natural choice is $\delta=1/n$, meaning that we only want to approximate $\vpi(s,t)$ if it is above $1/n$, which is the average value across all pairs. 
One may also choose $\delta$ to be smaller, e.g., $\delta=1/m$, in order to obtain high-precision results. 
In \cite{tnse_Chung14_survey} they suggest using $\delta=1/n^2$. 
In general, $\delta$ can be chosen from any value in $(0,1)$ based on real needs. It can even be set to $0$, meaning that we want a constant factor approximation of $\vpi(s,t)$ regardless of how small it is ($\vpi(s,t)$ can be as small as $n^{-\Theta(n)}$). 
However, handling very small values of $\delta$ is computationally expensive and requires dedicated techniques. For simplicity, in this paper, we assume that $\delta = n^{-O(1)}$, as is implicitly assumed in many prior works that set $\delta=1/n$, e.g.,~\cite{FastPPR, FORA, RBS}. We refer interested readers to another manuscript of ours~\cite{bertram2026instance_opt_ppr} for techniques that enable efficient estimation when $\delta$ is extremely small (e.g., $\delta=0$). ~\footnote{In that manuscript of ours~\cite{bertram2026instance_opt_ppr}, we present a vertex-freezing technique that improves the worst-case time complexity of single-pair, single-source, and single-target PPR estimation from $O(m\log(1/\delta))$ to $O(m\log n)$. This improvement is significant when estimating extremely small PPR values, e.g., $\vpi(s,t)=n^{-\Theta(n)}$. This $O(m\log{n})$ result also shows that the lower bound established in the present manuscript is tight up to a logarithmic factor in $n$ for all possible values of $\delta$. This vertex-freezing technique was originally presented in an earlier version of the present manuscript, but we have moved it to \cite{bertram2026instance_opt_ppr} where it fits more naturally.}

For the above-mentioned problem of estimating $\pi(s,t)$, 
the most obvious approach is to do Monte Carlo sampling of random walks from the source node $s$ estimating $\pi(s,t)$ as the fraction of walks that end in $t$~\cite{MC}. With an approximation threshold
of $\delta$, it suffices with
$O(1/\delta)$ samples. This works well if $\delta$ is large. 
For small $\delta$, an alternative is to compute $\pi(s,t)$ 
using a global deterministic iterative algorithm~\cite{page1999pagerank, pagerank98}, with a computational complexity of $O(m\log{(1/\delta)})$.  
Using this instead of Monte Carlo sampling for any $\delta<1/m$ gives an upper bound of $\tilde O(\min\{m, 1/\delta\})$ on the worst-case computational complexity of estimating $\pi(s,t)$, where $\tilde{O}(\cdot)$ hides all polylogarithmic factors. 
Improvements over this bound have only been demonstrated when considering the computational complexity averaged over all $n$ possible target nodes in $G$. 
By combining Monte Carlo sampling with a deterministic backward exploration approach, Lofgren, Banerjee, and Goel~\cite{BiPPR} establish an average case complexity of $O(\min\{m\log{(1/\delta)}, (d/\delta)^{1/2}, 1/\delta\})$, where $d = m/n$ is the average (in- or out-) degree of the graph.

On the lower bound side, the previous lower bound for the worst case is $\Omega(\min\{n, 1/\delta\})$, derived from a simple folklore graph construction (see~\Cref{subsec:kl-singlepair}), leaving a gap for $\delta < 1/n$. 
For the average case, the previously best lower bound is $\Omega(n^{1/2})$~\cite{FastPPR}, established only for $\delta = 1/n$. This lower bound is obtained by adapting the $\Omega(n^{1/2})$ lower bound of~\cite{GoldreichR02} for expansion testing on degree-bounded graphs, where the approximation of $\pi(s,t)$ is used to distinguish between two 3-regular graphs on $n$ nodes. This leaves a $\Theta(d^{1/2})$ gap to the upper bound of $O((d/\delta)^{1/2})$, and whether the dependence on the degree $d$ can be removed from the average-case complexity has remained unclear for a long time.

In this paper we provide tight bounds for both worst and average cases, and for different choices of graph queries. We assume that algorithms have query access to the adjacency lists of the graph. Formally, we require that algorithms can only access the underlying graph through a graph oracle, which supports the query operations $\DegIn(u)$ and $\DegOut(u)$ that return the in-degree and out-degree of a node $u$, respectively, as well as the queries $\In(u, i)$ and $\Out(u, i)$, which return the $i$th in-neighbor and out-neighbor of $u$, respectively.
Each of these queries can be performed in constant time. This model is commonly referred to as the {\em adjacency-list model}~\cite{books_Goldreich17}, which aligns well with real-world scenarios where massive-scale network structures are ubiquitous.

Going beyond this canonical model, we also consider additional graph access queries that have been studied in the literature and are often practical in real-world settings.  
These include $\Adj(u, v)$, which allows checking whether there exists an edge from $u$ to $v$ in constant time~\cite{adjacencylistmodel_Dana}; $\InOrd(u, i)$, which returns the $i$-th in-neighbor of $u$ sorted by out-degree~\cite{RBS,AGP}; 
and $\Jump()$, which returns a uniformly random node from graph~\cite{stoc/WangWW024, BPP23, soda_Thorup0W026}.

We provide tight lower bounds (up to logarithmic factors), showing that existing algorithms are in fact optimal for all query combinations except the combination of $\Adj$ and $\InOrd$. For this previously unexplored combination, we improve both the upper and lower bounds to achieve optimality. This is the first algorithm that leverages this query combination.
Understanding the impact of different types of queries on computational complexity is important for designing an application programming interface (API) for large graphs, as it helps determine which query types are worth supporting.

While the focus of this paper is the above \emph{single-pair} problem, estimating $\pi(s,t)$ for a given pair $(s,t)$, we will also consider some of the important related problems and provide tight bounds for these as well.
As we know it from e.g.\ shortest path problems, there is also a \emph{single-source}~\cite{FORA, SpeedPPR, MC} and \emph{single-target} problem~\cite{RBS, heavyhitters_WangT18, BackwardPush_WAW, BackwardPush_InternetMath, Lofgren_backpush}.
The single-source problem asks for approximations of $\pi(s,t) > \delta$ for a given source node $s$ and all $n$ possible target nodes $t$.
The single-target problem is defined analogously, asking for the approximation of $\vpi(s,t)>\delta$ for a given target node $t$ and all $n$ possible source nodes $s$. Finally, the \emph{single-node} problem~\cite{stoc/WangWW024, BPP18, BPP23, reversePageRank_Bar-YossefM08} asks for an approximation of $\pi(t) = \frac{1}{n}\sum_{s \in V}\pi(s,t)$ for a given $t$.
This quantity represents the probability that a random walk starting at a uniformly random source node $s$ will stop at $t$, and is also known as a type of \emph{graph centrality} of $t$ in $G$. We give tight bounds across all problems and query choices. 

\subsection{Our results}

We summarize our results in this subsection. We provide tight bounds in both worst and average cases across all query combinations. 

\paragraph{Tight lower bounds matching existing upper bounds.}
Our first result is a tight lower bound for the single-pair problem in the worst-case setting, as detailed below. The formal statement appears in \Cref{sp-wc-j-s-a}.
This result shows that the existing worst-case upper bound of $\tilde{O}(\min\{m, 1/\delta\})$~\cite{page1999pagerank, pagerank98, MC} is tight for all graph parameters $n$ and $m$, and for all values of the threshold parameter $\delta \in (0, 1)$.
In contrast, the previously known lower bound was the folklore result $\Omega(\min\{n, 1/\delta\})$.

\begin{result}[Informal]\label{result:wc-sp-lowerbound-basicqueries}
In the adjacency-list model with all the above graph access queries (and more), the expected computational complexity of estimating $\vpi(s,t)$ for arbitrary nodes $s$ and $t$ is $\Omega\p{\min\curly{m, 1/\delta}}$.
\end{result}

Our second result is a tight lower bound for the single-pair problem in the average case when $\InOrd$ and $\Adj$ are not simultaneously available, as detailed in~\Cref{result:ac-sp-lowerbound-basicqueries}. The formal statement is given in~\Cref{sp-ac-j-s-xor-a}.
This result shows that, in the absence of $\InOrd$ or $\Adj$, the known average-case upper bound of $\tilde{O}(\min\{m, (d/\delta)^{1/2}, \plainfrac{1}{\delta}\})$~\cite{page1999pagerank, pagerank98, BiPPR, FastPPR} is tight for all graph parameters $n$ and $m$, and for all values of $\delta \in (0, 1)$.

Previously, the best known average-case lower bound was $\Omega(n^{1/2})$~\cite{FastPPR}, assuming $\delta = 1/n$. In contrast, under the same setting, our new tight lower bound is $\Omega(m^{1/2})$, representing a quadratic improvement for dense graphs where $m = \Theta(n^2)$.

On the technical side, recall that the earlier $\Omega(n^{1/2})$ lower bound from~\cite{FastPPR} was derived by reducing the single-pair problem to the expansion testing problem on constant-degree graphs~\cite{GoldreichR02}, which left a gap in the degree dependence compared to the upper bound. 
In contrast, all of our tight lower bounds are based on direct constructions of concrete instance families for which no faster algorithm can exist. Given the importance of the problems considered, it is surprising that many of these polynomial gaps can be closed using such simple, direct constructions. 

\begin{result}[Informal]\label{result:ac-sp-lowerbound-basicqueries}
In the adjacency-list model with $\Jump$ and either $\InOrd$ or $\Adj$, but not both, the expected computational complexity averaged over all nodes s and $t$, of estimating $\pi(s,t)$, is $\Omega(\min\{m, (d/\delta)^{1/2}, 1/\delta\})$, where $d=m/n$ is the average degree of the graph.
\end{result}

\paragraph{A new algorithm with tight upper and lower bounds.}
It turns out to be no coincidence that~\Cref{result:ac-sp-lowerbound-basicqueries} does not hold when both $\InOrd$ and $\Adj$ are available: we present a faster algorithm that exploits the combination of these two queries, as detailed in~\Cref{sp-upperbound}. The formal version appears in~\Cref{sp-ac-j-s-a,thm:pair-alg-log-final}.

\begin{result}[Informal]\label{sp-upperbound}
In the adjacency-list model with $\InOrd$ and $\Adj$, the expected computational complexity averaged over all nodes $t$, of estimating $\pi(s,t)$ for an arbitrary node $s$ is $\tilde\Theta(\min\{m, (d/\delta)^{1/2}, (1/\delta)^{2/3}\})$, where $d=m/n$ is the average degree of the graph. The lower bound also holds when allowing $\Jump$ and averaging over all sources $s$. 
\end{result}
Our algorithm is the first to exploit the above query combination. It introduces a novel randomized bidirectional structure that combines randomized backward propagation with selective Monte Carlo estimation.
This result highlights that the combination of $\InOrd$ and $\Adj$ queries offers computational advantages worth considering when designing an application programming interface (API) for large graphs. 

\paragraph{A complete picture of all problems.} Table~\ref{tab:results} summarizes the complexity bounds for the single-pair problem under different query combinations. In addition, we analyze the complexity of related problems (single-source, single-target, and single-node) under both worst-case and average-case scenarios, considering various types of queries. All our results are included in \Cref{tab:results}, which also highlights the gaps between previously known upper and lower bounds. Logarithmic factors are generally omitted. This table provides a comprehensive understanding of the power of different query combinations for estimating random-walk probabilities in directed graphs. As many queries are useful in practice (e.g., $\InOrd$, by using radix sort the neighbor sorting can be performed during graph loading without asymptotically increasing the running time), but are not standard in canonical graph query models, there exists a gap between theory and practice. We hope our results provide insights into the understanding and selection of graph queries.

\begin{table}
    \centering
    \renewcommand{\arraystretch}{1.4}
    \begin{adjustbox}{center}
    \begin{tabular}{|c|c|p{0.2cm}|p{0.2cm}|p{0.2cm}|c|c|c|}
        \hline
        \multirow{2}{*}{\textbf{Problem}} & \multirow{2}{*}{\textbf{Case}} & \multicolumn{3}{c|}{\textbf{Model}} & \multirow{2}{*}{\textbf{Ours}} & \multicolumn{2}{c|}{\textbf{Previously best}} \\ \cline{3-5}\cline{7-8}
        & & \textbf{J} & \textbf{S} & \textbf{A} & & \textbf{Lower} & \textbf{Upper}\\ \hline\hline
        \multirow{4}{*}{\makecell{Single-\\pair}} & Worst & \cboth & \cboth & \cboth & $\tilde\Theta\np{\min\ncurly{m,\plainfrac{1}{\delta}}}$ & \makecell{$\Omega(\min\{n, 1/\delta\})~\star$} & \makecell{$\tilde O\np{\min\ncurly{m,\plainfrac{1}{\delta}}}$\\\cite{page1999pagerank, MC}} \\\cline{2-8}
        & \multirow{3}{*}{Avg.} & \cboth & \cno & \cboth &  \multirow{2}{*}{$\tilde\Theta\np{\min\ncurly{m, \np{\plainfrac{d}{\delta}}^{\plainfrac12}, \plainfrac{1}{\delta}}}$} & \multirow{3}{*}{\makecell{$\Omega(n^{1/2})$ if \\[2pt] $\delta = 1/n$~\cite{FastPPR}}} & \multirow{3}{*}{\makecell{$\tilde O\np{\min\ncurly{m, \np{\plainfrac{d}{\delta}}^{\plainfrac12}, \plainfrac{1}{\delta}}}$\\\cite{page1999pagerank, MC, BiPPR}}} \\ \cline{3-5}
        & & \cboth & \cboth & \cno &  & & \\\cline{3-6}
        & & \cboth & \cyes & \cyes & $\tilde\Theta\np{\min\ncurly{m, \np{\plainfrac{d}{\delta}}^{\plainfrac12}, \np{\plainfrac{1}{\delta}}^{\plainfrac23}}}$ & & \\ \hline\hline
        \makecell{Single-\\source} & N/A & \cboth & \cboth & \cboth & $\tilde\Theta\np{\min\ncurly{m, \plainfrac{1}{\delta}}}$ & 
        $\Omega\np{\min\ncurly{n, \plainfrac{1}{\delta}}}~\star$ 
        & \makecell{$\tilde O\np{\min\ncurly{m, \plainfrac{1}{\delta}}}$~\cite{page1999pagerank}\\ \cite{MC, FORAjournal}} \\ \hline\hline
        \multirow{7}{*}{\makecell{Single-\\target}} & \multirow{3}{*}{Worst} & \cno & \cno & \cboth & $\tilde\Theta\np{m}$ & \multirow{3}{*}{$\Omega(n)~\star$} & \multirow{2}{*}{\makecell{$\tilde O(m)$~\cite{stoc/WangWW024}\\ \cite{BackwardPush_WAW, page1999pagerank}}} \\ \cline{3-6}
        & & \cyes & \cno & \cboth & \multirow{2}{*}{$\tilde\Theta\np{\min\ncurly{m, \plainfrac{n}{\delta}}}$} & & \\ \cline{3-5}\cline{8-8}
        & & \cboth & \cyes & \cboth & & & \makecell{$\tilde O\np{\min\ncurly{m, \plainfrac{n}{\delta}}}$\\ \cite{page1999pagerank, RBS}} \\ \cline{2-8}
        & \multirow{3}{*}{Avg.} & \cno & \cno & \cboth & $\tilde\Theta\np{\min\ncurly{m, \plainfrac{d}{\delta}}}$ & \multirow{3}{*}{\makecell{$\Omega\np{\min\ncurly{n, \plainfrac{1}{\delta}}}$\\\cite{RBS}}} & \multirow{2}{*}{\makecell{$\tilde O\np{\min\ncurly{m, \plainfrac{d}{\delta}}}$\\ \cite{page1999pagerank, Lofgren_backpush}}} \\  \cline{3-6}
        & & \cyes & \cno & \cboth & $\tilde\Theta\np{\min\ncurly{m, \np{\plainfrac{m}{\delta}}^{\plainfrac12}, \plainfrac{d}{\delta}}}$ & & \\ \cline{3-6}\cline{8-8}
        & & \cboth & \cyes & \cboth & $\tilde\Theta\np{\min\ncurly{m, \plainfrac{1}{\delta}}}$ & & \makecell{$\tilde O\np{\min\ncurly{m, \plainfrac{1}{\delta}}}$\\ \cite{page1999pagerank, RBS}} \\ \hline\hline
        \multirow{9}{*}{\makecell{Single-\\node}} & \multirow{4}{*}{Worst} & \cno & \cno & \cboth & $\tilde\Theta(m)$ & $\Omega(n)$~\cite{reversePageRank_Bar-YossefM08} & \multirow{2}{*}{$\tilde O(m)$~\cite{page1999pagerank}} \\ \cline{3-7}
        & & \cno & \cyes & \cboth & $\tilde\Theta\np{n}$ & --- & \\ \cline{3-8}
        & & \cyes & \cno & \cboth & \multirow{2}{*}{$\Theta\np{n^{\plainfrac12}m^{\plainfrac14}}$} & \makecell{$\Omega\np{n^{\plainfrac12}m^{\plainfrac14}}$\\ ~\cite{stoc/WangWW024, soda_Thorup0W026}} & \multirow{2}{*}{\makecell{$O\np{n^{\plainfrac12}m^{\plainfrac14}}$\\ \cite{stoc/WangWW024, soda_Thorup0W026}}} \\ \cline{3-5}\cline{7-7}
        & & \cyes & \cyes & \cboth & & --- & \\ \cline{2-8}
        & \multirow{5}{*}{Avg.} & \cno & \cno & \cboth & $\tilde\Theta\np{m}$ & \multirow{5}{*}{---} & $\tilde{O}(m)$~\cite{page1999pagerank} \\ \cline{3-6}\cline{8-8}
        & & \cno & \cyes & \cboth & $\tilde\Theta\np{n}$ & & $\tilde{O}(n)$~\cite{RBS} \\ \cline{3-6}\cline{8-8}
        & & \cyes & \cboth & \cno & \multirow{2}{*}{$\tilde{\Theta}\np{m^{\plainfrac12}}$} & & \multirow{3}{*}{$\tilde{O}\np{m^{\plainfrac12}}$~\cite{BiPPR}}\\ \cline{3-5}
        & & \cyes & \cno & \cboth & & & \\ \cline{3-6}
        & & \cyes & \cyes & \cyes & $\tilde\Theta\np{\min\ncurly{m^{\plainfrac12}, n^{\plainfrac23}}}$ & & \\ \hline
    \end{tabular}
    \end{adjustbox}
    \caption{Overview of results.
    In the Case column, we indicate whether the given bounds are for a worst-case target node or averaged over all $n$ possible target nodes.
    In the Model column, circles indicate presence or absence of operations.
    The letters J, S, and A are abbreviations of $\Jump$, $\InOrd$, and $\Adj$, respectively.
    A full circle \cyes{} indicates that the operation is present in the model, and an empty circle \cno{} indicates that the operation is absent in the model.
    A half-full circle \cboth{} acts as a wildcard, indicating that the bounds hold both when the operation is present and absent.
    All possible combinations of presence and absence of operations are covered. 
    The results marked with $\star$ refer to folklore results. 
    Entries marked with “—” indicate that we did not find any explicit bounds in the literature. For the single-node problem, we follow the tradition of setting $\delta$ as the smallest possible value of $\pi(t)$ (i.e., $\delta=\alpha/n$). $\tilde{O}$ suppresses logarithmic factors in $n$ and $1/\delta$.
    }
    \label{tab:results}
\end{table}

\subsection{Other related work} \label{subsec:other-related}

In this paper, we focus on arbitrary directed graphs with $n$ vertices and $m$ edges, and parameterize the problem solely in terms of $n$ and $m$. Some prior work considers parameterizations beyond $n$ and $m$ and achieves more refined bounds. For example, a line of work~\cite{BPP18, BPP23, stoc/WangWW024, soda_Thorup0W026} studies the single-node problem parameterized by the maximum in-degree $\Deltain$ and maximum out-degree $\Deltaout$ of the graph. 
A very recent work~\cite{PageRank_TW} studies the instance-optimality of $\BiPPR$ for the single-node problem. 
It establishes the instance-wise complexity of a rich class of graphs whose in- and out-degrees are at most a constant fraction of $n$, and shows that $\BiPPR$ is instance-optimal on this class. We provide a brief description on these refined bounds in~\Cref{subsec:other-subsection}. 

We note that instance-optimality is orthogonal to, say, worst-case complexity. For instance-optimality we prove that an algorithm is best possible on all possible instances, but this does not mean that we know how bad it is in the worst-case or average-case which are complexities defined for the problem. A classic example is the deterministic minimum spanning tree algorithm by \cite{jacm_PettieR02} that they prove is optimal, but we still only know that its worst-case complexity is somewhere between $O(m \alpha(m,n))$\footnote{In this expression $\alpha$ denotes the inverse Ackermann function, whereas it in the rest of the paper is the damping factor of the random walks.} and $\Omega(m)$.

\section{Preliminaries}\label{subsec:average}

\subsection{Notations}
We denote the underlying directed graph as $G = (V, E)$, with $n = |V|$ and $m = |E|$. For each node $v \in V$, we use $\din(v)$ and $\dout(v)$ to denote its in-degree and out-degree, and $\Nin(v)$ and $\Nout(v)$ to denote its sets of in-neighbors and out-neighbors, respectively. The average (in- or out-) degree of $G$ is denoted as $d = m/n$. We define $\delta \in (0, 1)$ as the relative error threshold and use $\tilde{O}$ notation to suppress polylogarithmic factors in $n$ and $\delta$.
A summary of frequently used notation is provided in~\Cref{tbl:def-notation} in~\Cref{sec:table_notations}. Additional notation that appears only in specific sections will be introduced locally as needed.

\subsection{Paper Organization}
We organize the remainder of this paper as follows. In~\Cref{subsec:average}, we present a brief discussion on average-case complexity. In~\Cref{sec:sp_upper} and~\Cref{sec:sp_lower}, we present our main results on tight upper and lower bounds for the single-pair problem, respectively. Then, in~\Cref{sec:single-source},~\Cref{sec:single-target}, and~\Cref{sec:single-node}, we provide our results for the single-source, single-target, and single-node problems, respectively.

 \subsection{Average-case complexity}\label{subsec:average}

Normally we want algorithms that are fast in the worst case for any given graph $G=(V,E)$ with given source $s$ and given target $t$. 
However, interesting algorithms have been developed that are much more efficient when we look at the average running time over all targets $t\in V$. Note that $G$ and $s$ are still worst-case, and the algorithm still has to be probabilistically correct for every $t\in V$.
To distinguish the two cases, we shall refer to the normal case with given $s$ and $t$ as the \emph{worst-case complexity} and the one averaging the running time over all targets as the \emph{average-case complexity}. Formally, let $\mathcal{G}(n,m)$ be the family of all graphs on $n$ nodes and $m$ edges. Then the average-case running time of an algorithm is given by $\max_{\substack{G\in\mathcal{G}(n,m)}} \max_{s\in V} \sum_{t\in V} T_A(G,s,t,\delta)/n,$ where $T_A$ is the running time of an algorithm $A$.
Being efficient on average implies that for every graph, if we look at a random target $t$, then we expect a fast solution, and this may matter more than worst-case in practice.

One could similarly consider the average running time over all sources $s \in V$, either with a worst-case or average-case target $t \in V$. However,
when it comes to the source, it turns out that
the worst-case is no harder than the average-case.
\begin{lemma}
\label{ac-source}
 For the single-pair and single-source problems, the average complexity over all possible sources is the same as the complexity for a given worst-case source. This is for asymptotic complexity in terms of $n$, $m$, and $\delta$, in the  adjacency-list model with any subset of $\Jump$, $\InOrd$, and $\Adj$. In the single-pair case, the equivalence holds both if the target is worst-case and if the target is average-case. 
 
\end{lemma}
\begin{proof}
The proof is based on a simple reduction from the worst case to the average case. Suppose we have an instance of a graph $G$ with $n$ nodes and $m$ edges, a threshold $\delta$, and a worst-case source $s$. We will simulate an algorithm $A'$ with good source average case performance on an new graph $G'$ with a set $S'$ of $n$ new vertices, each with a single out-going edge to $s$. It thus has $n'=2n$ vertices and $m'=n+m$ edges. For any $s'\in S'$, the probability of moving to $s$ is $1-\alpha$ so for any target $t$, we have $\pi_G(s,t)=\pi_{G'}(s',t)/(1-\alpha)$. This also implies that we should use $A'$ with $\delta'=(1-\alpha)\delta$.

The basic idea is that we just pick a random new source $s'\in S'$ and simulate $A'$ on $G'$ with source $s'$. Since $S'$ has half the nodes in $G'$, the average run-time for $A'$ on sources in $S'$ is at most twice its average run-time over all vertices as sources. Using a random $s'\in S$ yields this expected run-time for our worst-case $s$ in $G$.

To get a fixed run-time with worst-case source $s$, let $T'(n',m',\delta')$ be the average run-time of $A'$ on $G'$ and assume that the error probability of $A'$ is independent of its actual running time. We know that $A'$ run at most 4 times slower than the overall average on half the vertices in $S'$. We now pick a random sample $U$ from $S'$, and run $A'$ on all $s'\in U$ in parallel, returning the first estimate found, or giving up after $4|U|T'(n',m',\delta')$ total time. For constant error probability it suffices that $U$ has constant size. 
\end{proof}

\section{Single-Pair Upper Bounds}\label{sec:sp_upper}

We study the single-pair problem in this section, focusing on the upper-bound analysis. 
Formally, we say that an algorithm solves \emph{the single-pair problem} if for every given graph $G=(V,E)$, source $s \in V$, target $t \in V$, and approximation threshold $\delta \in (0,1]$, the algorithm outputs an estimate $\epi(s,t)$ of $\vpi(s,t)$ that is \emph{probabilistically correct} in the sense that 
\begin{align}\label{eqn:def-singlepair}
    \Pr\left\{ \left| \epi(s,t) - \vpi(s,t) \right| \ge \rela \max\{\vpi(s,t), \delta\} \right\} \leq \pf,
\end{align}
where $\rela$ and $\pf$ are small constants.

\subsection{Known upper bounds}\label{subsec:ku-singlepair}

We first review known algorithms for the single-pair problem. The purpose is twofold: one reason is that we are going to use some of their techniques as a starting point for our own upper-bound algorithm, and the other is to showcase the algorithms that we will match with our lower bounds. \Cref{thm:wc-single-pair} and~\Cref{thm:avg-single-pair} summarizes the known upper bounds established by existing algorithms. We will cover these results in this subsection.

\begin{lemma}[\cite{page1999pagerank, MC}]\label{thm:wc-single-pair}
In the adjacency-list model, the worst-case complexity of the single-pair problem is $\tilde{O}\left(\min\{m, 1/\delta\}\right)$.
\end{lemma}

\begin{lemma}[\cite{BiPPR, MC, page1999pagerank}]\label{thm:avg-single-pair}
In the adjacency-list model, the average-case complexity of the single-pair problem is $\tilde{O}(\min\{\sqrt{d/\delta}, 1/\delta, m\})$.
\end{lemma}

\subsubsection{Monte Carlo simulation}\label{subsubsec:MC}
A canonical approach to estimating $\pi(s, t)$ is Monte Carlo simulation, which generates multiple $\alpha$-discounted random walks starting from $s$ and uses the fraction of walks that terminate at $t$ as an estimate of $\pi(s, t)$. 
By standard concentration bounds, we can estimate $\pi(s,t)$ with constant relative error using $\Theta(1/\pi(s,t))$ independent walks.
Since the expected length of each walk is $1/\alpha = O(1)$, and we only need constant relative error when $\vpi(s,t)>\delta$, this method achieves running time $O(1/\delta)$.

Monte Carlo simulation can also be used to address the \emph{single-source} problem: given a target node $t$, estimate $\pi(s,t)$ for all source nodes $s$.

\subsubsection{The $\Push$ operation}\label{subsubsec:backpush}

Whereas Monte Carlo simulation explores the graph in a \emph{forward} direction from $s$, another line of research~\cite{BackwardPush_WAW, BackwardPush_InternetMath, Lofgren_backpush, RBS, page1999pagerank, pagerank98} estimates $\vpi(s,t)$ by exploring the graph in a \emph{backward} direction from the given target node $t$.
A key operation used in these works is called $\Push$~\cite{BackwardPush_InternetMath}.
It maintains two variables for each node $v \in V$: the \emph{reserve} $p(v)$ and the \emph{residue} $r(v)$.
Here, the reserve $p(u)$ serves as an underestimate of $\pi(u,t)$, while the residues collectively capture the error of these estimates. Formally $\Push$ maintains the invariant for all $u \in V$.
\begin{equation}\label{eqn:push_invariant}
    \pi(u,t) = p(u)+\sum_{v\in V}\pi(u,v)r(v). 
\end{equation}
At initialization, $p(v)$ and $r(v)$ are set to $0$ for all $v \in V$, except for $r(t) = 1$, so the invariant is trivially satisfied. A $\Push$ operation on a node $v$ transfers an $\alpha$-fraction of the residue $r(v)$ to the reserve $p(v)$, pushes the remaining residue to the residues of $v$’s in-neighbors, and sets $r(v)$ to $0$, as detailed in~\Cref{alg:push}. 
One can verify that the invariant in~\eqref{eqn:push_invariant} is consistently maintained before and after every $\Push$ on any node $v$, using the following property of discounted random walks: 
\begin{align}\label{eqn:onestep_walkproperty}
    \pi(u, v)
    = \sum_{x \in \Nout(u)}\frac{(1-\alpha)\pi(x,v)}{\dout(u)} + \alpha\Ind{u = v}
    = \sum_{y \in \Nin(v)}\frac{(1-\alpha)\pi(u, y)}{\dout(y)} + \alpha\Ind{u = v}, 
\end{align}
where $\Ind{u = v}$ is the indicator variable that takes the value $1$ when $u = v$ and 0 otherwise.

\begin{algorithm}[h]
    \caption{$\Push(v)$ }
    \label{alg:push}
    \textbf{Input:} node $v$\\
    \textbf{Output:} updated $p()$ and $r()$
    \begin{algorithmic}[1]
        \State $\residue \gets \residue(v)$
        \State $\residue(v)\gets 0$
        \State $\reserve(v)\gets \reserve(v) + \alpha \residue$
        \For{each $u\in \Nin(v)$}
            \State $\residue(u) \gets \residue(u) + (1-\alpha)\residue / \dout(u)$
        \EndFor
    \State \Return $\reserve()$ and $\residue()$
    \end{algorithmic}
\end{algorithm}

\subsubsection{$\Push$ with threshold}
According to the invariant~\eqref{eqn:push_invariant}, if we continue performing $\Push$ on all nodes with $r(v) < \pushthreshold$, where $\pushthreshold \in (0, 1)$ is a predefined threshold parameter, then after terminating $\Push$, the reserve $p(s)$ will be an approximation of $\pi(s,t)$ within an additive error of $\pushthreshold$:
\[
\pi(s,t) - p(s) = \sum_{v \in V} \pi(s,v) r(v) < \pushthreshold \sum_{v \in V} \pi(s,v) = \pushthreshold,
\]
where we used the fact that $\sum_{v \in V} \pi(s,v) = 1$. By setting $\pushthreshold = \rela \delta$, we ensure that $p(s)$ is within an $\rela$ relative error when $\pi(s,t) \ge \delta$, deterministically.

This approach was introduced in~\cite{BackwardPush_InternetMath}, called $\bpush$, originally proposed to address the \emph{single-target} problem: given a target node $t$, estimate $\pi(s,t)$ for all source nodes $s$.
Its running time is bounded by $O\left(\sum_{v\in V}\frac{\pi(s,t)\din(v)}{\pushthreshold}\right)$. 
A recent work~\cite{stoc/WangWW024} shows that the running time can be further bounded as $O\left(\frac{n\pi(t)m^{1/2}}{\pushthreshold}\right)$, where $\pi(t)=\frac{1}{n}\sum_{u\in V}\pi(u,t)$.
However, 
$\pi(t)$ can be as large as a constant, making this bound no better than $O(m/\pushthreshold)$.
But when we shift focus to the average running time over $t$, the bound becomes:
\begin{align}\label{eqn:avg_time_push}
    \frac{1}{n} \sum_{t \in V} \sum_{v \in V} \frac{\vpi(v,t)\din(v)}{\alpha \pushthreshold} 
    = \frac{1}{\alpha n \pushthreshold} \sum_{v \in V} \din(v)\sum_{t \in V} \vpi(v,t)
    = \frac{1}{\alpha n \pushthreshold} \sum_{v \in V} \din(v)
    = \frac{m}{\alpha n \pushthreshold} = O\left(\frac{d}{\pushthreshold}\right).
\end{align}
This establishes the average-case complexity of $O(d/\delta)$ when setting $\pushthreshold =\rela \delta$~\cite{Lofgren_backpush}. 

\subsubsection{$\Push$ with Monte Carlo simulation}
A set of methods, e.g., $\BiPPR$~\cite{BiPPR} and $\FastPPR$~\cite{FastPPR}, estimate $\vpi(s,t)$ by combining $\Push$ with Monte Carlo simulations based on the invariant~\eqref{eqn:push_invariant}. These methods first perform $\Push$ on all vertices $v$ with $r(v)\ge \pushthreshold$ for some predefined threshold parameter $\pushthreshold \in (0,1)$. When no such vertex exists, Monte Carlo simulations are invoked to compute an approximation $\tpi(s,v)$ for $\pi(s,v)$ for all $v\in V$. The approximation $\epi(s,t)$ for $\pi(s,t)$ is then computed as follows:
\begin{align*}
\epi(s,t)=p(s) + \sum_{v\in V}\tpi(s,v) r(v),
\end{align*}
where $r(v)$ and $p(v)$ represent the residues and reserves after the $\Push$ phase. It was shown in~\cite{BiPPR} that $O(\pushthreshold/\delta)$ random walks are sufficient to ensure that $\epi(s,t)$ satisfies~\Cref{eqn:def-singlepair}. Combining this with the $O(d/\pushthreshold)$ average running time of $\Push$ (as given in~\Cref{eqn:avg_time_push}), we establish the average-case complexity of $O(\pushthreshold/\delta + d/\pushthreshold) = O(\sqrt{d/\delta})$ as shown in~\Cref{thm:avg-single-pair}, when setting $\pushthreshold = \sqrt{\delta d}$.

\subsubsection{$\Push$ by level}\label{subsubsec:pw}
The $\Push$ operations can also be performed synchronously~\cite{page1999pagerank}, rather than selectively on the vertices $v$ with $r(v) \ge \pushthreshold$. In this case, $\Push$ is applied to every vertex $v \in V$, and this process is repeated for $L$ rounds. The estimate $\epi(s,t)$ for $\pi(s,t)$ is then computed as the sum of reserves of $s$ in all the $L$ rounds, i.e., $\epi(s,t)=\sum_{i=0}^L p_i(s)$, where $p_i(s)$ denotes the reserve of $s$ after round $i$. This approach is referred to as $\pw$, and its pseudocode is provided in~\Cref{alg:push_by_level} 
in~\Cref{subsec:app_notation_code}. 
We note that the values of the residues used in each round decrease geometrically, and after $L$ rounds, the residues for all vertices become smaller than $(1-\alpha)^L$. Therefore, setting $L = \log_{1-\alpha} \rela \delta = O(\log(1/\delta))$ suffices to ensure that the estimate satisfies~\Cref{eqn:def-singlepair}. Since each round can touch all edges in the graph, this approach establishes a linear complexity bound of $O(m \log(1/\delta)) = \tilde{O}(m)$, as shown in~\Cref{thm:wc-single-pair} and~\Cref{thm:avg-single-pair}.

\subsubsection{Randomized $\Push$ by level}\label{subsubsec:RBS}
\cite{RBS} proposes a randomized variant of the aforementioned $\pw$ method called the $\RBS$ method, which utilizes the $\InOrd$ query. At the core of $\RBS$ is a randomized $\Push$ operation. 
For a vertex $v$ in round $i$ of the push process, this randomized $\Push$ increases $\residue_i(u)$ deterministically only when its increment, $\frac{(1-\alpha)\residue_{i-1}(v)}{\dout(u)}$, exceeds a predefined threshold $\rmax\in (0,1)$. Otherwise, it increases $\residue_i(u)$ by $\theta$ with probability $\frac{(1-\alpha)\residue_{i-1}(v)}{\dout(u)\theta}$.
Making a random decision for each edge would take time $\Omega(\din(v))$ which we want to avoid.
But since the increment to $\residue(u)$ is inversely proportional to $u$’s out-degree, the algorithm uses the $\InOrd$ query to scan the sorted in-adjacency list of $v$ from top to bottom and terminate the scan upon encountering an in-neighbor $u$ for which $\frac{(1-\alpha)\residue(v)}{\dout(u)}$ falls below a random threshold, pushing $\theta$ for each scanned edge, which was not deterministically pushed.
This approach was originally proposed for the single-target problem, and for the single-pair it establishes a worst-case complexity of $\tilde{O}(n/\delta)$ and an average-case complexity of $\tilde{O}(1/\delta)$, both of which are weaker than the $O(1/\delta)$ established by the Monte Carlo simulation. 
Nonetheless, this randomized $\Push$ operation will serve as the starting point for our upper-bound algorithm, as detailed in~\Cref{subsec:upperbound_single-pair}.

\subsection{Our upper bounds}\label{subsec:upperbound_single-pair}
\label{sec:new-pair-alg}
This subsection presents our algorithm for solving the single-pair problem in the adjacency-list model with both $\InOrd$ and $\Adj$ queries.
Our algorithm runs in $\tilde{O}\p{(1/\delta)^{2/3}}$ expected time averaged over all possible targets $t$. By combining this with the $O\p{(d/\delta)^{1/2}}$ complexity achieved by combining $\Push$ with Monte-Carlo simulations~\cite{BiPPR} and the $\tilde{O}(m)$ complexity achieved by $\Push$ by level (both are described in~\Cref{subsec:ku-singlepair}), we obtain the following theorem: 
\begin{theorem}
\label{thm:pair-alg-log-final}
There exists an algorithm estimating $\vpi(s,t)$ in $\tilde{O}(\min\{m, (d/\delta)^{1/2}, (1/\delta)^{2/3}\})$ average expected time in the adjacency-list model with $\InOrd$ and $\Adj$.
\end{theorem}
We note that the above upper bound (ignoring logarithmic factors) matches the lower bound established in~\Cref{sp-ac-j-s-a}, demonstrating the optimality of our result. 

\subsubsection{Technical Overview}
Let us first take an overview of the main ideas and techniques we used in our algorithm and proofs. For ease of understanding, \Cref{alg:meta} shows a (overly) simplified structure of our algorithm. In the remaining subsections, our novel ideas and techniques will be combined into the basic structure step by step.
The pseudocode of the complete algorithm can be found in~\cref{subsec:app_notation_code}.
Briefly speaking, our algorithm is a novel combination of the randomized $\Push$ technique and the Monte Carlo simulation. 

Our randomized $\Push$ process is divided into $L$ levels, where $L = O(\log(1/\delta))$. Different from previous algorithms, only when the residue $\residue_i(v)$ of a node $v$ at level $i$ is larger than a predefined threshold $\rmax_i$ for level $i$, we push it and update the residues of nodes $u \in \Nin(v)$ at level $i+1$. We utilize the $\InOrd$ operation to sample in-neighbors $u$ with probabilities inversely proportional to their out-degrees. At each level $i$, we update $\residue_i(v)$ only for the sampled $u$. 

We show that our randomized $\Push$ maintains a “pseudo-invariant”, which in expectation matches the invariant~\eqref{eqn:push_invariant} maintained by the deterministic $\Push$. This “pseudo-invariant” serves as the basis for combining the randomized $\Push$ with Monte Carlo simulations. 

Importantly, the approximation error introduced by our randomized $\Push$ is too large, if we want to combine it with Monte Carlo sampling. To address this, we design a better estimator $R(v)$ for each $v$, which is a partially derandomized version of the (randomized) residues of vertex $v$ derived in randomized $\Push$. Substituting $R(v)$ into the pseudo-invariant, we can more tightly control the approximation error of the estimate for $\vpi(s,t)$ using standard concentration inequalities.

However, it is difficult to compute $R(v)$ directly, as this would introduce a degree term $d$ in the complexity bound.
Instead, we are going to do the following.
We will sample $O((1/\delta)^{1/3})$ nodes in the 
Monte Carlo simulation. For each sampled node $v$, we actually only compute an estimator $\erh(v)$ for $R(v)$. For this, we use the $\Adj$ query to collect the contributions from the $O((1/\delta)^{1/3})$ heaviest vertices, i.e. the vertices with highest reserves. As each other vertex only contributes a small proportion to our final estimator, sampling another $O((1/\delta)^{1/3})$ out-neighbors will be enough to bound the error within the acceptable threshold. In total, we spend $O((1/\delta)^{1/3})$ time on each of the $O((1/\delta)^{1/3})$ Monte Carlo sampled vertices, so the total time is $O((1/\delta)^{2/3})$.

\begin{algorithm}[H]
    \caption{$\spppr(s, t, L, \nr, \rmax_i, \gap_{i})$}
    \label{alg:meta}
    \begin{algorithmic}[1]
        \State $\rh_0(t) \gets 1$.
        \For{$i=0,1,2,\dots,L-1$}
            \For{each $v\in V$ with $\rh_i(v)>\rmax_i$}
                % \State $\ranPush(\rh_i(v), \rmax_i, \gap_{i})$.
                \State $\ranPushthres(v, i, \rmax_i, \gap_{i})$ $\quad$ \textcolor{gray}{// invoke~\Cref{alg:random-push}.}
            \EndFor
        \EndFor
        \State Start a random walk from $s$, and suppose it ends at some vertex $u$. 
        \State $q(s,t)\gets \ph(s)+\rh(u)$. 
        \textcolor{gray}{// $\rh$ will eventually be substituted with a partly derandomized version, $\erh$, to ensure bounded approximation error.}
        \State \Return the average of $\nr$ independent copies of $q(s,t)$ as the final estimate of $\vpi(s,t)$.
    \end{algorithmic}
\end{algorithm}

\header{\bf Notation.} We now define some additional notations to analyze our upper-bound algorithm. Specifically, we define several variables with subscript $i$ denoting level $i$ (e.g. $\rh_i(v)$ and $\rmax_i$). For simplicity, for each of them, we use the same symbol without the subscript to denote the sum of that variable over all levels. For example, $\rh(v)=\sum_{i=0}^L\rh_i(v)$ and $\rmax=\sum_{i=0}^L\rmax_i$. Unless otherwise specified, all variables used in this subsection are initialized to $0$.

\subsubsection{Randomized $\Push$ with threshold}
\label{subsec:random-push}

The first part of our algorithm is performing randomized $\Push$ on every node $v$ with $\rh_i(v) > \rmax_i$ at every level $i \in \{0, 1, \ldots, L-1\}$, where $\rmax_i$ is a predefined threshold parameter. 
In each randomized $\Push$ on $v$ at level $i$, we update $\rh_{i+1}(u)$ for a node $u \in \Nin(v)$ only if its increment $\dr_{i+1}(u,v)= \frac{(1-\alpha)\rh_i(v)}{\dout(u)}$ exceeds a predefined threshold $\gap_{i+1}\rmax_{i+1}$, where $\gap_{i+1}$ is a small enough parameter to be determined later. Otherwise, each $u \in \Nin(v)$ is sampled with probability $\frac{\dr_{i+1}(u,v)}{\gap_{i+1}\rmax_{i+1}}$, and only for those sampled $u$, $\rh_{i+1}(u)$ is increased by $\gap_{i+1}\rmax_{i+1}$. The pseudocode of performing a randomized $\Push$ operation on node $v$ at level $i$ is given in~\Cref{alg:random-push}. 
We assume $\dr_0(t,t)=1$ for simplicity of analysis.

\begin{algorithm}[h]
    \caption{$\ranPushthres(v, i, \rmax_i, \gap_{i})$}
    \label{alg:random-push}
    \begin{algorithmic}[1]
        \For{each $u\in \Nin(v)$}
            \State $\dr_{i+1}(u,v)\gets \frac{(1-\alpha)\rh_i(v)}{\dout(u)}$.
            \If{$\dr_{i+1}(u,v)\ge\gap_{i+1}\rmax_{i+1}$}
                \State $\rh_{i+1}(u)\gets \rh_{i+1}(u)+\dr_{i+1}(u,v)$.
            \Else
                \State $\rh_{i+1}(u)\gets \rh_{i+1}(u)+\gap_{i+1}\rmax_{i+1}$ with probability $\frac{\dr_{i+1}(u,v)}{\gap_{i+1}\rmax_{i+1}}$.
            \EndIf
        \EndFor
        \State $\ph(v)\gets \ph(v)+\alpha \rh_i(v)$.
        \State $\rh_i(v)\gets 0$.
    \end{algorithmic}
\end{algorithm}

The following pseudo-invariant is maintained before and after each randomized push. We will use this pseudo-invariant as the basis to combine the results from randomized $\Push$ and Monte Carlo simulation. 
\begin{align*}
\ph(s) + \sum_{u \in V} \pi(s,u) \rh(u) = \pi(s,t).
\end{align*}
We refer to this as a pseudo-invariant because it only holds in expectation. We formalize this result in~\Cref{lem:exp-inv}.

\begin{lemma}
\label{lem:exp-inv}
    For each $w \in V$, the following equality holds consistently before and after each invocation of~\Cref{alg:random-push}.
\begin{align}\label{eqn:exp-invariant}
\E{\ph(w)+\sum_{u\in V}\pi(w,u)\rh(u)}=\pi(w,t).    
\end{align}
\end{lemma}

\begin{proof}
The equality holds in the initial state, where $\rh(t) = 1$ and $\rh(u) = 0$ for all $u \ne t$. Our goal is to show that this equality remains valid after each invocation of~\Cref{alg:random-push}.
Let us consider the change of the left-hand side of equation~\eqref{eqn:exp-invariant} after executing~\Cref{alg:random-push} from node $v$ at level $i$. We note that $\ph(v)$ increases by $\alpha\rh_i(v)$, $\rh(v)$ decreases by $\rh_i(v)$, and for all $u\in\Nin(v)$, $\rh(u)$ increases by $\frac{(1-\alpha)\rh_i(v)}{\dout(u)}$ in expectation. As a result, the left-hand side of equation~\eqref{eqn:exp-invariant} changes by
\begin{align*}
\Ind{w = v} \alpha \rh_i(v) + \sum_{u \in V} \pi(w, u) \frac{(1 - \alpha)\rh_i(v)}{\dout(u)} - \pi(w, v) \rh_i(v),
\end{align*}
which is equal to zero by~\Cref{eqn:push_invariant}. This shows that equation~\eqref{eqn:exp-invariant} is preserved in expectation after each call to~\Cref{alg:random-push}.
\end{proof}

In the following, we analyze the expected time cost of~\Cref{alg:random-push}. We show that, with access to the $\InOrd$ query, a randomized push can be executed from a node $v$ in time proportional to the actual number of sampled nodes $u \in \Nin(v)$, rather than $\din(v)$ as required by the standard $\Push$ operation. A formal statement is provided in~\Cref{lem:random-push-time}. While similar time-cost analysis has appeared in~\cite{RBS}, our proof incorporates the threshold parameters $\rmax_i$ and $\gap_i$. We present our proof here for completeness.

\begin{lemma}
\label{lem:random-push-time}
    \cref{alg:random-push} can be implemented in $O\p{\frac{\sum_{u\in \Nin(v)}\dr_{i+1}(u,v)}{\gap_{i+1}\rmax_{i+1}}+1}$ expected time.
\end{lemma}

\begin{proof}
Let us consider a randomized push operation from a node $v$ at level $i$. We observe that $\dr_{i+1}(u,v)$ for each $u \in \Nin(v)$ is inversely proportional to $\dout(u)$. To implement the sampling, we first generate a uniformly random number $\rannum \in [0, \gap_{i+1}\rmax_{i+1}]$, and then use the $\InOrd$ query to visit the in-neighbors of $v$ in non-decreasing order of their out-degrees $\dout(u)$, stopping once we encounter a node $u \in \Nin(v)$ with $\dr_{i+1}(u,v) \le \rannum$. In this way, we visit only the sampled nodes $u \in \Nin(v)$ and one additional node to terminate the process. The lemma then follows directly. 
\end{proof}

It is worth noting that the above implementation guarantees unbiasedness in sampling, but not independence. Each increment to $\rh_{i+1}(u)$ for $u \in \Nin(v)$ is unbiased, with an expected value equal to $\dr_{i+1}(u,v)$. However, since all increments are determined using a shared random number $\rannum$, they are not mutually independent. Nevertheless, we will show that this sampling scheme is sufficient for our subsequent analysis. 

Furthermore, \Cref{lem:pushback-time} provides an upper bound on the total time cost of performing randomized $\Push$ in~\Cref{alg:meta} (i.e., Lines 1–4). The proof of~\Cref{lem:pushback-time} is deferred to~\cref{app:pushback-time}. 

\begin{lemma}
\label{lem:pushback-time}
Let $\rmaxx$ denote a lower bound such that $\gap_i\rmax_i\ge\rmaxx$ for all $i$. The expected time cost of performing the backward exploration in~\Cref{alg:meta} is upper bounded by $O\left(\frac{n\pi(t)}{\alpha \rmaxx}\right)$. 
\end{lemma}

By~\Cref{lem:pushback-time}, we observe that achieving the anticipated $\tilde{O}\left((1/\delta)^{2/3}\right)$ time complexity stated in~\Cref{thm:pair-alg-log-final} requires setting $\rmaxx \ge \delta^{2/3}$. However, in a randomized $\Push$ operation on a node $v$ at level $i$, the increment to $\rh_{i+1}(u)$ may deviate from its expected value by up to $\gap_i\rmax_i$. This can lead to an additive error of $O\left(\gap_i\rmax_i\right)$ between the estimated value $\epi(s,t)$ computed by~\Cref{alg:meta} and the true value $\vpi(s,t)$ in the worst case. As a result, to ensure a $(1 \pm O(1))$-multiplicative approximation when $\vpi(s,t) = \delta$, as required by equation~\eqref{eqn:def-singlepair}, we would need to set $\rmaxx \le \delta$, which contradicts the earlier requirement.

To resolve this conflict, in the following subsection, we introduce a substitute variable $\er(u)$ for $\rh(u)$ and show that the approximation error can be reduced by replacing $\rh(u)$ with $\er(u)$ in the computation of $\epi(s,t)$ in~\Cref{alg:meta}.

\subsubsection{An ideal estimator}
\label{subsec:ideal-estimator}

As shown in~\Cref{alg:meta}, after completing the backward exploration phase (i.e., Lines 1–4), we compute $\rh(u)$ for the terminal node $u$ of each of the $\nr$ random walks. To reduce the approximation error introduced by $\rh(u)$, we construct a “derandomized” version $\er(u)$ of $\rh(u)$ as follows.
\begin{definition}
For each $u\in V$, 
\begin{align*}
\er(u)&=\sum_{i=0}^{L}\nopush_{i}(u)\er_i(u),\\
\text{where} \quad \er_i(u)&=\sum_{v\in\Nout(u)}\dr_{i}(u,v). 
\end{align*}
\end{definition}
In the above, $\dr_{i}(u,v)$ is the value computed by~\Cref{alg:random-push} from node $v$ at level $i-1$, and $\nopush_i(u) = [\rh_{i}(u) \le \rmax_i]$ is an indicator variable that equals $1$ if we never perform randomized $\Push$ on $u$ at level $i$ during the entire push process (i.e., the condition $\rh_i(u) \le \rmax_i$ holds at the checkpoint shown in Line 3 of~\Cref{alg:meta}).
Intuitively speaking, $\er_i(u)$ is the expectation of $\rh_i(u)$ before $\rh_i(u)$ was pushed. After pushing $\rh_i(u)$, we would set $\rh_i(u)$ to $0$, and $\nopush_i(u)$ would also be $0$.

Ideally, we would like to ensure that $\er(u) = \E{\rh(u)}$. However, this equality does not hold because $\nopush_i(u)$ and $\rh_i(u)$ are not independent. 
To resolve this issue, each time we invoke~\Cref{alg:random-push} from a node $v$ at level $i-1$, we additionally generate an independent copy $\rhh_i(u)$ of $\rh_i(u)$, and use $\rhh_i(u)$, rather than $\rh_i(u)$, to determine whether to push $\rh_i(u)$ (i.e., substituting the push condition in Line 3 of~\Cref{alg:meta} from $[\rh_i(u) > \rmax_i]$ to $[\rhh_i(u) > \rmax_i]$). 
Consequently, the definition of $\nopush_i(u)$ is updated as:
\begin{align*}
\nopush_i(u)=[\rhh_{i}(u)\le\rmax_{i}]. 
\end{align*}
In this way, we have $R(u)= \E{\rh(u)}$ for any $u\in V$, and the following invariant holds for $R(u)$. 

\begin{lemma}\label{lem:invariant_R}
    The following equality holds consistently before and after each invocation of~\Cref{alg:random-push}: 
    $$\E{\ph(s)+\sum_{u\in V}\pi(s,u)\er(u)}=\pi(s,t).$$
\end{lemma}

\begin{proof}
    Given \Cref{lem:exp-inv}, it suffices to show that $\E{\nopush_i(u)R_i(u)}=\E{\rh_i(u)}$ for any $u$ and $i$. Note that $\rh_i(u)=0$ when $\nopush_i(u)=0$. On the other hand, given $\nopush_i(u)=1$ and $\rh_{i-1}(v)$ for all $v\in V$, we have
    $$\E{\rh_i(u)}=\sum_{v\in\Nout(u)}\dr_i(u,v).$$
    Comparing it with the definition of $R_i(u)$ completes the proof.
\end{proof}

In $\er(u)$, there is still some randomness in $\dr_i(u,v)$ from previous rounds. However, this randomness is actually on (the out-edges of) $\rh_i(v)$ which has been pushed (that means $\rhh_i(v)>\rmax_i$). Since the random variables are bounded by $\gap_i\rmax_i$, if $\gap_i$ is small enough, $\rhh_i(v)>\rmax_i$ infers that $\er_i(u), \rh_i(u)$ and $\rhh_i(u)$ are close to each other with high probability. Then, all errors introduced during the randomized $\Push$ process can be viewed as small relative errors independent of $\rmax_i$. See \cref{app:gap-1} for the detailed proof.

\begin{lemma}
\label{lem:gap-1}
    There exists a constant $C$ such that, for any $\rela\le1$, if $\gap_i\le C\rela^2/\log(nL)$ for all $i$, then with high probability, throughout the whole backward exploration process, whenever we decide to push $\rh_i(u)$, we have $|\rh_i(u)-\er_i(u)|\le \rela \er_i(u)$. 
\end{lemma}

Based on \cref{lem:gap-1}, we can obtain the following concentration bound by examining how the value changes from rounds to rounds. See \cref{app:err-1} for the detailed proof.

\begin{lemma}
\label{lem:err-1}
    There exists a constant $C$ such that, for any $\rela\le1$, if $\gap_i\le C\rela^2/(L^2\log(nL))$ for all $i$, then with high probability, $|\ph(s)+\sum_{u\in V}\pi(s,u)\er(u)-\pi(s,t)|\le \rela\pi(s,t)$.
\end{lemma}

\subsubsection{The number of random walks}
\label{subsec:concentration-2}

Now we move to the random walk part. Let's temporarily pretend that we can compute the exact $R(u)$ and see how many random walks we need. Recall that for each random walk, if we stop at vertex $u$, we estimate $\pi(s,t)$ by $q(s,t)=\ph(s)+R(u)$. We take the average of $\nr$ independent copies of $q(s,t)$ as the final estimator $\pit(s,t)$.
In this subsection, we are going to show that $\pit(s,t)$ is a good estimator of our invariant.

First, It is easy to see that $\pit(s,t)$ is unbiased.

\begin{lemma}
\label{lem:pit-exp}
    $\E{\pit(s,t) \mid \ph(s),\{\er(u)\}_{u\in V}}=\ph(s)+\sum_{u\in V}\pi(s,u)\er(u)$.
\end{lemma}

\begin{proof}
    Each $q(s,t)$ is unbiased since the random walk stops at each vertex $u$ with probability $\pi(s,u)$. Then $\pit(s,t)$ is unbiased.
\end{proof}

When we finish the push process, we know that $\rhh_i(u)\le\rmax_i$ for any $u\in V$ and $0\le i<L$, because otherwise it should be pushed. Similar to \cref{lem:gap-1}, as long as $\gap_i$ is small, it also indicates that $\er_i(u)$ is bounded with high probability. The detailed proof is given in \cref{app:gap-2}.

\begin{lemma}
\label{lem:gap-2}
    There exists a constant $C$ such that, if $\gap_i\le C/\log(nL)$ for all $i$, then with high probability, for all $u\in V$ and $0\le i<L$ such that $\rh_i(u)$ is not pushed, we have $R_i(u)\le 2\rmax_i$.
\end{lemma}

On the other hand, notice that the residues are multiplied by $(1-\alpha)$ at each level when pushing, which means even though we never push at level $L$, $R_L(u)$ can still be bounded. The detailed proof is given in \cref{app:last-level}.

\begin{lemma}
\label{lem:last-level}
    There exist constants $C_1,C_2$ such that, if $L\ge C_1\log(1/\rmax_L)/\alpha$ and $\gap_i\le C_2/(L^2\log(nL))$ for all $i$, then with high probability, $R_L(u)\le\rmax_L$ for all $u\in V$.
\end{lemma}

Combining the above lemmas, we know that with high probability, all $R(u)$ can be bounded by $2\rmax$, which means $q(s,t)-\ph(s)$ is a random variable in $[0,2\rmax]$. Then we can obtain the following concentration bound by applying Chernoff bounds. The detailed proof is given in \cref{app:err-2}.

\begin{lemma}
\label{lem:err-2}
    There exist constants $C_1,C_2,C_3$ such that, for any $\rela\le1$, if $L\ge C_1\log(1/\rmax_L)/\alpha$, $\gap_i\le C_2/(L^2\log(nL))$ for all $i$ and $\nr\ge C_3\rmax\log(1/\pf)/(\rela\delta)$, then with probability $1-\pf$, $|\pit(s,t)-\p{\ph(s)+\sum_{u\in V}\pi(s,u)\er(u)}| \le \rela\max\{\delta,\ph(s)+\sum_{u\in V}\pi(s,u)\er(u)\}$. 
\end{lemma}

\subsubsection{The real estimator}
\label{subsec:concentration-3}

Finally, the only missing part is how to compute $\er(u)$. Note that $\pit(s,t)$ can be written as:
$$\pit(s,t)=\ph(s)+\frac{1}{\nr}\sum_{k=1}^{\nr}\er(u_k),$$
where $u_k$ is the destination of the $k$-th random walk. If we directly compute $R(u)$ by definition, we need to go through its out-neighbors. To avoid introducing $d$ to our time complexity, we actually compute an estimator $\erh(u_k)$ of each\footnote{We may have $u_{k_1}=u_{k_2}$. In this case we still compute $\erh(u_{k_1})$ and $\erh(u_{k_2})$ separately to make sure they are independent (given $\{R(u)\}_{u\in V}$).} $\er(u_k)$, resulting in:
$$\pitt(s,t)=\ph(s)+\frac{1}{\nr}\sum_{k=1}^{\nr}\erh(u_k).$$
The idea is, each out-neighbor of $u_k$ has some contribution to $\er(u_k)$. For the out-neighbors whose contributions are small, we only need to sample some of them to estimate their total contribution. On the other hand, if a neighbor $v$ has a large contribution, it must have a large $\ph(v)$, since we must have pushed a lot of residue from $v$. Since $\ph(v)$ is at most $\pi(v,t)$, the number of such vertices can be bounded. 
Therefore, we first leverage $\Adj$ to efficiently compute the contributions from out-neighbors $v$ with $\ph(v) > \pth$, where $\pth$ is a predefined threshold parameter. We then sample $\ns$ nodes from the remaining out-neighbors to estimate their total contributions. The pseudocode for computing $\erh(u_k)$ is provided in~\Cref{alg:estimate-R}.

\begin{algorithm}[H]
    \caption{Compute $\erh(u_k)$}
    \label{alg:estimate-R}
    \begin{algorithmic}[1]
        \State $\erh(u_k) \gets 0$.
        \For{each $v\in \Vp$} $\quad$ \textcolor{gray}{\small // The set $\Vp$ contains all nodes $v$ in $G$ with $\ph(v)>\pth$.}
            \If{$(u_k,v)\in E$}
                \State $\erh(u_k)\gets \erh(u_k)+\sum_i \nopush_i(u_k)\dr_i(u_k,v)$.
            \EndIf
        \EndFor
        
        \For{$j=1,2,\dots,\ns$}
            \State $v_j\gets$ a uniformly random vertex in $\Nout(u_k)\setminus \Vp$.
            \State $\erh(u_k)\gets \erh(u_k)+\frac{|\Nout(u_k)\setminus \Vp|}{\ns} \sum_i \nopush_i(u_k)\dr_i(u_k,v_j)$.
        \EndFor

        \State \Return $\erh(u_k)$.
    \end{algorithmic}
\end{algorithm}

\begin{lemma}
\label{lem:estimate-R-time}
    Each $\erh(u_k)$ can be computed in $O\p{\ns L+\frac{n\pi(t)L}{\pth}}$ time.
\end{lemma}

\begin{proof}
    $\Vp$ can be easily computed as a list during backward exploration, and $|\Vp|\le \frac{n\pi(t)}{\pth}$ since
    $$\sum_{v\in V}\ph(v) \le \sum_{v\in V}\pi(v,t) = n\pi(t).$$
    Line $6$ can be simply done in constant time if $|\Nout(u)\setminus \Vp|\ge|\Vp|$; Otherwise we can traverse $\Nout(u)$ in $O(|\Vp|)$ time. In total, we visit $O\p{\ns+\frac{n\pi(t)}{\pth}}$ out-neighbors, and for each of them, we use $O(L)$ time to go through all levels.
\end{proof}

Here is our last concentration bound. The proof is again basically Chernoff bounds. See \cref{app:err-3} for the detailed proof.

\begin{lemma}
\label{lem:err-3}
    There exists a constant $C$ such that, for any $\rela\le 1$, if $\nr\ns/\pth \ge C\log(1/\pf)/(\alpha\min\{\delta,\rela\})$, then with probability $1-\pf$, $|\pitt(s,t)-\pit(s,t)| \le \rela\max\{\delta,\pit(s,t)\}$.
\end{lemma}

\subsubsection{Putting everything together}
\label{subsec:pair-alg-together}

Now we have everything we need for an $\tilde{O}((1/\delta)^{2/3})$ time algorithm. \cref{lem:err-1,lem:err-2,lem:err-3} guarantees the error probability, and \cref{lem:pushback-time,lem:estimate-R-time} tells us the time complexity.

\begin{theorem}
\label{thm:pair-alg-log}
    In $\tilde{O}((1/\delta)^{2/3})$ time, we can compute $\pitt(s,t)$ such that with probability at least $1-\pf$, $|\pitt(s,t)-\pi(s,t)| \le \rela\max\{\delta,\vpi(s,t)\}$, for any constants $\pf,\rela\in(0,1)$.
\end{theorem}

\begin{proof}
    Combining \cref{lem:err-1,lem:err-2,lem:err-3,lem:pushback-time,lem:estimate-R-time}, we can get the desired concentration bound in time
    $$O\p{\frac{n\pi(t)}{\alpha\rmaxx}+\nr\p{\frac{1}{\alpha}+\ns L+\frac{n\pi(t)L}{\pth}}},$$
    with the following constraints for the parameters:
    \begin{enumerate}
        \item $\gap_i\rmax_i \ge \rmaxx$ for each level $i$;
        \item $\gap_i=O(\rela^2/(L^2\log(nL)))$ for each level $i$;
        \item $L=\Omega(\log(1/\rmax_L)/\alpha)$;
        \item $\nr=\Omega(\rmax\log(1/\pf)/(\rela\delta)$;
        \item $\nr\ns/\pth=\Omega(\log(1/\pf)/(\alpha\rela\delta))$.
    \end{enumerate}
    Recall that $\E{n\pi(t)}=1$ for a uniformly random target node $t$.

    Setting $\rmax_i=\Theta\p{\delta^{2/3}}$ for all $i$ and $L=\Theta\p{\frac{\log(1/\delta)}{\alpha}}$ satisfies the third constraint. Then, the second constraint suggests that $\gap_i=\Theta\p{\frac{\rela^2\alpha^2}{\log^2(1/\delta)\log(nL)}}$ for all $i$. The first constraint is satisfied by $\rmaxx=\Theta\p{\frac{\delta^{2/3}\rela^2\alpha^2}{\log^2(1/\delta)\log(nL)}}$. On the other hand, $\rmax=\sum_{i}\rmax_i=\Theta\p{\frac{\delta^{2/3}\log(1/\delta)}{\alpha}}$, so $\nr=\Theta\p{\frac{\log(1/\delta)\log(1/\pf)}{\delta^{1/3}\rela\alpha}}$ satisfies the fourth constraint. Finally, the fifth constraint is satisfied by $\ns=1/P=\Theta\p{\frac{1}{\delta^{1/3}}}$.
    Then the expected time complexity is
    $$O\p{\frac{\log^2(1/\delta)\log(nL/\pf)}{\delta^{2/3}\rela^2\alpha^3}} = \tilde{O}\p{(1/\delta)^{2/3}}$$
    for a uniformly random target node $t$.
\end{proof}

\section{Single-Pair Lower Bounds}\label{sec:sp_lower}
We establish our lower bounds for the single-pair problem in this section. 

\subsection{Known lower bounds}\label{subsec:kl-singlepair}

In this subsection, we will briefly review known lower bounds for the single pair problem. The previously best worst-case lower bound is $\Omega(\min\{n, 1/\delta\})$, derived by the following simple folklore argument.
We construct a graph consisting of a source node $s$ with $\min\{n,1/\delta\}$ out-neighbors and a target node $t$ with $\min\{n,1/\delta\}$ in-neighbors and a self-loop, as in \Cref{fig:sp-folklore}. 
With probability $1/2$, we add an edge from a random out-neighbor of $s$ to a random in-neighbor of $t$.
If the extra edge is added, we have $\pi(s,t) \geq \delta$, and otherwise $\pi(s,t) = 0$.
The algorithm must therefore determine whether the extra edge was added, so a deterministic algorithm must look at a constant fraction of the nodes.
We conclude by applying Yao's minimax principle~\cite{DBLP:conf/focs/Yao77}.

\begin{figure}[h]
    \centering
    \begin{tikzpicture}[
  style1/.style={circle, draw, inner sep=0, minimum size=12pt, line width=0.5pt},
  style2/.style={-latex, line width=0.5pt},
  >=latex
  ]

    \node[style1] (s) at (0, 0) {$s$};
    \node[style1] (v1) at (-0.75, -0.75) {};
    \node[style1] (v2) at (-0.25, -0.75) {};
    \node at (0.25, -0.75) {$\cdots$};
    \node[style1] (v3) at (0.75, -0.75) {};
    \foreach \i in {1,...,3} {
        \draw[style2] (s) -- (v\i);
    }

    \node[style1] (u1) at (-0.75, -1.5) {};
    \node[style1] (u2) at (-0.25, -1.5) {};
    \node at (0.25, -1.5) {$\cdots$};
    \node[style1] (u3) at (0.75, -1.5) {};
    \node[style1] (t) at (0, -2.25) {$t$};
    \draw[style2] (t) edge [loop below] (t);
    \foreach \i in {1,...,3} {
        \draw[style2] (u\i) -- (t);
    }
    \draw[style2, color=blue] (v1) -- (u2);

\end{tikzpicture}
    \caption{Current best hard instance for the worst-case single-pair problem.}
    \label{fig:sp-folklore}
\end{figure}

The previous average-case lower bound is $\Omega(n^{1/2})$~\cite[Theorem 4]{FastPPR}, which assumes $\delta = 1/n$~\cite{FastPPR}.
The proof is based on a reduction to the property testing problem of distinguishing between an expander graph and a graph consisting two disjoint expanders.

\subsection{Our lower bounds}\label{subsec:lowerbound_single-pair}

We are now ready to present our lower bounds for the single-pair problem, starting with the worst-case lower bound.
Loosely speaking, we construct a graph with an upper and lower component, where the upper component makes forward exploration costly, and the lower component makes backward exploration costly.
This result proves that the previously best upper bound of $O(\min\{m, 1/\delta\})$ described in \Cref{thm:wc-single-pair} is already tight.

\begin{theorem}\label{sp-wc-j-s-a}
    Consider the adjacency-list model with $\Jump$, $\InOrd$ and $\Adj$.
    For any $n$ and $m$ with $n \leq m \leq n^2$ and any $\delta \in (0,1]$, there exists a graph $G=(V,E)$ with $\Theta(n)$ nodes, $\Theta(m)$ edges, and nodes $s,t \in V$, such that for any algorithm solving the single-pair problem, the expected running time on $G$ with source $s$, target $t$, and approximation threshold $\delta$ is $\Omega(\min\{m, 1/\delta\})$.
\end{theorem}
\begin{proof}
    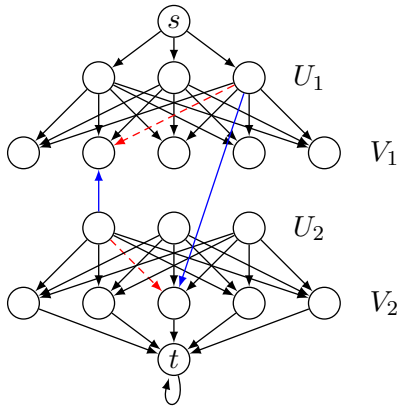
\begin{figure}[h]
        \centering
        \begin{tikzpicture}[
  style1/.style={circle, draw, inner sep=0, minimum size=12pt, line width=0.5pt},
  style2/.style={-latex, line width=0.5pt},
  >=latex
  ]
  \pgfmathsetmacro{\n}{9};
  \pgfmathsetmacro{\L}{3}; % Should be divisor of \n
  \pgfmathsetmacro{\D}{5}; % Prettiest if odd
  \pgfmathsetmacro{\labelOffset}{.3}
  \pgfmathsetmacro{\layerZero}{0}
  \pgfmathsetmacro{\layerOne}{\layerZero-0.75}
  \pgfmathsetmacro{\layerTwo}{\layerOne-1}
  \pgfmathsetmacro{\layerThree}{\layerTwo-1}
  \pgfmathsetmacro{\layerFour}{\layerThree-1}
  \pgfmathsetmacro{\layerFive}{\layerFour-0.75}
  \node[style1] (s) at (0, \layerZero) {$s$};
  \node[style1] (t) at (0, \layerFive) {$t$};
  \foreach \i in {1,...,\L} {
    \pgfmathsetmacro{\xi}{\i-1-(\L-1) / 2}
    \node[style1] (u1\i) at (\xi, \layerOne) {};
    \node[style1] (u2\i) at (\xi, \layerThree) {};
    \draw[style2] (s) -- (u1\i);
  }
  \foreach \i in {1,...,\D} {
    \pgfmathsetmacro{\xi}{\i-1-(\D-1) / 2}
    \node[style1] (v1\i) at (\xi, \layerTwo) {};
    \node[style1] (v2\i) at (\xi, \layerFour) {};
    \draw[style2] (v2\i) -- (t);
  }
  \foreach \i in {1,...,\L} {
    \foreach \j in {1,...,\D} {
      \ifthenelse{\i = 3 \AND \j = 2}{\draw[style2, densely dashed, color=red] (u1\i) -- (v1\j)}{\draw[style2] (u1\i) -- (v1\j)};
      \ifthenelse{\i = 1 \AND \j = 3}{\draw[style2, densely dashed, color=red] (u2\i) -- (v2\j)}{\draw[style2] (u2\i) -- (v2\j)};
    }
  }
  \draw[style2, color=blue] (u13) -- (v23);
  \draw[style2, color=blue] (u21) -- (v12);
  \draw[style2] (t) edge [loop below] (t);
  \node (U1) at (\L/2+\labelOffset, \layerOne) {$U_{1}$};
  \node (V1) at (\D/2+\labelOffset, \layerTwo) {$V_{1}$};
  \node (u2) at (\L/2+\labelOffset, \layerThree) {$U_{2}$};
  \node (v2) at (\D/2+\labelOffset, \layerFour) {$V_{2}$};
\end{tikzpicture}
        \caption{Hard instance for the worst-case single-pair problem.
        With the {\color{red} red} edge pair, $s$ does not reach $t$, but with the  {\color{blue} blue} edge pair, $s$ does reach $t$.
        An algorithm has to distinguish between these two cases, and because of the regular structure, this essentially means that it has to check a constant fraction of the edges.}
        \label{fig:sp-wc}
    \end{figure}

    The proof is sketched in Figure \ref{fig:sp-wc}.
    Technically, the proof will flow roughly as follows.
    We will construct, independent of the algorithm, a graph $G$ and a family of similar modified graphs.
    To be correct on both $G$ and all the modified graphs (together with a certain source and target), the algorithm must distinguish $G$ from all these modified graphs.
    No two modified graphs will differ from $G$ in the same location, so the algorithm, when given $G$, must perform at least one query per modified graph, yielding us the desired lower bound.
    Note that this deviates from the more usual approach of constructing an input distribution and applying Yao's minimax principle (e.g.\ as sketched in~\Cref{subsec:kl-singlepair}).
    Notably, unlike Yao's minimax principle, our approach provides a single instance $G$, that is hard for all correct algorithms.\footnote{One can convert our proof into a (weaker) argument based on Yao's minimax principle.
    To see this, choose an input distribution that with constant probability sends $G$, and otherwise sends a random choice from the family of modified graphs.
    We avoid such an approach, as it seems to fail for the average-case construction (\Cref{sp-ac-j-s-xor-a,sp-ac-j-s-a}), where the input distribution is not allowed to depend on $t$.}
    
    In more detail, let us construct the graph $G = (V, E)$.
    First, we let the node set $V$ be the disjoint union of sets $\{s\}$, $U_{1}$, $V_{1}$, $U_{2}$, $V_{2}$, and $\{t\}$.
    We give these sets sizes $\abs{U_{1}} = \abs{U_{2}} = L$ and $\abs{V_{1}}=\abs{V_{2}} = D$, where $L$ and $D$ are parameters to be set later.
    We construct the edge set $E$ as follows: $s$ has an edge to every node in $U_{1}$; each node in $U_{1}$ has an edge to every node in $V_{1}$; each node in $U_{2}$ has an edge to every node in $V_{2}$; each node in $V_{2}$ has an edge to $t$; and $t$ has a self-loop.
    See~\Cref{fig:sp-wc} for an illustration, which also includes a \emph{swap} as introduced below.
    Let $E_{i}$ denote the subset of edges from $U_{i}$ to $V_{i}$ for $i \in \{1, 2\}$.
    To ensure a well-defined construction, we will ensure $L \geq 1$ and $D \geq 1$ when setting $L$ and $D$.
    To satisfy $\abs{V} = O(n)$ and $\abs{E} = O(m)$, we will ensure $L \leq n$, $D \leq n$, and $LD \leq m$.
    To satisfy $\abs{V} = \Omega(n)$ and $\abs{E} = \Omega(m)$, we add an isolated subgraph with $n$ nodes and $m$ edges.

    Note that $\InOrd$ is no different from $\In$ in $G$, since for every node $v$, the in-neighbors of $v$ all have the same out-degree.
    
    Let $A$ be a deterministic algorithm, deriving an estimate $\hat\pi(s,t)$ of $\pi(s,t)$.
    We say that $A$ is \emph{correct} if the estimate has error $\abs{\hat\pi(s,t)-\pi(s,t)} < \epsilon\max\{\pi(s,t), \delta\}$.
    In particular, if $\pi(s,t) = 0$, it must hold that $\hat\pi(s,t) < \epsilon\delta$.
    If on the other hand $\pi(s,t) \geq \delta$, it must hold that $\hat\pi(s,t) > (1-\epsilon)\delta$.
    Since $\rela$ is a small constant as mentioned in equation~\eqref{eqn:def-singlepair}, we assume $\rela \le 1/2$. This means that $A$ distinguishes $\pi(s,t) = 0$ from $\pi(s,t) \geq \delta$ if $A$ is correct. 
    This is the only property of the estimate, that our lower bound will employ.
    
    Clearly, $\pi(s,t) = 0$ in $G$, and we will now introduce a modified graph $G'$ where $\pi(s,t) \geq \delta$.
    We construct $G'$ by performing what we call a \emph{swap} on two edges $e_{1}=(u_{1},v_{1}) \in E_{1}$ and $e_{2}=(u_{2},v_{2}) \in E_{2}$.
    We will pick these two edges in the next paragraph.
    To perform the swap, we delete $e_{1}$ and $e_{2}$, and insert the edges $(u_{1}, v_{2})$ and $(u_{2}, v_{1})$ instead.
    The resulting graph $G'$ is illustrated in~\Cref{fig:sp-wc}, where the deleted edges $e_{1}$ and $e_{2}$ are drawn as red, dashed arrows, and the inserted edges $(u_1, v_2)$ and $(u_2, v_1)$ are drawn as blue arrows.
    We now have $\pi(s,t) = (1-\alpha)^{3}/(LD)$ in $G'$, as can be verified using equation~\eqref{eqn:onestep_walkproperty}.
    We will later set $L$ and $D$ such that $\pi(s,t) \geq \delta$ in $G'$.
    Note that the number of vertices and edges, as well as the out-degree and in-degree of each node is the same before and after the swap.
    We can also preserve the ordering of neighbors in the adjacency lists.
    This means that if $A$ does not query any of the edges of the swap in $G$, (through an $\In$, $\Out$, or $\Adj$ query) then $A$ will also not query any edges of the swap in $G'$.
    If so, the behavior of $A$ is unchanged whether it is given $G$ or $G'$, and in particular, the output will be the same.
    As a correct algorithm must distinguish between $G$ and $G'$, we get that $A$ is incorrect on $G$ or $G'$, unless it queries an edge of the swap.

    The general idea of this proof is that an algorithm must determine whether a swap has been performed, and with the models considered, this means that the algorithm either has to check a constant fraction of the edges in $E_1$ or $E_2$.
    This will now be formalized.    
    Let $R$ be a randomized algorithm deriving an estimate $\hat\pi(s,t)$ of $\pi(s,t)$.
    Formally, $R$ is a random variable over deterministic algorithms.
    We assume that $R$ is incorrect with probability at most $p_{f} < 1/2$.
    Let $Q$ be the set of edges and non-edge node pairs queried by $R$ through $\In$, $\Out$ and $\Adj$ queries. 
    For $e_1 = (u_1, v_1) \in E_1$ and $e_2 = (u_2, v_2) \in E_2$, define $q(e_1, e_2) = \{(u_1, v_1), (u_2, v_2), (u_1, v_2), (u_2, v_1)\}$.
    Then $q(e_1, e_2)$ represents the ``quadrangle'' of edges deleted or inserted during a swap on $e_1$ and $e_2$ (the quadrangle formed by red and blue edges in~\Cref{fig:sp-wc}).
    Assume for the sake of contradiction, that there exist edges $e_{1} \in E_{1}$ and $e_{2} \in E_{2}$ such that $\P{q(e_1, e_2) \cap Q \neq \emptyset} < 1/2$.
    Then pick these edges for our swap when constructing $G'$ above.
    Denote by $R(H)$ the output of $R$ on a graph $H$.
    Then $\P{R(G) = R(G')} \geq \P{q(e_1, e_2) \cap Q = \emptyset} \geq 1/2$.
    This contradicts $R$ being incorrect with probability at most $p_{f} < 1/2$, so we can assume that $\P{q(e_1, e_2) \cap Q \neq \emptyset} \geq 1/2$ for every $e_{1} \in E_{1}$ and $e_{2} \in E_{2}$.
    Enumerating $U_1$ and $V_2$, let $\varphi \colon E_1 \to E_2$ be the injection sending the $j$th out-edge of the $i$th node of $U_1$ to the $i$th in-edge of the $j$th node of $V_2$.
    Note that the sets $q(e, \varphi(e))$ are disjoint for different $e \in E_1$.
    We now have
    \begin{equation*}
        \E{\abs{Q}}
        \geq \sum_{(u,v) \in V \times V}\P{(u,v) \in Q}
        \geq \sum_{e \in E_1}\P{q(e, \varphi(e)) \cap Q \neq \emptyset}
        \geq \abs{E_1}/2
        = LD/2.
    \end{equation*}
    So $R$ uses $\Omega(LD)$ queries in expectation.
    
    We now set the parameters $L$ and $D$.
    In future proofs, we will give $L$ and $D$ separate values, but for now, set $L = D = ((1-\alpha)^{3}\min\{m, 1/\delta\})^{1/2}$.
    We can assume $(1-\alpha)^{3}\min\{m, 1/\delta\} \geq 1$, as otherwise the~\lcnamecref{sp-wc-j-s-a} is trivial. 
    Note that $1 \leq L = D \leq m^{1/2} \leq n$, $1 \leq LD \leq m$, and $\pi(s,t) \geq \max\{1/m, \delta\} \geq \delta$, as promised.
    We conclude a lower bound of $\Omega(LD) = \Omega(\min\{m, 1/\delta\})$ queries.
\end{proof}

We now present an average-case lower bound for the single pair problem, i.e. averaging over all $n$ possible target nodes.
Our construction will be similar to our worst-case construction, although now with $n$ possible targets joined in a number of groups.
Increasing the group size will increase the cost of backward exploration, but also decrease the probability of terminating at the target.
Likewise, increasing the cost of forward exploration will decrease the probability of terminating at the target.
This leads to a bidirectional tradeoff in our lower bound, which was not present in the worst case, interestingly matching the tradeoff between forward and backward exploration in bidirectional algorithms like \FastPPR~\cite{FastPPR} and \BiPPR~\cite{BiPPR}---algorithms which we hereby show are optimal, unless both $\InOrd$ and $\Adj$ are available.

\begin{theorem}\label{sp-ac-j-s-xor-a}
    Consider the adjacency-list model with $\Jump$ and either $\InOrd$ or $\Adj$, but not both.
    For any $n$ and $m$ with $n \leq m \leq n^2$ and any $\delta \in (0,1]$, there exists a graph $G=(V,E)$ with $\Theta(n)$ nodes and $\Theta(m)$ edges, such that for any algorithm solving the single pair problem, the expected running time on $G$ with approximation threshold $\delta$, averaging over all sources $s \in V$ and targets $t \in V$, is $\Omega(\min\{m, (d/\delta)^{1/2}, 1/\delta\})$, where $d=m/n$.
\end{theorem}

\begin{proof}
    \begin{figure}[h]
        \centering
        \begin{tikzpicture}[
    style1/.style={circle, draw, inner sep=0, minimum size=12pt, line width=0.5pt},
    style2/.style={-latex, line width=0.5pt},
    >=latex
    ]
    \pgfmathsetmacro{\n}{9};
    \pgfmathsetmacro{\L}{3}; % Should be divisor of \n
    \pgfmathsetmacro{\D}{5}; % Prettiest if odd
    \pgfmathsetmacro{\labelOffset}{.3}
    \pgfmathsetmacro{\layerOne}{0};
    \pgfmathsetmacro{\layerTwo}{\layerOne-0.75};
    \pgfmathsetmacro{\layerThree}{\layerTwo-1};
    \pgfmathsetmacro{\layerFive}{\layerThree-1};
    \pgfmathsetmacro{\layerSix}{\layerFive-2};
    \pgfmathsetmacro{\layerSeven}{\layerSix-0.75};
    \pgfmathsetmacro{\layerEight}{\layerSeven-0.75};
    \node[style1] (s) at (0, \layerOne) {$s$};
    \foreach \i in {1,...,\L} {
        \pgfmathsetmacro{\xi}{\i-1-(\L-1) / 2}
        \node[style1] (v1\i) at (\xi, \layerTwo) {};
        \draw[style2] (s) -- (v1\i);
    }
    \foreach \i in {1,...,\D} {
        \pgfmathsetmacro{\xi}{\i-1-(\D-1) / 2}
        \node[style1] (w1\i) at (\xi, \layerThree) {};
    }
    \foreach \i in {1,...,\n} {
        \pgfmathsetmacro{\xi}{\i-1-(\n-1) / 2}
        \node[style1] (w2\i) at (\xi, \layerFive) {};
        \node[style1] (v2\i) at (\xi, \layerSix) {};
        \node[style1] (u2\i) at (\xi, \layerEight) {};
        \draw[style2] (u2\i) edge [loop below] (u2\i);
    }
    \pgfmathsetmacro{\g}{int(\n / \L)}
    \foreach \i in {1,...,\g} {
        \pgfmathsetmacro{\xi}{(\i-1/2)*\L - \n/2}
        \node[style1] (g\i) at (\xi, \layerSeven) {};
        \foreach \j in {1,...,\L} {
            \pgfmathsetmacro{\k}{int((\i-1)*\L+\j)}
            \draw[style2] (v2\k) -- (g\i);
            \draw[style2] (g\i) -- (u2\k);
        }
    }
    \foreach \i in {1,...,\L} {
        \foreach \j in {1,...,\D} {
            \ifthenelse{\i = 2 \AND \j = 4}{
                \draw[style2, densely dashed, color=red] (v1\i) -- (w1\j)
            }{
                \draw[style2] (v1\i) -- (w1\j)
            };
        }
    }
    \foreach \i in {1,...,\n} {
        \foreach \j in {1,...,\D} {
            \pgfmathsetmacro{\k}{int(mod(\i+\j-1+\n-\D/2, \n)+1)}
            \ifthenelse{\i = 8 \AND \k = 6}{
                \draw[style2, densely dashed, color=red] (w2\i) -- (v2\k)
            }{
                \draw[style2] (w2\i) -- (v2\k)
            };
        }
    }
    \draw[style2, color=blue] (v12) -- (v26);
    \draw[style2, color=blue] (w28) -- (w14);

    \node (V1) at (\L/2+\labelOffset, \layerTwo) {$U_{1}$};
    \node (W1) at (\D/2+\labelOffset, \layerThree) {$V_{1}$};
    \node (W1) at (\n/2+\labelOffset, \layerFive) {$U_{2}$};
    \node (V2) at (\n/2+\labelOffset, \layerSix) {$V_{2}$};
    \node (X)  at (\n/2+\labelOffset, \layerSeven) {$X~$};
    \node (U2) at (\n/2+\labelOffset, \layerEight) {$W_{2}$};
\end{tikzpicture}
        \caption{Hard instance for the average-case single-pair problem.
        With the {\color{red} red} edge pair, $s$ does not reach any $t \in W_2$, but with the  {\color{blue} blue} edge pair, $s$ does reach every $t$ in the appropriate group of $W_2$.
        An algorithm has to distinguish between these two cases, and because of the regular structure, this essentially means that it has to check a constant fraction of the edges from the upper component or a constant fraction of the edges into the appropriate group of $V_2$.}
        \label{fig:sp-ac}
    \end{figure}
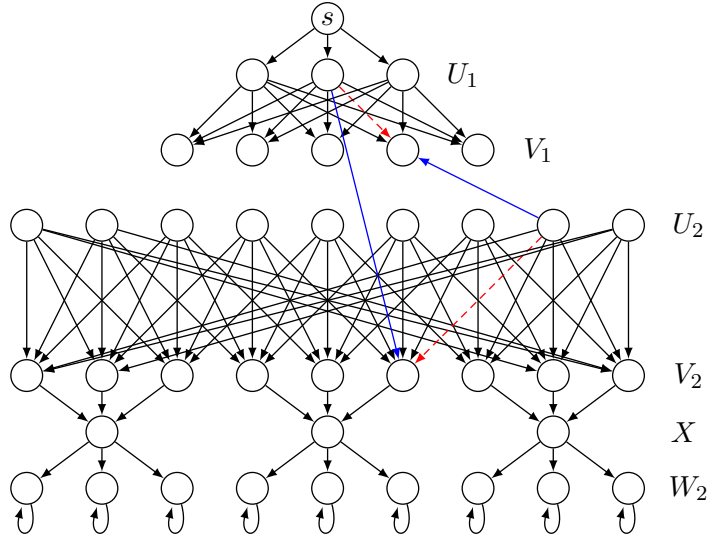

    By~\Cref{ac-source}, it suffices to prove the lower bound for a worst-case source $s$, averaging only over targets $t$.
    The proof is sketched in Figure \ref{fig:sp-ac}.
    Let us construct the graph $G = (V, E)$.
    We will actually use $\Theta(n)$ nodes and $\Theta(m)$ edges.
    First, we let the node set $V$ be the disjoint union of sets $\{s\}$, $U_{1}$, $V_{1}$, $U_{2}$, $V_{2}$, $X$, and $W_{2}$. 
    We give these sets sizes $\abs{U_{1}} = L$, $\abs{V_{1}} = D$, $\abs{U_{2}} = \abs{V_{2}} = \abs{W_{2}} = n$ and $\abs{X} = n/L$ where $L$ and $D$ are parameters to be set later.
    We form a family of subsets $\{\mathcal V_{1}, \ldots, \mathcal V_{n/L}\}$ (resp. $\{\mathcal W_{1}, \ldots, \mathcal W_{n/L}\}$) partitioning $V_{2}$ (resp. $W_{2}$) into subsets of size $L$, and enumerate the nodes of $X = \{x_{1}, \ldots, x_{n/L}\}$.
    For each $i \in \{1, \ldots, n/L\}$, we refer to $\mathcal V_{i} \cup \{x_{i}\} \cup \mathcal W_{i}$ as a \emph{group}.
    We construct the edge set $E$ as follows: $s$ has an edge to every node in $U_{1}$; each node in $U_{1}$ has an edge to every node in $V_{1}$; each node in $U_{2}$ has $D$ edges to $V_{2}$, such that each node in $V_{2}$ has in-degree $D$; for each $i \in \{1, \ldots, n/L\}$ each node in $\mathcal V_{i}$ has an edge to $x_i$ which has an edge to every node in $\mathcal W_{i}$; and each node in $W_{2}$ has a self-loop.
    See~\cref{fig:sp-ac} for an illustration, which also includes a swap, as in the proof of~\Cref{sp-wc-j-s-a}.
    Note that the upper component is the same as in our worst-case construction.
    To ensure a well-defined construction, we will ensure $L \geq 1$ and $D \geq 1$.
    To satisfy $\abs{V} = O(n)$ and $\abs{E} = O(m)$, we will ensure $L \leq n$ and $D \leq d$.
    To satisfy $\abs{V} = \Omega(n)$ and $\abs{E} = \Omega(m)$, we add an isolated subgraph with $n$ nodes and $m$ edges.

    Since $W_2$ contains a constant fraction of the nodes in $G$, it suffices to show the claimed lower bound for the graph $G$, averaging over all targets $t$ in $W_2$.
    So fix a target $t \in \mathcal W_g$ for some $g$.
    Let $E_1$ be the set of edges from $U_1$ to $V_1$, and let $E_2$ be the set of all edges from $U_2$ to $\mathcal V_g$.
    If we perform a swap on any $e_1 \in E_1$ and $e_2 \in E_2$ as in the proof of~\Cref{sp-wc-j-s-a}, we get a modified graph $G'$, where $\pi(s,t) = (1-\alpha)^4/(L^2D)$. 
    When setting $L$ and $D$, we will ensure that $\pi(s,t) \geq \delta$, so an algorithm must distinguish between $G$ and $G'$. 

    We start by handling the case where $\InOrd$ is present and $\Adj$ is absent.
    Note that $\InOrd$ is no different from $\In$ in $G$, since for every node $v$, the in-neighbors of $v$ all have the same out-degree.
    Let $R$ be a randomized algorithm solving the single pair problem with failure probability $p_f < 1/2$.
    Let $Q$ be the set of edges queried by $R$ through $\In$ and $\Out$ queries.
    Then for any $e_1 \in E_1$ and $e_2 \in E_2$, we get analogously to the proof of~\Cref{sp-wc-j-s-a}, that assuming $\P{\{e_1, e_2\} \cap Q \neq \emptyset} < 1/2$ leads to a contradiction by performing a swap on $e_1$ and $e_2$.
    So we have $\P{\{e_1, e_2\} \cap Q \neq \emptyset} \geq 1/2$ for all $e_1 \in E_1$ and $e_2 \in E_2$.
    Note that while we considered the quadrangle $q(e_1, e_2)$ in~\Cref{sp-wc-j-s-a}, we only worry about $\{e_1, e_2\}$ here, as the algorithm does not have access to $\Adj$ here.
    Enumerating $U_1$ and $\mathcal V_i$, let $\varphi \colon E_1 \to E_2$ be the injection sending the $j$th out-edge of the $i$th node of $U_1$ to the $j$th in-edge of the $i$th node of $\mathcal V_g$.
    Note that the sets $\{e, \varphi(e)\}$ are disjoint for different $e \in E_1$.
    We now have
    \begin{equation*}
        \E{\abs{Q}}
        \geq \sum_{(u,v) \in V \times V}\P{(u,v) \in Q}
        \geq \sum_{e \in E_1}\P{\{e, \varphi(e)\} \cap Q \neq \emptyset}
        \geq \abs{E_1}/2
        = LD/2.
    \end{equation*}
    So $R$ uses $\Omega(LD)$ queries in expectation. 

    Before setting our parameters $L$ and $D$, let us also show a lower bound of $\Omega(LD)$ for the case when $\InOrd$ is absent and $\Adj$ is present.
    In this case, we modify our construction of $G$, setting instead $\abs{U_1} = D$ and $\abs{V_1} = L$.
    Now $\InOrd$ is not the same as $\In$, but we need not worry in this case.
    This change does not affect $\pi(s,t)$ in $G$ or $G'$.
    Let $\varphi \colon E_1 \to E_2$ be the injection sending the $j$th in-edge of the $i$th node of $V_1$ to the $j$th in-edge of $i$th node of $\mathcal V_g$. 
    Defining $Q$ and $q$ as in the proof of~\Cref{sp-wc-j-s-a}, note that the sets $q(e, \varphi(e))$ are again disjoint for different $e \in E_1$, so we again get $\E{\abs{Q}} \geq \sum_{e\in E_1}\P{q(e, \varphi(e)) \cap Q \neq \emptyset} \geq \abs{E_1}/2 = LD/2$, i.e.\ a lower bound of $\Omega(LD)$. 

    We now set our parameters, casing on the minimum term among $m$, $(d/\delta)^{1/2}$ and $1/\delta$.
    In each case, it is easy to check that $1 \leq L \leq n$, $1 \leq D \leq d$, and $\pi(s,t) \geq \delta$, as promised.
    Let $c = (1-\alpha)^4 = O(1)$  and note that we can assume $cn \geq 1$ and $c/\delta \geq 1$ as otherwise the~\lcnamecref{sp-ac-j-s-xor-a} is trivial. 

    \emph{Case 1:} For $0 < \delta \leq \frac{1}{nm}$, set $L = cn$ and $D = d$, giving a lower bound of $\Omega(m)$.

    \emph{Case 2:} For $\frac{1}{nm} \leq \delta \leq \frac{c}{d}$, set $L = (c/(d\delta))^{1/2}$ and $D = d$, giving a lower bound of $\Omega((d/\delta)^{1/2})$.

    \emph{Case 3:} For $\frac{c}{d} \leq \delta \leq 1$, set $L = 1$ and $D = c/\delta$, giving a lower bound of $\Omega(1/\delta)$.
\end{proof}

Comparing the above lower bound with~\Cref{thm:avg-single-pair} reveals that our lower bound is tight. Finally, when both $\InOrd$ and $\Adj$ are available, we derive the following lower bound, which we will later show to be tight.

\begin{theorem}\label{sp-ac-j-s-a}
    Consider the adjacency-list model with $\Jump$, $\InOrd$ and $\Adj$.
    For any $n$ and $m$ with $n \leq m \leq n^2$ and any $\delta \in (0,1]$, there exists a graph $G=(V,E)$ with $\Theta(n)$ nodes and $\Theta(m)$ edges, such that for any algorithm solving the single pair problem, the expected running time on $G$ with approximation threshold $\delta$, averaging over all sources $s \in V$ and targets $t \in V$, is $\Omega(\min\{m, (d/\delta)^{1/2}, (1/\delta)^{2/3}\})$, where $d=m/n$.
\end{theorem}
\begin{proof}
    By~\Cref{ac-source}, it suffices to prove the lower bound for a worst-case source $s$, averaging only over targets $t$.
    Construct $G$ as in the proof of~\Cref{sp-ac-j-s-xor-a}, and note again that $\InOrd$ is no different than $\In$.
    Once again, it suffices to show the lower bound for a given $t \in \mathcal W_g$ for a given $g$.
    Enumerate each set $U_1$, $V_1$, $U_2$, $V_2$ and $\mathcal V_g$ from 0 to the size of the set minus one.
    For each $i$, write $U_1(i)$ for the $i$th node in $U_1$, and write similarly for the other sets. 
    Our enumeration of $V_2$ and $\mathcal V_g$ should respect $\mathcal V_g(i) = V_2((g-1)L+i)$ for all $i \in \{0, \ldots, n/L-1\}$.
    In our construction $G$, explicitly set $\Nin(V_2(i)) = \{U_2(i), U_2((i+1)\bmod n), \ldots, U_2((i+D-1)\bmod n)\}$ for each $i$.
    This allows us to define $\varphi \colon E_1 \to E_2$ by $\varphi((U_1(i), V_1(j)) = (U_2(((g-1)L+((i+j)\bmod L)+j) \bmod n), \mathcal V_g((i+j) \bmod L))$ for each $i$ and $j$. 
    Let $E_1' = U_1 \times V_1'$, where $V_1'$ is the set of the first $\min\{L, D\}$ nodes of $V_1$.
    Define $Q$ and $q$ as in the proof of~\Cref{sp-wc-j-s-a}.
    Noting that the sets $q(e, \varphi(e))$ are disjoint for different $e \in E_1'$, we similarly get
    \begin{align*} 
        \E{\abs{Q}}
        \geq \sum_{(u,v) \in V \times V}\P{(u,v) \in Q}
        \geq \sum_{e \in E_1'}\P{q(e, \varphi(e)) \cap Q \neq \emptyset}
        \geq \frac{1}{2}\min\curly{LD, L^2}.
    \end{align*}
    So we have a lower bound of $\Omega(\min\{LD, L^2\})$. 

    As in the proof of~\Cref{sp-ac-j-s-xor-a}, let $c = (1-\alpha)^4=O(1)$ and note that we can assume $cn \geq 1$ and $c/\delta \geq 1$ as otherwise the \lcnamecref{sp-ac-j-s-a} is trivial.
    We set our parameters as follows:
    
    \emph{Case 1:} For $0 < \delta \leq \frac{1}{nm}$, set $L = cn$ and $D = d$, giving a lower bound of $\Omega(m)$.

    \emph{Case 2:} For $\frac{1}{nm} \leq \delta \leq \frac{c}{d^3}$, set $L = (c/(d\delta))^{1/2}$ and $D = d$, giving a lower bound of $\Omega((d/\delta)^{1/2})$.

    \emph{Case 3:} For $\frac{c}{d^3} \leq \delta \leq 1$, set $L = D = (c/\delta)^{1/3}$, giving a lower bound of $\Omega\p{(1/\delta)^{2/3}}$.
\end{proof}

In~\Cref{sec:new-pair-alg}, we prove that this lower bound is tight, by introducing a novel algorithm exploiting its access to $\InOrd$ and $\Adj$.
We thus achieve optimal bounds for both the worst and average case of the single pair problem under all models combining inclusion and exclusion of $\Jump$, $\InOrd$, and $\Adj$.

\section{The Single-Source Problem}\label{sec:single-source}

This section presents our results for the single-source problem, where we are interested in estimating $\vpi(s, t)$ for every possible target $t \in V$. The error requirement for each $\vpi(s, t)$ is the same as in the single-pair case (i.e.,~\Cref{eqn:def-singlepair}).

Recall from~\Cref{ac-source} that for the single-source problem, the average-case complexity (averaged over all $n$ possible sources $s$) is the same as the worst-case complexity. Therefore, we will only consider the problem for a worst-case source.

Prior work~\cite{MC, page1999pagerank, FORA, FORAjournal} shows that the single-source problem can be solved in $\tilde{O}\left(\min\{1/\delta, m\}\right)$ time in the adjacency-list model. This bound is obtained by combining the $O(1/\delta)$ complexity achieved by Monte Carlo sampling~\cite{MC} from the given source $s$, with the $\tilde{O}(m)$ complexity achieved by $\pw$~\cite{page1999pagerank, SpeedPPR} (in its forward version, which complements the global backward exploration approach described in~\Cref{subsubsec:backpush}).

On the lower bound side, the previous result is $\Omega(\min\{1/\delta, n\})$, derived simply by considering the worst-case output size of the single-source problem. 
In the following theorem, we show that the lower bound can be improved to the matching $\Omega(\min\{m, 1/\delta\})$, even in the adjacency-list model augmented with $\Jump$, $\InOrd$, and $\Adj$ queries. This lower bound matches the previous upper bound, establishing that the complexity of the single-source problem is $\tilde{\Theta}(\min\{m, 1/\delta\})$. \footnote{A concurrent work~\cite{PODS_ssppr} presents the same lower bound for the single-source problem on multi-graphs.}

\begin{theorem}\label{ss-j-s-a}
Consider the adjacency-list model with $\Jump$, $\InOrd{}$ and $\Adj{}$.
For any $n$ and $m$ with $n \leq m \leq n^2$ and any $\delta \in (0,1]$, there exists a graph $G=(V,E)$ with $\Theta(n)$ nodes, $\Theta(m)$ edges, and a node $s \in V$, such that for any algorithm solving the single source problem, the expected running time on $G$ with source $s$ and approximation threshold $\delta$ is $\Omega(\min\{m, 1/\delta\})$.
\end{theorem}
\begin{proof}
The single-source problem is harder than the single-pair problem, as it requires estimating $\pi(s, t)$ for all $t \in V$. Thus, the lower bound follows from Theorem~\ref{sp-wc-j-s-a}.
\end{proof}

\section{The Single-Target Problem}\label{sec:single-target}
This section focuses on the single-target problem: estimating $\vpi(s, t)$ for a given target $t \in V$ and all $n$ possible sources $s \in V$. The error requirement for each $\vpi(s, t)$ is also the same as that in the single-pair problem, as specified in equation~\eqref{eqn:def-singlepair}. 

\subsection{Known upper bounds}\label{subsec:ku-single-target}
Prior work for solving the single-target problem is mainly based on $\Push$ operations. 
Among them, the global $\pw$ method~\cite{page1999pagerank} as described in~\Cref{subsubsec:pw} can solve the single-target problem in $\tilde{O}(m)$ time in the adjacency-list model. 

Moreover, recall that the $\RBS$ method~\cite{RBS} introduces randomness into the original $\Push$ operations (see \Cref{subsubsec:RBS} for descriptions). It solves the single-target problem in $\tilde{O}\left(\sum_{v \in V} \frac{\vpi(v,t)}{\delta}\right)$ expected time in the adjacency-list model with access to the $\InOrd$ query. This running time becomes $\tilde{O}(n / \delta)$ in the worst case. 
Combining these results gives~\Cref{lem:single-target-known-upper}.

\begin{lemma}[\cite{page1999pagerank, RBS}]\label{lem:single-target-known-upper}
The single-target problem can be solved in $\tilde{O}(m)$ time in the adjacency-list model. If $\InOrd$ is also available, the problem can be solved in $\tilde{O}(\min\{m, n/\delta\})$ expected time.
\end{lemma}

When considering the average running time over all targets $t \in V$, the $\Push$ with threshold method $\bpush$~\cite{BackwardPush_InternetMath} solves the single-target problem in $O(d/\delta)$ average time in the adjacency-list model. The $\RBS$ algorithm\cite{RBS} solves it in $\tilde{O}\left(\frac{1}{n}\sum_{t\in V} \sum_{v \in V} \frac{\vpi(v,t)}{\delta}\right) = \tilde{O}(1 / \delta)$ time with the help of the $\InOrd$ query. Together with the $\tilde{O}(m)$ complexity achieved by $\pw$, we derive~\Cref{lem:st-ac-old}.

\begin{lemma}[\cite{page1999pagerank,BackwardPush_InternetMath,RBS}]\label{lem:st-ac-old}
In the adjacency-list model, the single-target problem can be solved in $\tilde{O}(\min\{m, d/\delta\})$ expected time, averaged over all possible $t$. If the $\InOrd$ query is also available, then the problem can be solved in $\tilde{O}(\min\{m, 1/\delta\})$ expected time, averaged over all possible targets $t$.
\end{lemma}

\subsection{Our upper bounds}
Below, we establish the upper bound for solving the single target problem in the adjacency-list model with $\Jump$. 

\begin{theorem}\label{upper-ac-st-ours}
In the adjacency-list model with $\Jump$, the single-target problem can be solved in $\tilde{O}(\min\{m, n/\delta\})$ time in the worst case, or in $\tilde{O}(\min\{m, (m/\delta)^{1/2}, d/\delta\})$ average time over all targets $t$. 
\end{theorem}

\begin{proof}
In the worst case, we can first use the $\Jump$ operation to jump to a node $s$, and then perform Monte Carlo sampling~\cite{MC} from $s$ to estimate $\vpi(s,t)$. The expected running time of Monte Carlo sampling for estimating $\vpi(s,t)$ is upper bounded by $O(1/\delta)$. To ensure that any node $s \in V$ is visited with constant probability via $\Jump$, we need $\Theta(n)$ $\Jump$ operations. As a result, the single-target problem can be solved in $O(n/\delta)$ time. Combining this $O(n/\delta)$ bound with the $\tilde{O}(m)$ bound achieved by $\pw$~\cite{page1999pagerank}, we conclude that the single-target problem can be solved in $\tilde{O}(\min\{m, n/\delta\})$ time in the adjacency-list model with $\Jump$.

In the average case, we adopt the bidirectional algorithm structure introduced in~\cite{BiPPR}, which combines Monte Carlo simulation with \(\Push\) from the target t. In particular, during each Monte Carlo simulation, we first use \(\Jump\) to uniformly sample a node from the graph, and then simulate random walks from the sampled node. 
It was shown in~\cite{BiPPR} that to estimate $\vpi(s,t)$ for a single node pair $(s,t)$ under the requirement defined in equation~\eqref{eqn:def-singlepair}, it is sufficient to simulate $O(\pushthreshold/\delta)$ random walks, along with a $\bpush$ computation requiring $O(d/\pushthreshold)$ expected time on average. Therefore, to solve the single-target problem, the total expected time for the Monte Carlo simulation becomes $O(n \pushthreshold/\delta)$. Balancing this cost with the $O(d/\pushthreshold)$ time of $\bpush$ gives an optimal setting of $\pushthreshold = (d \delta / n)^{1/2}$, resulting in a total time of $O((m/\delta)^{1/2})$.
Combining this $O((m/\delta)^{1/2})$ bound with the $O(d/\delta)$ bound achieved by $\bpush$ and the $\tilde{O}(m)$ bound achieved by $\pw$, we obtain the final bound of $\tilde{O}(\min\{m, (m/\delta)^{1/2}, d/\delta\})$, as claimed in~\Cref{upper-ac-st-ours}. This concludes the proof.
\end{proof}

\subsection{Known lower bounds}
To solve the single-target problem, an algorithm must output a nonzero estimate for each node $s$ satisfying $\pi(s,t)\geq\delta$, with probability $1-\pf$. Thus, constructing a graph with $k$ such nodes yields a lower bound of $\Omega(k)$. Existing lower bounds~\cite{RBS}~\footnote{The lower bound is stated in the final paragraph of Section 1.1 in~\cite{RBS}. } are based on this approach and establish a lower bound on the worst-case output size of
\begin{align*}
\Omega\left(\min\left\{n, \sum_{s\in V}\frac{\vpi(s,t)}{\delta}\right\}\right) = \Omega\left(\min\left\{n, \frac{n\vpi(t)}{\delta}\right\}\right), 
\end{align*}
for solving the single-target problem. This yields an $\Omega(n)$ lower bound for the worst-case computational complexity, and an $\Omega(\min\{n, 1/\delta\})$ lower bound for the average case. However, formal proofs, especially for the average case, are omitted in previous works. For completeness, we provide formal proofs of the two lower bounds below. 

\begin{figure}[h]
    \centering
    \begin{subfigure}{0.4\textwidth}
        \centering
        \begin{tikzpicture}[
  style1/.style={circle, draw, inner sep=0, minimum size=12pt, line width=0.5pt},
  style2/.style={-latex, line width=0.5pt},
  >=latex
  ]
  \node[style1] (u1) at (-0.75, 0) {};
  \node[style1] (u2) at (-0.25, 0) {};
  \node at (0.25, 0) {$\cdots$};
  \node[style1] (u3) at (0.75, 0) {};
  \node[style1] (t) at (0, -0.75) {$t$};
  \draw[style2] (t) edge [loop below] (t);
  \foreach \i in {1,...,3} {
    \draw[style2] (u\i) -- (t);
  }
\end{tikzpicture}
        \caption{Worst-case single-target.}
        \label{subfig:output-size-st-wc}
    \end{subfigure}
    \begin{subfigure}{0.4\textwidth}
        \centering
        \begin{tikzpicture}[
  style1/.style={circle, draw, inner sep=0, minimum size=12pt, line width=0.5pt},
  style2/.style={-latex, line width=0.5pt},
  >=latex
  ]
  \node[style1] (u1) at (-0.75, 0) {};
  \node[style1] (u2) at (-0.25, 0) {};
  \node at (0.25, 0) {$\cdots$};
  \node[style1] (u3) at (0.75, 0) {};

  \node[style1] (g) at (0, -0.75) {$g$};

  \node[style1] (v1) at (-0.75, -1.5) {};
  \node[style1] (v2) at (-0.25, -1.5) {};
  \node at (0.25, -1.5) {$\cdots$};
  \node[style1] (v3) at (0.75, -1.5) {};

  \foreach \i in {1,...,3} {
    \draw[style2] (u\i) -- (g);
    \draw[style2] (g) -- (v\i);
    \draw[style2] (v\i) edge [loop below] (v\i);
  }
\end{tikzpicture}
        \caption{Average-case single-target.}
        \label{subfig:output-size-st-ac}
    \end{subfigure}
    \caption{Output-size lower bound constructions.}
    \label{fig:output-size}
\end{figure}
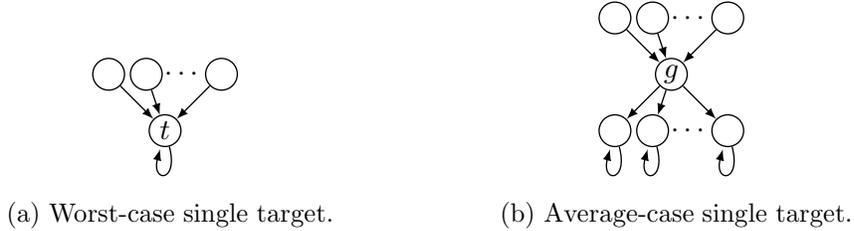

We construct a graph consisting of a target node $t$ with a self-loop and $n$ in-neighbors, as in~\Cref{subfig:output-size-st-wc}. Any algorithm must output an estimate for each in-neighbor $u$ of $t$, since $\pi(u,t) = 1-\alpha \geq \delta$. We assume $(1-\alpha)/\delta \ge 1$ as otherwise the case is trivial. 
This yields the $\Omega(n)$ lower bound for the worst-case computational complexity. For the average case, two constructions can both give a lower bound of $\Omega(\min\{n, 1/\delta\})$. For the first construction, let $g$ be a node with $n$ in-neighbors and $\min\{n, 1/\delta\}$ out-neighbors each with a self-loop, as in~\Cref{subfig:output-size-st-ac}. Here, we get output size $\Theta(n)$ when the target is any of the $\min\{n, 1/\delta\}$ out-neighbors of $g$, so averaged over all $n$ possible targets, we get output size $\Omega(\min\{n, 1/\delta\})$. For the second construction, we consider the disjoint union of $\max\{1, n\delta\}$ copies of the graph consisting a node $g$ with $\min\{n, 1/\delta\}$ in-neighbors and $\min\{n, 1/\delta\}$ out-neighbors each with a self-loop. Here, we get output size $\Theta(\min\{n, 1/\delta\})$ when the target is any of the $n$ out-neighbors.  

Notably, we cannot improve the above lower bounds using the output-size technique, since each node $u$ can have $\pi(u, v) \geq \delta$ for at most $\min\{n, 1/\delta\}$ nodes $v$. In other words, this technique can never yield a lower bound better than $\Omega(n)$. In the next subsection, we will show how to go beyond these limitations and obtain stronger lower bounds. 

\subsection{Our lower bounds}

This subsection improves previous lower bounds. Our results are tight under the adjacency-list model in both the worst and average cases, for any subset of $\Jump$, $\InOrd$, and $\Adj$. 
Our approach builds on the lower bounds of the single-pair problem, where the hard instance includes a lower part that makes exploration from the target node expensive. We push this lower part for the single-target setting, modifying the hard instance, and in doing so, obtain optimal lower bounds. Note that the tight bounds show that having access to the $\Adj$ operation does not change the complexity of the problem. Therefore, when considering the different models, we assume that $\Adj$ is always included for the lower bounds. 
Throughout the proofs in this section, we will assume that $\delta\leq (1-\alpha)^3$. 

Starting with the worst-case complexity for the adjacency-list model, we get a lower bound of $\Omega(m)$, showing that the $\pw$ algorithm is optimal up to logarithmic factors.

\begin{theorem}\label{st-wc-a}
    Consider the adjacency-list model with $\Adj$.
    For any $n$ and $m$ with $n \leq m \leq n^2$ and any $\delta \in (0,1]$, there exists a graph $G=(V,E)$ with $\Theta(n)$ nodes, $\Theta(m)$ edges, and a node $t \in V$, such that for any algorithm solving the single-target problem, the expected running time for $G$ with target $t$ and approximation threshold $\delta$ is $\Omega(m)$.
\end{theorem}
\begin{proof}
\begin{figure}[h]
    \centering
    \begin{tikzpicture}[
  style1/.style={circle, draw, inner sep=0, minimum size=12pt, line width=0.5pt},
  style2/.style={-latex, line width=0.5pt},
  >=latex
  ]

  \pgfmathsetmacro{\n}{9};
  \pgfmathsetmacro{\L}{3}; % Should be divisor of \n
  \pgfmathsetmacro{\D}{5}; % Prettiest if odd
  \pgfmathsetmacro{\d}{6};
  \pgfmathsetmacro{\labelOffset}{.3}
  \pgfmathsetmacro{\layerZero}{0}
  \pgfmathsetmacro{\layerOne}{\layerZero - 1}
  \pgfmathsetmacro{\layerTwo}{\layerOne - 2}
  \pgfmathsetmacro{\layerThree}{\layerTwo - 1}

  \node[style1] (s) at (0, \layerZero) {$u$};
  \node[style1] (t) at (0, \layerThree) {$t$};

  \foreach \i in {1,...,\n} {
    \pgfmathsetmacro{\xi}{\i - 1 - (\n - 1)/2}
    \node[style1] (u2\i) at (\xi, \layerOne) {};
    \node[style1] (v2\i) at (\xi, \layerTwo) {};
  }

  % Self-loops
  \draw[style2, densely dashed, color=red] (s) edge [loop above] (s);
  \draw[style2] (t) edge [loop below] (t);

  % Edges U2 -> V2
  \foreach \i in {1,...,\n} {
    \foreach \j in {1,...,\d} {
        \pgfmathsetmacro{\k}{int(mod(\i+\j-1+\n-\d/2, \n)+1)}
        \ifthenelse{\i = 2 \AND \k = 4}{
            \draw[style2, densely dashed, color=red] (u2\i) -- (v2\k)
        }{
            \draw[style2] (u2\i) -- (v2\k)
        };
    }
  }

  % Edges V2 -> t
  \foreach \i in {1,...,\n} {
    \draw[style2] (v2\i) -- (t);
  }

  % Labels
  \draw[style2, color=blue] (u22) -- (s);
  \draw[style2, color=blue] (s) -- (v24);
  \node at (\n/2 + \labelOffset, \layerOne) {$U_{2}$};
  \node at (\n/2 + \labelOffset, \layerTwo) {$V_{2}$};
\end{tikzpicture}
    \caption{Hard instance for the worst-case single-target problem with $\Adj$.}
    \label{fig:st-wc}
\end{figure}
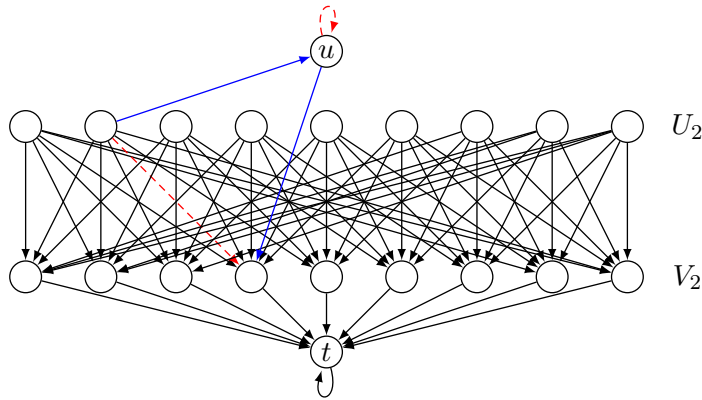

Let us construct the graph $G = (V, E)$. First, we let the node set $V$ be the disjoint union of sets $\{u\}$, $U_{2}$, $V_{2}$, and $\{t\}$. We give these sets sizes $\abs{U_{2}} = \abs{V_{2}} = n$.
We construct the edge set $E$ as follows: $u$ has a self-loop; each node in $U_2$ has $d$ edges to $V_2$, such that each node in $V_2$ has in-degree $d$; each node in $V_{2}$ has an edge to $t$; and $t$ has a self-loop. Let $e_1$ denote the self-loop of $u$, and let $E_2$ denote the subset of edges from $U_2$ to $V_2$.
See~\Cref{fig:st-wc} for an illustration, which also includes a \emph{swap}. Note that $|V|=\Theta(n)$ and $|E|=\Theta(m)$.

If we perform a swap on $e_1$ and any $e_2 \in E_2$ as in the proof of \cref{sp-wc-j-s-a}, we get a modified graph $G^\prime$, where $\pi(u,t)=(1-\alpha)^2\geq \delta$.
Thus, an algorithm must distinguish between $G$ and $G'$.
An algorithm cannot distinguish $G$ from $G'$ without querying $e_2$, since it cannot find $u$ (without \Jump{}).
To achieve constant failure probability, an algorithm must thus query $e_2$ with constant probability.
Since $e_2$ was chosen arbitrarily from $E_2$, we get a lower bound of $\Omega(\abs{E_2}) = \Omega(m)$.
\end{proof}

This result shows that local methods are not useful in this model. 
Furthermore, for the stronger model that also includes $\Jump$ and $\Adj$, we get a lower bound of $\Omega(\min\{m,n/\delta\})$, as shown in the below theorem. 

\begin{theorem}\label{st-wc-j-s-a}
    Consider the adjacency-list model with $\Jump$, $\InOrd$ and $\Adj$.
    For any $n$ and $m$ with $n \leq m \leq n^2$ and any $\delta \in (0,1]$, there exists a graph $G=(V,E)$ with $\Theta(n)$ nodes, $\Theta(m)$ edges, and a node $t \in V$, such that for any algorithm solving the single-target problem, the expected running time on $G$ with target $t$ and approximation threshold $\delta$ is $\Omega(\min\{m, n/\delta\})$.
\end{theorem}
\begin{proof}
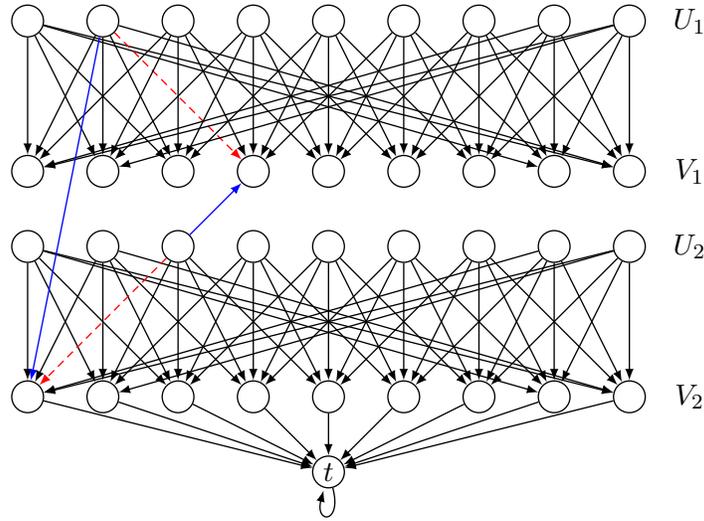
\begin{figure}[h]
    \centering
    \begin{tikzpicture}[
  style1/.style={circle, draw, inner sep=0, minimum size=12pt, line width=0.5pt},
  style2/.style={-latex, line width=0.5pt},
  >=latex
  ]

  \pgfmathsetmacro{\n}{9};
  \pgfmathsetmacro{\L}{3}; % Should be divisor of \n
  \pgfmathsetmacro{\D}{5}; % Prettiest if odd
  \pgfmathsetmacro{\d}{6};
  \pgfmathsetmacro{\labelOffset}{.3}
  \pgfmathsetmacro{\layerZero}{0}
  \pgfmathsetmacro{\layerOne}{\layerZero - 2}
  \pgfmathsetmacro{\layerThree}{\layerOne - 1}
  \pgfmathsetmacro{\layerFour}{\layerThree - 2};
  \pgfmathsetmacro{\layerFive}{\layerFour - 1};

  \node[style1] (t) at (0, \layerFive) {$t$};

  \foreach \i in {1,...,\n} {
    \pgfmathsetmacro{\xi}{\i - 1 - (\n - 1)/2}
    \node[style1] (u1\i) at (\xi, \layerZero) {};
    \node[style1] (v1\i) at (\xi, \layerOne) {};
  }
  \foreach \i in {1,...,\n} {
    \pgfmathsetmacro{\xi}{\i - 1 - (\n - 1)/2}
    \node[style1] (u2\i) at (\xi, \layerThree) {};
    \node[style1] (v2\i) at (\xi, \layerFour) {};
  }
  % Self-loops
  \draw[style2] (t) edge [loop below] (t);

  % Edges U1 -> V1
  \foreach \i in {1,...,\n} {
    \foreach \j in {1,...,\D} {
        \pgfmathsetmacro{\k}{int(mod(\i+\j-1+\n-\D/2, \n)+1)}
        \ifthenelse{\i = 2 \AND \k = 4}{
            \draw[style2, densely dashed, color=red] (u1\i) -- (v1\k)
        }{
            \draw[style2] (u1\i) -- (v1\k)
        };
    }
  }
  % Edges U2 -> V2
  \foreach \i in {1,...,\n} {
    \foreach \j in {1,...,\D} {
        \pgfmathsetmacro{\k}{int(mod(\i+\j-1+\n-\D/2, \n)+1)}
        \ifthenelse{\i = 3 \AND \k = 1}{
            \draw[style2, densely dashed, color=red] (u2\i) -- (v2\k)
        }{
            \draw[style2] (u2\i) -- (v2\k)
        };
    }
  }

  % Edges V2 -> t
  \foreach \i in {1,...,\n} {
    \draw[style2] (v2\i) -- (t);
  }
  
  % Labels
  \draw[style2, color=blue] (u12) -- (v21);
  \draw[style2, color=blue] (u23) -- (v14);
  \node at (\n/2 + \labelOffset, \layerZero) {$U_{1}$};
  \node at (\n/2 + \labelOffset, \layerOne) {$V_{1}$};
  \node at (\n/2 + \labelOffset, \layerThree) {$U_{2}$};
  \node at (\n/2 + \labelOffset, \layerFour) {$V_{2}$};
\end{tikzpicture}
    \caption{Hard instance for the worst-case single-target problem in the in the adjacency-list model with $\InOrd$, $\Jump$ and $\Adj$.}
    \label{fig:st-wc-j-d}
\end{figure}

Let us construct the graph $G = (V, E)$. First, we let the node set $V$ be the disjoint union of sets $U_1$, $V_1$, $U_{2}$, $V_{2}$, and $\{t\}$.
We give these sets sizes $\abs{U_{1}} = \abs{V_{1}} = \abs{U_{2}} = \abs{V_{2}} = n$. Let $D$ be a parameter that will be set later. We construct the edge set $E$ as follows: for each $i\in \{1,2\}$, each node in $U_i$ has $D$ edges to $V_i$, such that each node in $V_i$ has in-degree $D$; each node in $V_2$ has an edge to $t$; and $t$ has an self-loop. See~\Cref{fig:st-wc-j-d} for an illustration, which also includes a \emph{swap}. Let $E_i$ denote the subset of edges from $U_i$ to $V_i$ for $i \in \{1, 2\}$. To ensure a well-defined construction, we will ensure that $D \geq 1$ when setting $D$. To satisfy $\abs{E}=O(m)$ we will ensure that $D\leq d$. To satisfy $\abs{E} = \Omega(m)$, we add an isolated subgraph with $m$ edges. Note that we always have $|V|=\Theta(n)$.

If we perform a swap on any $(u_1,v_1)\in E_1$ and any $(u_2,v_2) \in E_2$ as in the proof of \cref{sp-wc-j-s-a}, we get a modified graph $G^\prime$, where $\pi(u_1,t)=(1-\alpha)^2/D$. When setting $D$, we will ensure that $\pi(u_1,t)\geq \delta$, so an algorithm must distinguish between $G$ and $G^\prime$. Analogously to previous proofs, we get a lower bound of $\Omega(nD)$.

We now set our parameters, casing on the minimum term among $m$ and $n/\delta$. In each case, it is easy to check that $1\leq D\leq d$, and $\pi(u_1,t)\geq \delta$, as promised. Let $c=(1-\alpha)^2$ and recall our assumption that $\delta \leq c$.

    \emph{Case 1:} For $0 < \delta \leq \frac{c}{d}$, set $D=d$, giving a lower bound of $\Omega(m)$.

    \emph{Case 2:} For $\frac{c}{d} \leq \delta \leq 1$, set $D=c/\delta$, giving a lower bound of $\Omega(n/\delta)$.
\end{proof}

In the average-case setting, we begin with the adjacency-list model with $\Adj$. We establish a tight lower bound of $\Omega(\min{m, d/\delta})$, improving upon the previous result of $\Omega(\min\{n, d/\delta\})$.
% }
\begin{theorem}\label{st-ac-a}
    Consider the adjacency-list model with $\Adj$.
    For any $n$ and $m$ with $n \leq m \leq n^2$ and any $\delta \in (0,1]$, there exists a graph $G = (V, E)$ with $\Theta(n)$ nodes and $\Theta(m)$ edges, such that for any algorithm solving the single target problem, the expected running time on $G$ with approximation threshold $\delta$, averaging over all targets $t \in V$, is $\Omega(\min\{m, d/\delta\})$, where $d=m/n$.
\end{theorem}
\begin{proof}
\begin{figure}[h]
    \centering
    \begin{tikzpicture}[
    style1/.style={circle, draw, inner sep=0, minimum size=12pt, line width=0.5pt},
    style2/.style={-latex, line width=0.5pt},
    >=latex
    ]
    \pgfmathsetmacro{\n}{9};
    \pgfmathsetmacro{\L}{3}; % Should be divisor of \n
    \pgfmathsetmacro{\D}{5}; % Prettiest if odd
    \pgfmathsetmacro{\d}{6};
    \pgfmathsetmacro{\labelOffset}{.3}
    \pgfmathsetmacro{\layerThree}{0};
    \pgfmathsetmacro{\layerFive}{\layerThree-1};
    \pgfmathsetmacro{\layerSix}{\layerFive-2};
    \pgfmathsetmacro{\layerSeven}{\layerSix-0.75};
    \pgfmathsetmacro{\layerEight}{\layerSeven-0.75};
    % Layer 2 = U1
    % Layer 3 = V1
    % Layer 5 = U2
    % Layer 6 = V2
    % Layer 7 = x
    % Layer 8 = W2
    \node[style1] (s) at (0, \layerThree) {$u$};

    \draw[style2, densely dashed, color=red] (s) edge [loop above] (s);

    \foreach \i in {1,...,\n} {
        \pgfmathsetmacro{\xi}{\i-1-(\n-1) / 2}
        \node[style1] (u2\i) at (\xi, \layerFive) {};
        \node[style1] (v2\i) at (\xi, \layerSix) {};
        \node[style1] (w2\i) at (\xi, \layerEight) {};
        \draw[style2] (w2\i) edge [loop below] (w2\i);
    }
    \pgfmathsetmacro{\g}{int(\n / \L)}
    \foreach \i in {1,...,\g} {
        \pgfmathsetmacro{\xi}{(\i-1/2)*\L - \n/2}
        \node[style1] (g\i) at (\xi, \layerSeven) {};
        \foreach \j in {1,...,\L} {
            \pgfmathsetmacro{\k}{int((\i-1)*\L+\j)}
            \draw[style2] (v2\k) -- (g\i);
            \draw[style2] (g\i) -- (w2\k);
        }
    }
    \foreach \i in {1,...,\n} {
        \foreach \j in {1,...,\d} {
            \pgfmathsetmacro{\k}{int(mod(\i+\j-1+\n-\d/2, \n)+1)}
            \ifthenelse{\i = 8 \AND \k = 6}{
                \draw[style2, densely dashed, color=red] (u2\i) -- (v2\k)
            }{
                \draw[style2] (u2\i) -- (v2\k)
            };
        }
    }
    \draw[style2, color=blue] (s) -- (v26);
    \draw[style2, color=blue] (u28) -- (s);

    \node (U1) at (\n/2+\labelOffset, \layerFive) {$U_{2}$};
    \node (V2) at (\n/2+\labelOffset, \layerSix) {$V_{2}$};
    \node (X) at (\n/2+\labelOffset, \layerSeven) {$X~$};
    \node (w2) at (\n/2+\labelOffset, \layerEight) {$W_{2}$};
\end{tikzpicture}
    \caption{Hard instance for the average-case single-target problem with $\Adj$.}
    \label{fig:st-ac}
\end{figure}
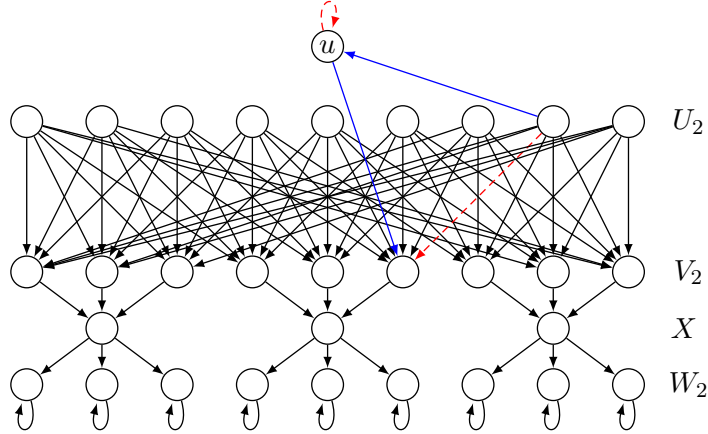

Let us construct the graph $G = (V, E)$. First, we let the node set $V$ be the disjoint union of sets $\{u\}$, $U_{2}$, $V_{2}$, $X$ and $W_2$. We give these sets sizes $\abs{U_{2}} = \abs{V_{2}} = \abs{W_2} = n$. Let $L$ be a parameter to be set later. We form a family of subsets $\{\mathcal V_{1}, \ldots, \mathcal V_{n/L}\}$ (resp. $\{\mathcal W_{1}, \ldots, \mathcal W_{n/L}\}$) partitioning $V_{2}$ (resp. $W_{2}$) into subsets of size $L$, and enumerate the nodes of $X = \{x_{1}, \ldots, x_{n/L}\}$.
We construct the edge set $E$ as follows: $u$ has a self-loop; each node in $U_{2}$ has $d$ edges to $V_{2}$, such that each node in $V_{2}$ has in-degree $d$; for each $i \in \{1, \ldots, n/L\}$ each node in $\mathcal V_{i}$ has an edge to $x_{i}$ which has an edge to every node in $\mathcal W_{i}$; and each node in $W_{2}$ has a self-loop.
See~\cref{fig:st-ac} for an illustration, which also includes a swap.
To ensure a well-defined construction, we will ensure $1\leq L \leq n$.
Note that $|V|=\Theta(n)$ and $|E|=\Theta(m)$.

Since $W_2$ contains a constant fraction of the nodes in $G$, it suffices to show the claimed lower bound for the graph $G$, averaging over all targets $t$ in $W_2$.
So fix a target $t \in \mathcal W_g$ for some $g$.
Let $e_1$ denote the self-loop of $u$, and let $E_2$ denote the subset of edges from $U_2$ to $\mathcal V_g$.
If we perform a swap on $e_1$ and any $e_2 \in E_2$ as in the proof of~\Cref{sp-ac-j-s-xor-a}, we get a modified graph $G'$, where $\pi(u,t) = (1-\alpha)^3/L$. This can be verified using equation~\eqref{eqn:onestep_walkproperty}.
When setting $L$ we will ensure that $\pi(u,t) \geq \delta$, so an algorithm must distinguish between $G$ and $G'$.
An algorithm cannot distinguish $G$ from $G'$ without querying $e_2$, since it cannot find $u$ (without \Jump{}).
To achieve constant failure probability, an algorithm must thus query $e_2$ with constant probability.
Since $e_2$ was chosen arbitrarily in $E_2$, we get a lower of $\Omega(\abs{E_2}) = \Omega(Ld)$.

We now set our parameters, casing on the minimum term among $m$ and $d/\delta$. In each case, it is easy to check that $1\leq L\leq n$, and $\pi(u,t)\geq \delta$, as promised. Let $c=(1-\alpha)^3$ and recall our assumption that $\delta \leq c$.

\emph{Case 1:} For $0 < \delta \leq \frac{1}{n}$, set $L=c n$, giving a lower bound of $\Omega(m)$.

\emph{Case 2:} For $\frac{1}{n} \leq \delta \leq 1$, set $L=c/\delta$, giving a lower bound of $\Omega(d/\delta)$.
\end{proof}

\sloppy Moreover, when $\Jump$ is also available, we obtain a lower bound of $\Omega(\min\{m, (m/\delta)^{1/2}, d/\delta\})$ as shown in the below theorem. 

\begin{theorem}\label{st-ac-j-a}
    Consider the adjacency-list model with $\Jump$ and $\Adj$.
    For any $n$ and $m$ with $n \leq m \leq n^2$ and any $\delta \in (0,1]$, there exists a graph $G = (V, E)$ with $\Theta(n)$ nodes and $\Theta(m)$ edges, such that for any algorithm solving the single-target problem, the expected running time on $G$ with approximation threshold $\delta$, averaging over all targets $t \in V$, is $\Omega(\min\{m, (m/\delta)^{1/2}, d/\delta\})$, where $d=m/n$.
\end{theorem}
\begin{proof}
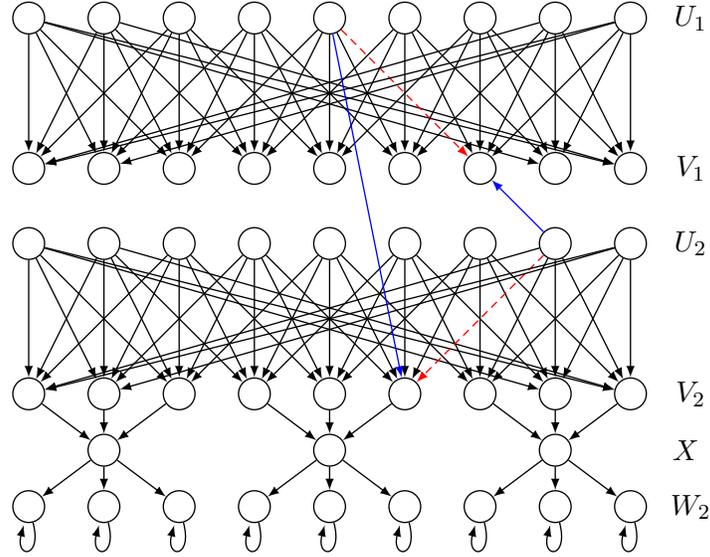
\begin{figure}[h]
    \centering
    \begin{tikzpicture}[
    style1/.style={circle, draw, inner sep=0, minimum size=12pt, line width=0.5pt},
    style2/.style={-latex, line width=0.5pt},
    >=latex
    ]
    \pgfmathsetmacro{\n}{9};
    \pgfmathsetmacro{\L}{3}; % Should be divisor of \n
    \pgfmathsetmacro{\D}{5}; % Prettiest if odd
    \pgfmathsetmacro{\d}{6};
    \pgfmathsetmacro{\labelOffset}{.3}
    \pgfmathsetmacro{\layerTwo}{0};
    \pgfmathsetmacro{\layerThree}{\layerTwo-2};
    \pgfmathsetmacro{\layerFive}{\layerThree-1};
    \pgfmathsetmacro{\layerSix}{\layerFive-2};
    \pgfmathsetmacro{\layerSeven}{\layerSix-0.75};
    \pgfmathsetmacro{\layerEight}{\layerSeven-0.75};
    % Layer 2 = U1
    % Layer 3 = V1
    % Layer 5 = U2
    % Layer 6 = V2
    % Layer 7 = x
    % Layer 8 = W2
    \foreach \i in {1,...,\n} {
        \pgfmathsetmacro{\xi}{\i-1-(\n-1) / 2}
        \node[style1] (u1\i) at (\xi, \layerTwo) {};
    }
    \foreach \i in {1,...,\n} {
        \pgfmathsetmacro{\xi}{\i-1-(\n-1) / 2}
        \node[style1] (v1\i) at (\xi, \layerThree) {};
    }
    \foreach \i in {1,...,\n} {
        \pgfmathsetmacro{\xi}{\i-1-(\n-1) / 2}
        \node[style1] (u2\i) at (\xi, \layerFive) {};
        \node[style1] (v2\i) at (\xi, \layerSix) {};
        \node[style1] (w2\i) at (\xi, \layerEight) {};
        \draw[style2] (w2\i) edge [loop below] (w2\i);
    }
    \pgfmathsetmacro{\g}{int(\n / \L)}
    \foreach \i in {1,...,\g} {
        \pgfmathsetmacro{\xi}{(\i-1/2)*\L - \n/2}
        \node[style1] (g\i) at (\xi, \layerSeven) {};
        \foreach \j in {1,...,\L} {
            \pgfmathsetmacro{\k}{int((\i-1)*\L+\j)}
            \draw[style2] (v2\k) -- (g\i);
            \draw[style2] (g\i) -- (w2\k);
        }
    }
    \foreach \i in {1,...,\n} {
        \foreach \j in {1,...,\D} {
            \pgfmathsetmacro{\k}{int(mod(\i+\j-1+\n-\D/2, \n)+1)}
            \ifthenelse{\i = 5 \AND \k = 7}{
                \draw[style2, densely dashed, color=red] (u1\i) -- (v1\k)
            }{
                \draw[style2] (u1\i) -- (v1\k)
            };
        }
    }
    \foreach \i in {1,...,\n} {
        \foreach \j in {1,...,\D} {
            \pgfmathsetmacro{\k}{int(mod(\i+\j-1+\n-\D/2, \n)+1)}
            \ifthenelse{\i = 8 \AND \k = 6}{
                \draw[style2, densely dashed, color=red] (u2\i) -- (v2\k)
            }{
                \draw[style2] (u2\i) -- (v2\k)
            };
        }
    }
    \draw[style2, color=blue] (u15) -- (v26);
    \draw[style2, color=blue] (u28) -- (v17);

    \node (V1) at (\n/2+\labelOffset, \layerTwo) {$U_{1}$};
    \node (W1) at (\n/2+\labelOffset, \layerThree) {$V_{1}$};
    \node (W1) at (\n/2+\labelOffset, \layerFive) {$U_{2}$};
    \node (V2) at (\n/2+\labelOffset, \layerSix) {$V_{2}$};
    \node (X) at (\n/2+\labelOffset, \layerSeven) {$X~$};
    \node (w2) at (\n/2+\labelOffset, \layerEight) {$W_{2}$};
\end{tikzpicture}
    \caption{Hard instance for the average-case single-target problem with $\Jump$ and $\Adj$.}
    \label{fig:st-ac-j-d}
\end{figure}

Let us construct the graph $G = (V, E)$. First, we let the node set $V$ be the disjoint union of sets $U_1$, $V_1$, $U_{2}$, $V_{2}$, $X$ and $W_2$. Let $L$ and $D$ be parameters to be set later. We give these sets sizes $\abs{U_{1}} = \abs{V_{1}} = \abs{U_{2}} = \abs{V_{2}} = \abs{W_2} = n$. We form a family of subsets $\{\mathcal V_{1}, \ldots, \mathcal V_{n/L}\}$ (resp. $\{\mathcal W_{1}, \ldots, \mathcal W_{n/L}\}$) partitioning $V_{2}$ (resp. $W_{2}$) into subsets of size $L$, and enumerate the nodes of $X = \{x_{1}, \ldots, x_{n/L}\}$.
We construct the edge set $E$ as follows: each node in $U_{1}$ has $D$ edges to $V_{1}$, such that each node in $V_{1}$ has in-degree $D$; each node in $U_{2}$ has $d$ edges to $V_{2}$, such that each node in $V_{2}$ has in-degree $d$; for each $i \in \{1, \ldots, n/L\}$ each node in $\mathcal V_{i}$ has an edge to $x_{i}$ which has an edge to every node in $\mathcal W_{i}$; and each node in $W_{2}$ has a self-loop. Let $E_i$ denote the subset of edges from $U_i$ to $V_i$ for $i \in \{1, 2\}$. See~\cref{fig:st-ac-j-d} for an illustration, which also includes a swap.
To ensure a well-defined construction, we will ensure $1\leq L \leq n$ and $D \geq 1$.
To satisfy $\abs{E} = O(m)$, we will ensure $D \leq d$. Observe that we always have $\abs{V} = \Theta(n)$ and $\abs{E}=\Omega(m)$. 

If we perform a swap on any $(u_1,v_1)\in E_1$ and $(u_2,v_2) \in E_2$ as in the proof of~\Cref{sp-ac-j-s-xor-a}, we get a modified graph $G'$, where $\pi(u_1,t) = (1-\alpha)^3/ (LD)$. This can be verified using equation~\eqref{eqn:onestep_walkproperty}.
When setting $L$ and $D$ we will ensure that $\pi(u,t) \geq \delta$, so an algorithm must distinguish between $G$ and $G'$. 
Analogously to previous proofs, we get a lower bound of $\Omega(\min\{n D, L d\})$, where the $Ld$ time cost is incurred by scanning backward from $t$, while the $nD$ time cost comes from jumping to a node in $U_1$ and then locating the swapped edge.

We now set our parameters, casing on the minimum term among $m, (m/\delta)^{1/2}$ and $d/\delta$. In each case, it is easy to check that $1\leq L\leq n$, $1\leq D\leq d$, and $\pi(u_1,t)\geq \delta$, as promised. Let $c=(1-\alpha)^3$ and recall our assumption that $\delta \leq c$.

\emph{Case 1:} For $0 < \delta \leq \frac{1}{m}$, set $L=cn$ and $D=d$, giving a lower bound of $\Omega(m)$.

\emph{Case 2:} For $\frac{1}{m} \leq \delta \leq \frac{d c}{n}$, set $L=(nc/(d\delta))^{1/2}$ and $D=(dc/(n\delta))^{1/2}$, giving a lower bound of $\Omega((m/\delta)^{1/2})$.

\emph{Case 3:} For $\frac{d c}{n} \leq \delta \leq 1$, set $L=c/\delta$ and $D=1$, giving a lower bound of $\Omega(d/\delta)$.
\end{proof}

By including the $\InOrd$ query, we get a lower bound of $\Omega(\min\{m, 1/\delta\})$ as presented below. 

\begin{theorem}\label{st-ac-j-s-a}
    Consider the adjacency-list model with $\Jump$, $\InOrd$, and $\Adj$.
    For any $n$ and $m$ with $n \leq m \leq n^2$ and any $\delta \in (0,1]$, there exists a graph $G = (V, E)$ with $\Theta(n)$ nodes and $\Theta(m)$ edges, such that for any algorithm solving the single-target problem, the expected running time on $G$ with approximation threshold $\delta$, averaging over all targets $t \in V$, is $\Omega(\min\{m, 1/\delta\})$, where $d=m/n$.
\end{theorem}
\begin{proof}
    The hard instance is nearly identical to the one presented in the proof of \cref{st-ac-j-a}, with a single modification: both \( U_1 \) and \( U_2 \) now have \( D \) edges to \( V_1 \) and \( V_2 \), respectively. After the swap is performed, we still have \( \pi(u_1, t) = {(1 - \alpha)^3}/({LD}) \). 
    We will ensure $1\leq L\leq n$, $1\leq D\leq d$, and $\pi(u_1,t)\geq \delta$. The lower bound then becomes \( \Omega(LD) \).
    
    We now set our parameters, casing on the minimum term among $m$, $1/\delta$. In each case, it is easy to check that $1\leq L\leq n$, $1\leq D\leq d$, and $\pi(u_1,t)\geq \delta$, as promised (in \cref{st-ac-j-a}). Let $c=(1-\alpha)^3$ and recall our assumption that $\delta \leq c$.

\emph{Case 1:} For $0 < \delta \leq \frac{1}{m}$, set $L=cn$ and $D=d$, giving a lower bound of $\Omega(m)$.

\emph{Case 2:} For $\frac{1}{m} \leq \delta \leq \frac{c}{d}$, set $L=c/(d\delta)$ and $D=d$, giving a lower bound of $=\Omega(1/\delta)$.

\emph{Case 3:} For $\frac{c}{d} \leq \delta \leq 1$, set $L=1$ and $D=c/\delta$, giving a lower bound of $\Omega(1/\delta)$.
\end{proof}

\section{The Single-Node Problem}\label{sec:single-node}
We focus on the single-node problem in this section: given a target node $t$, we wish to compute an estimate $\epi(t)$ of $\vpi(t)$, such that
\begin{align}\label{eqn:def-singlenode}
\Pr \curly{ \left| \epi(t) - \vpi(t) \right| \ge \rela \vpi(t)} \leq \pf,
\end{align}
where $\rela$ and $\pf$ are small constants. 
We note that for any $t\in V$,  $\vpi(t)=\frac{1}{n}\sum_{s\in V}\vpi(s,t)$, and $\vpi(t,t)\ge \alpha$ by equation~\eqref{eqn:onestep_walkproperty}. Thus, we have $\vpi(t)\ge \alpha/n$ for every $t\in V$. Compared to the single-pair, single-source, and single-target problem, the relative error threshold $\delta$ in the single-node problem is implicitly set to the smallest possible value of $\pi(t)$, i.e., $\delta=\alpha/n$. 

We again consider the complexity of this problem both for a worst-case target, and when averaging the running time over all possible targets.
To the best of our knowledge, average-case complexity for the single-node problem has not been explicitly studied before, but some results for other problem types can be naturally adapted to this setting. For example, \cite{BiPPR} provides an average-case upper bound of $O((d/\delta)^{1/2})$ for the single-pair problem. This bound can be extended to $O(m^{1/2})$ as the average-case upper bound for the single-pair problem, as noted in~\cite{BPP23}. 
For completeness, we also establish tight complexity bounds for the single-node problem in the average case under all query combinations.

\subsection{Known upper bounds}

Recall that by definition, $\pi(t)=\frac{1}{n}\sum_{s\in V}\pi(s,t)$, so algorithms designed for the single-target problem can be used to solve the single-node problem. As illustrated in~\Cref{subsec:ku-single-target}, the $\pw$ method~\cite{page1999pagerank} solves the single-target problem in $O(m\log{1/\delta})$ time in the adjacency-list model. With additional access to $\InOrd$, the $\RBS$ method~\cite{RBS} solves the single-target problem in $\tilde{O}(1/\delta)$ expected time, averaged over all possible $t$. By setting $\delta=\alpha/n$, we obtain the following lemma. 

\begin{lemma}[\cite{page1999pagerank, RBS}]\label{upper-sn-old-wc-ac}
In the adjacency-list model, the worst-case and average-case complexity of the single-node problem are $\tilde{O}(m)$. 
\end{lemma}

\begin{lemma}[\cite{RBS}]\label{upper-sn-old-rbs-ac}
In the adjacency-list model with access to $\InOrd$, the average-case complexity of the single-node problem is $\tilde{O}(n)$. 
\end{lemma}

Beyond these, the single-node problem has a rich history in the context of PageRank centrality estimation~\cite{reversePageRank_Bar-YossefM08, BPP18, BPP23, stoc/WangWW024, ChenGS04_pagerank}, in which the commonly adopted query model is the adjacency-list model with access to $\Jump$ (referred to as the arc-centric graph-access model in this line of research). In this model, \cite{BPP18} establishes an upper bound of $\tilde{O}(\min\{n^{3/4}\Deltaout^{1/4}, n^{5/7}m^{1/7}\})$, where $\Deltaout$ denotes the maximum out-degree of the graph. The bound is later improved to $O(n^{2/3} \min\{\Deltaout^{1/3}, m^{1/6}\})$~\cite{BPP23}. Furthermore, \cite{stoc/WangWW024} establishes an upper bound of $O(n^{1/2} \min\{\Deltain^{1/2}, \Deltaout^{1/2}, m^{1/4}\})$, where $\Deltain$ denotes the maximum in-degree of the graph. 
Moreover, a recent work~\cite{soda_Thorup0W026} shows that, by exploiting knowledge of $\Deltain$, the upper bound can be further improved to $\tilde{\Theta}\left(n^{1/2}\min\left\{ \Deltain^{1/2} \big/ n^{\smallexpo},\Deltaout^{1/2} \big/ n^{\smallexpo},m^{1/4}\right\}\right)$, where $\smallexpo = \frac{1}{2} \left(2\max\left\{\log_{1/(1-\alpha)}\Deltain,1\right\}-1\right)^{-1}$. When $\Deltain=n^{o(1)}$, this result is polynomially better than the previous bounds. 
However, without prior knowledge of $\Deltain$, the computational complexity for the single-node problem remains $O(n^{1/2} \min\{\Deltain^{1/2}, \Deltaout^{1/2}, m^{1/4}\})$. In this paper, we parameterize complexity solely by $n$ and $m$, leading to the following lemma.

\begin{lemma}[\cite{stoc/WangWW024, soda_Thorup0W026}]\label{upper-sn-old}
In the adjacency-list model with access to $\Jump$, the worst-case complexity of the single-node problem is $O(n^{1/2} m^{1/4})$.  
\end{lemma}

It is worth noting that the $\RBS$ algorithm \cite{RBS} can also be applied to the single-node problem by interpreting $\vpi(t)$ as the average of $\vpi(u,t)$ over all $u \in V$. Its time complexity becomes $\tilde{O}(n \vpi(t)/\delta) = \tilde{O}(n^2 \vpi(t))$ when setting $\delta = \alpha/n$ as smallest possible value of $\vpi(t)$ for any node $t$. However, since $\vpi(t)$ can be as large as $\alpha = \Theta(1)$, this complexity may not improve upon the $\tilde{O}(m)$ bound achieved by $\pw$. In the next subsection, we show that by adaptively setting $\delta = \vpi(t)$, the complexity can be improved to $\tilde{O}(n \vpi(t)/\delta) = \tilde{O}(n)$.

Additionally, when $\Jump$ is available, algorithms for the single-pair problem can be used to solve the single-node problem. A folklore bound of $\tilde{O}(m^{1/2})$ can be obtained by extending the average-case upper bound of $O((d/\delta)^{1/2})$ established for solving single-pair problem~\cite{BiPPR}. We provide a formal proof below for completeness.

\begin{lemma}[\cite{BiPPR}]
\label{thm:sn-upper-avg-jump}
    In the adjacency-list model with $\Jump$, the average-case complexity of the single-node problem is $\tilde{O}(\sqrt{m})$.
\end{lemma}

\begin{proof}
    Consider any graph $G$ in the single-node problem with $\Jump$. Let $G'$ be the graph by adding a special node $s$ to $G$ which has an outgoing edge to every original node. Let $\pi'$ denote the random walk probability in the graph $G'$. It's easy to see that $\pi(t)=\pi'(s,t)/(1-\alpha)$. Therefore, it suffices for us to simulate the algorithm in \cref{thm:avg-single-pair}, which has a time complexity of $O(\sqrt{d/\delta})$. Since we know $\pi(t)=\Omega(1/n)$, we can set $\delta=\Omega(1/n)$, so that the time complexity for the single-pair algorithm becomes $O(\sqrt{m})$. Then we simulate this algorithm in $G$ while manually dealing with the special node $s$ as follows. For each node $v \neq s$, when we visit in-neighbors, we pretend that $s$ is one of them. When we are at $s$ and need to visit a new out-neighbor, we use $\Jump$ to generate it. Note that generating $x$ different nodes needs at most $O(x\log n)$ $\Jump$ operations in expectation. So our total time complexity is $\tilde{O}(\sqrt{m})$.
\end{proof}

\subsection{Our upper bounds}
We now prove our new upper bound for the single-node problem, in both worst and average cases. 

\begin{theorem}\label{upper-sn-s}
The single-node problem can be solved in  $\tilde{O}(n)$ time in the adjacency-list model with $\InOrd$.
\end{theorem}

\begin{proof}
    As described in~\Cref{subsec:ku-single-target}, the $\RBS$ algorithm \cite{RBS} can solve the single-target problem in $\tilde{O}\p{\frac{n\pi(t)}{\delta}}$ time, such that $|\pih(u,t)-\pi(u,t)| \le \frac{\eps}{2}\max\{\pi(u,t),\delta\}$ holds for all $u$ with probability at least $1-\pf/\log n$. If we can set $\delta=\pi(t)$ and run the $\RBS$ algorithm, then we can collect the output of the single-target problem to compute the answer of the single-node problem within an additive error $\eps\pi(t)$. The only issue is that we don't know $\pi(t)$ in advance, of course. However, we know that $\pi(t)\in[\Omega(1/n),1]$.
    Our algorithm is, we first try $\delta=1$ and compute an estimate $\pih(t)$. Then, if $\delta>1/n$ and $\pih(t)>(1+\eps)\delta$, we stop and output it. Otherwise, we repeat with $\delta/2$.

    When $\delta>\pi(t)$, the probability that the additive error is larger than $\eps\delta$ is at most $\pf/\log n$, so the probability that we stop in this round is at most $\pf/\log n$. When $\delta\le\pi(t)$, the probability that the additive error is larger than $\eps\pi(t)$ is at most $\pf/\log n$, so the probability that we get an incorrect estimator in this round is $\pf/\log n$. Since there are at most $\log n$ rounds, by a union bound, the probability that we stop and output an incorrect estimator is at most $\pf$.

    The (expected) total time we spent in the rounds with $\delta=\Omega(\pi(t))$ is $\tilde{O}(n)$, since $\delta$ decreases exponentially. On the other hand, when $\delta=O(\pi(t))$, in each round we will stop with probability at least $1-\pf/\log n$. So, the probability that we reach the $i$-th round after the $\Theta(\pi(t))$ threshold is $O((\log n)^{-i})$, while the expected time we spend in this round (given that we reach this round) is only $\tilde{O}(n2^i)$. So the total time complexity is $\tilde{O}(n)$.
\end{proof}

For the average case, recall that \Cref{thm:sn-upper-avg-jump} provides a complexity of $\tilde{O}(m^{1/2})$ when $\Jump$ is available. 
If the model also supports $\InOrd$ and $\Adj$, the average-case complexity can be further improved to $\tilde{O}(\min\{m^{1/2},n^{2/3}\})$.

\begin{theorem}\label{upper-sn-ac-j-s-a}
    The single-node problem can be solved in $\tilde{O}(\min\{m^{1/2},n^{2/3}\})$ average expected time in the adjacency-list model with $\Jump$, $\InOrd$ and $\Adj$.
\end{theorem}

\begin{proof}
    The proof is analogous to the proof of \cref{thm:sn-upper-avg-jump}. The only difference is that we simulate the algorithm presented in~\Cref{sec:new-pair-alg}, whose running time is bounded by~\cref{thm:pair-alg-log-final}.
\end{proof}

\subsection{Known lower bounds}

For the worst-case complexity in the adjacency-list model with $\Jump$, let $\Deltain$ and $\Deltaout$ be the maximum in- and out-degree of the graph. \cite{BPP23} establishes a lower bound of $\Omega(n^{1/2}\Deltaout^{1/2}, n^{1/3}m^{1/3})$, which is later improved to $\Omega(n^{1/2} \min\{\Deltaout^{1/2}, m^{1/4}\})$~\cite{stoc/WangWW024}. A recent paper~\cite{soda_Thorup0W026} further improves this bound to  $\Omega\left(n^{1/2}\min\left\{ \Deltain^{1/2} \big/ n^{\smallexpo},\Deltaout^{1/2} \big/ n^{\smallexpo},m^{1/4}\right\}\right)$, where $\smallexpo = \frac{1}{2} \left(2\max\left\{\log_{1/(1-\alpha)}\Deltain,1\right\}-1\right)^{-1}$. When parameterizing only by $n$ and $m$, this lower bound is $\Omega(n^{1/2} m^{1/4})$ (i.e., by setting $\min\{\Deltain, \Deltaout\}=\Omega(m^{1/2})$ in the lower-bound construction).

The basic idea of the lower bound proof given in~\cite{stoc/WangWW024} is to construct a graph where the target $t$ has $\Omega\left(n^{1/2}m^{-1/4}\right)$ in-neighbors each with $m^{1/2}$ in-neighbors, one of which is denoted $u_*$, while ensuring $\pi(t) = n^{1/2}m^{1/4}$.
If $u_*$ is further given a large in-degree, $\pi(t)$ will increase by a constant.
So an algorithm must find this special node $u_*$ hiding at the end of one of the $n^{1/2}m^{1/4}$ edges, as it has to distinguish whether or not $u_*$ was given a large in-degree.
Since the edges are similar, an algorithm with constant failure probability must in expectation look through a constant fraction of them to find $u_*$.

\subsection{Our lower bounds}
This subsection presents all of our new lower bounds for the single-node problem. By combining these lower bounds with the upper bounds discussed above, we show that all of our bounds are tight—both in the worst case and the average case—across all graph access models. 

First, we show that in the adjacency-list model with $\Adj$, it is not possible to perform better than the basic $\tilde O(m)$ bound of $\pw$.

\begin{theorem}\label{sn-a}
    Consider the adjacency-list model with $\Adj$.
    For any $n$ and $m$ with $n \leq m \leq n^2$, there exists a graph $G = (V, E)$ with $\Theta(n)$ nodes and $\Theta(m)$ edges, such that for any algorithm solving the single-node problem, the expected running time on $G$, averaging over all targets $t \in V$, is $\Omega(m)$.
    In particular, this bound holds for a worst-case target $t \in V$.
\end{theorem}
\begin{proof}
    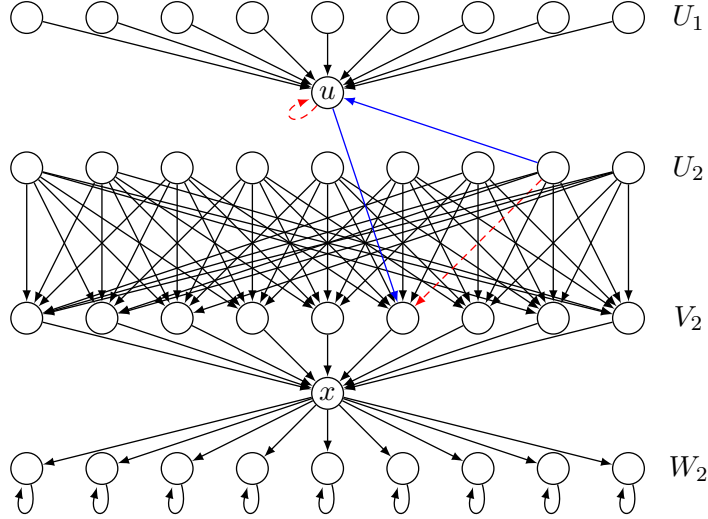
\begin{figure}[h]
        \centering
        \begin{tikzpicture}[
    style1/.style={circle, draw, inner sep=0, minimum size=12pt, line width=0.5pt},
    style2/.style={-latex, line width=0.5pt},
    >=latex
    ]
    \pgfmathsetmacro{\n}{9};
    \pgfmathsetmacro{\L}{3}; % Should be divisor of \n
    \pgfmathsetmacro{\D}{5}; % Prettiest if odd
    \pgfmathsetmacro{\d}{6};
    \pgfmathsetmacro{\labelOffset}{.3}
    \pgfmathsetmacro{\layerTwo}{0};
    \pgfmathsetmacro{\layerThree}{\layerTwo-1};
    \pgfmathsetmacro{\layerFive}{\layerThree-1};
    \pgfmathsetmacro{\layerSix}{\layerFive-2};
    \pgfmathsetmacro{\layerSeven}{\layerSix-1};
    \pgfmathsetmacro{\layerEight}{\layerSeven-1};
    \node[style1] (u) at (0, \layerThree) {$u$};
    \node[style1] (x) at (0, \layerSeven) {$x$};

    \draw[style2, densely dashed, color=red] (u) edge [in=200,out=230,loop] (u);

    \foreach \i in {1,...,\n} {
        \pgfmathsetmacro{\xi}{\i-1-(\n-1) / 2}
        \node[style1] (u1\i) at (\xi, \layerTwo) {};
        \draw[style2] (u1\i) -- (u);
        \node[style1] (u2\i) at (\xi, \layerFive) {};
        \node[style1] (v2\i) at (\xi, \layerSix) {};
        \draw[style2] (v2\i) -- (x);
        \node[style1] (w2\i) at (\xi, \layerEight) {};
        \draw[style2] (x) -- (w2\i);
        \draw[style2] (w2\i) edge [loop below] (w2\i);
    }
    \foreach \i in {1,...,\n} {
        \foreach \j in {1,...,\d} {
            \pgfmathsetmacro{\k}{int(mod(\i+\j-1+\n-\d/2, \n)+1)}
            \ifthenelse{\i = 8 \AND \k = 6}{
                \draw[style2, densely dashed, color=red] (u2\i) -- (v2\k)
            }{
                \draw[style2] (u2\i) -- (v2\k)
            };
        }
    }
    \draw[style2, color=blue] (u) -- (v26);
    \draw[style2, color=blue] (u28) -- (u);

    \node (U1) at (\n/2+\labelOffset, \layerTwo) {$U_{1}$};
    \node (U2) at (\n/2+\labelOffset, \layerFive) {$U_{2}$};
    \node (V2) at (\n/2+\labelOffset, \layerSix) {$V_{2}$};
    \node (w2) at (\n/2+\labelOffset, \layerEight) {$W_{2}$};
\end{tikzpicture}
        \caption{Hard instance for the average-case single-node problem with $\Adj$.}
        \label{fig:sn-ac-a}
    \end{figure}

    Let us construct the graph $G = (V, E)$.
    First, we let the node set $V$ be the disjoint union of sets $U_1$, $\{u\}$, $U_2$, $V_2$, $\{x\}$, and $W_2$.
    We give these sets size $\abs{U_1} = \abs{U_2} = \abs{V_2} = \abs{W_2} = n$.
    We construct the edge set $E$ as follows: each node in $U_1$ has an edge to $u$; $u$ has a self-loop; each node in $U_2$ has $d$ edges to $V_2$, such that each node in $V_2$ has in-degree $d$; each node in $V_2$ has an edge to $x$; $x$ has an edge to every node in $W_2$; and each node in $W_2$ has a self-loop.
    Let $e_1$ denote the self-loop of $u$, and let $E_2$ denote the subset of edges from $U_2$ to $V_2$.
    See~\Cref{fig:sn-ac-a} for an illustration, which also includes a swap.
    Note that $\abs{V} = \Theta(n)$ and $\abs{E} = \Theta(m)$.

    It suffices to show the lower bound for a fixed $t \in W_2$.
    Note that $\pi(t) = \Theta(1/n)$ in $G$.
    If we perform a swap on $e_1$ and any $e_2 \in E_2$ as in the proof of~\Cref{sp-wc-j-s-a}, we get a modified graph $G'$, where $\pi(t)$ has increased by $\Theta(1/n)$, i.e. by a constant fraction.
    So an algorithm must distinguish between $G$ and $G'$.
    An algorithm cannot distinguish $G$ from $G'$ without querying $e_2$, since it cannot find $u$ (without \Jump{}).
    To achieve constant failure probability, an algorithm must thus query $e_2$ with constant probability.
    Since $e_2$ was chosen arbitrarily from $E_2$, we get a lower bound of $\Omega(\abs{E_2}) = \Omega(m)$.
\end{proof}

In the adjacency-list model with $\InOrd$ and $\Adj$, it is not possible to perform better than the $\tilde O(n)$ bound of~\Cref{upper-sn-s}.

\begin{theorem}\label{sn-s-a}
    Consider the adjacency-list model with $\InOrd$ and $\Adj$.
    For any $n$ and $m$ with $n \leq m \leq n^2$, there exists a graph $G = (V, E)$ with $\Theta(n)$ nodes and $\Theta(m)$ edges, such that for any algorithm solving the single-node problem, the expected running time on $G$, averaging over all targets $t \in V$, is $\Omega(n)$.
    In particular, this bound holds for a worst-case target $t \in V$.
\end{theorem}
\begin{proof}
    \begin{figure}[h]
        \centering
        \begin{tikzpicture}[
    style1/.style={circle, draw, inner sep=0, minimum size=12pt, line width=0.5pt},
    style2/.style={-latex, line width=0.5pt},
    >=latex
    ]
    \pgfmathsetmacro{\n}{9};
    \pgfmathsetmacro{\L}{3}; % Should be divisor of \n
    \pgfmathsetmacro{\D}{5}; % Prettiest if odd
    \pgfmathsetmacro{\d}{6};
    \pgfmathsetmacro{\labelOffset}{.3}
    \pgfmathsetmacro{\layerTwo}{0};
    \pgfmathsetmacro{\layerThree}{\layerTwo-1};
    \pgfmathsetmacro{\layerFour}{\layerThree-1};
    \pgfmathsetmacro{\layerFive}{\layerFour-1};
    \pgfmathsetmacro{\layerSix}{\layerFive-1};
    \node[style1] (u) at (0, \layerThree) {$u$};
    \node[style1] (x) at (0, \layerFive) {$x$};

    \draw[style2, densely dashed, color=red] (u) edge [in=200,out=230,loop] (u);

    \foreach \i in {1,...,\n} {
        \pgfmathsetmacro{\xi}{\i-1-(\n-1) / 2}
        \node[style1] (u1\i) at (\xi, \layerTwo) {};
        \draw[style2] (u1\i) -- (u);
        \node[style1] (v2\i) at (\xi, \layerFour) {};
        \ifthenelse{\i = 6}{
            \draw[style2, densely dashed, color=red] (v2\i) -- (x)
        }{
            \draw[style2] (v2\i) -- (x)
        };
        \node[style1] (w2\i) at (\xi, \layerSix) {};
        \draw[style2] (x) -- (w2\i);
        \draw[style2] (w2\i) edge [loop below] (w2\i);
    }
    \draw[style2, color=blue] (u) edge [bend right] (x);
    \draw[style2, color=blue] (u26) -- (u);

    \node (U1) at (\n/2+\labelOffset, \layerTwo) {$U_{1}$};
    \node (V2) at (\n/2+\labelOffset, \layerFour) {$V_{2}$};
    \node (w2) at (\n/2+\labelOffset, \layerSix) {$W_{2}$};
\end{tikzpicture}
        \caption{Hard instance for the average-case single-node problem with $\InOrd$ and $\Adj$.}
        \label{fig:sn-ac-s-a}
    \end{figure}

    Let us construct the graph $G = (V, E)$.
    First, we let the node set $V$ be the disjoint union of sets $U_1$, $\{u\}$, $V_2$, $\{x\}$, and $W_2$.
    We give these sets size $\abs{U_1} = \abs{V_2} = \abs{W_2} = n$.
    We construct the edge set $E$ as follows: each node in $U_1$ has an edge to $u$; $u$ has a self-loop; each node in $V_2$ has an edge to $x$; $x$ has an edge to every node in $W_2$; and each node in $W_2$ has a self-loop.
    Let $e_1$ denote the self-loop of $u$, and let $E_2$ denote the subset of edges from $V_2$ to $x$.
    See~\Cref{fig:sn-ac-s-a} for an illustration, which also includes a swap.
    Note that $\abs{V} = \Theta(n)$ and $\abs{E} = \Theta(m)$.

    It suffices to show the lower bound for a fixed $t \in W_2$.
    Note that $\pi(t) = \Theta(1/n)$ in $G$.
    If we perform a swap on $e_1$ and any $e_2 \in E_2$ as in the proof of~\Cref{sp-wc-j-s-a}, we get a modified graph $G'$, where $\pi(t)$ has increased by $\Theta(1/n)$, i.e. by a constant fraction.
    Note that $\InOrd$ is no more useful than $\In$, as every node other than $x$ has out-degree one, so analogously to the proof of~\Cref{sn-a}, we get a lower bound of $\Omega(\abs{E_2}) = \Omega(n)$.
\end{proof}

The work of~\cite{stoc/WangWW024} establishes an $\Omega(n^{1/2}m^{1/4})$ lower bound under the adjacency-list model with $\Jump$ and $\Adj$. The following theorem shows that this $\Omega(n^{1/2}m^{1/4})$ lower bound also holds even when $\InOrd$ is available.

\begin{theorem}\label{sn-wc-j-s-a}
    Consider the adjacency-list model with $\Jump$, $\InOrd$, and $\Adj$.
    For any $n$ and $m$ with $n \leq m \leq n^2$, there exists a graph $G=(V,E)$ with $\Theta(n)$ nodes, $\Theta(m)$ edges, and a node $t \in V$, such that for any algorithm solving the single-node problem, the expected running time on $G$ with target $t$ is $\Omega(n^{1/2}m^{1/4})$.
\end{theorem}
\begin{proof}
    \begin{figure}[h]
        \centering
        \begin{tikzpicture}[
    style1/.style={circle, draw, inner sep=0, minimum size=12pt, line width=0.5pt},
    style2/.style={-latex, line width=0.5pt},
    >=latex
    ]
    \pgfmathsetmacro{\n}{9};
    \pgfmathsetmacro{\sqrtm}{4};
    \pgfmathsetmacro{\T}{2};
    \pgfmathsetmacro{\L}{3}; % Should be divisor of \n
    \pgfmathsetmacro{\D}{5}; % Prettiest if odd
    \pgfmathsetmacro{\d}{6};
    \pgfmathsetmacro{\offsetT}{0.7}
    \pgfmathsetmacro{\labelOffset}{.3}
    \pgfmathsetmacro{\layerOne}{0};
    \pgfmathsetmacro{\layerTwo}{\layerOne-1};
    \pgfmathsetmacro{\layerThree}{\layerTwo-0.75};
    \pgfmathsetmacro{\layerFour}{\layerThree-1.75};
    \pgfmathsetmacro{\layerFive}{\layerFour-1};
    \pgfmathsetmacro{\layerSix}{\layerFive-1.25};
    \pgfmathsetmacro{\layerSeven}{\layerSix-0.5};
    \pgfmathsetmacro{\layerEight}{\layerSeven-0.75};

    \node[style1] (x) at (0, \layerTwo) {$x$};
    \node[style1] (t) at (0, \layerEight) {$t$};

    % X nodes
    \foreach \i in {1,...,\n} {
        \pgfmathsetmacro{\xi}{\i-1-(\n-1) / 2}
        \node[style1] (x\i) at (\xi, \layerOne) {};
        \draw[style2] (x\i) -- (x);
    }
    % U1 and V2 nodes
    \foreach \i in {1,...,\L} {
        \pgfmathsetmacro{\xi}{\i-1-(\L-1) / 2}
        
        \node[style1] (u1\i) at (\xi, \layerThree) {};
        \draw[style2] (x) -- (u1\i);
        
        \node[style1] (v2\i) at (\xi, \layerSeven) {};
        \draw[style2] (v2\i) -- (t);
    }
    % V1 and U2 nodes
    \foreach \i in {1,...,\sqrtm} {
        \pgfmathsetmacro{\xi}{\i-1-(\sqrtm-1) / 2}
        \node[style1] (v1\i) at (\xi, \layerFour) {};
        \node[style1] (u2\i) at (\xi, \layerFive) {};
    }
    % Edges for U1 - V1
    \foreach \i in {1,...,\L} {
        \foreach \j in {1,...,\sqrtm} {
            \ifthenelse{\i = 1 \AND \j = 3}{
                \draw[style2, densely dashed, color=red] (u1\i) -- (v1\j)
            }{
                \draw[style2] (u1\i) -- (v1\j)
            };
        }
    }
    % Edges for U2 - V2
    \foreach \i in {1,...,\sqrtm} {
        \foreach \j in {1,...,\L} {
            \ifthenelse{\i = 4 \AND \j = 1}{
                \draw[style2, densely dashed, color=red] (u2\i) -- (v2\j)
            }{
                \draw[style2] (u2\i) -- (v2\j)
            };
        }
    }
    % T nodes
    \foreach \i in {1,...,\T} {
        \pgfmathsetmacro{\xi}{\i-1 + (\sqrtm-1) / 2 + \offsetT}
        \node[style1] (t\i) at (\xi, \layerSix) {};
    }
    % Edges for U2 - T
    \foreach \i in {1,...,\sqrtm} {
        \foreach \j in {1,...,\T} {
            \draw[style2] (u2\i) -- (t\j);
        }
    }
    \draw[style2] (t) edge [loop below] (t);

    \draw[style2, color=blue] (u11) -- (v21);
    \draw[style2, color=blue] (u24) -- (v13);

    \node (X)  at (\n/2+\labelOffset, \layerOne) {$X~$};
    \node (U1) at (\L/2+\labelOffset, \layerThree) {$U_{1}$};
    \node (V1) at (\sqrtm/2+\labelOffset, \layerFour) {$V_{1}$};
    \node (U2) at (\sqrtm/2+\labelOffset, \layerFive) {$U_{2}$};
    \node (T) at (\sqrtm/2+\T-1+\offsetT+\labelOffset, \layerSix) {$T$};
    \node (V2) at (\L/2+\labelOffset, \layerSeven) {$V_{2}$};
\end{tikzpicture}
        \caption{Hard instance for the worst-case single-node problem with $\Jump, \InOrd$ and $\Adj$.}
        \label{fig:sn-wc-j-d}
    \end{figure}

Let us construct the graph $G = (V, E)$. First, we let the node set $V$ be the disjoint union of sets $X$, $\{x\}$, $U_1$, $V_1$, $U_2$, $V_2$, $T$, and, $\{t\}$. We give these sets size $\abs{X} = n$, $\abs{U_1}=\abs{V_2}=L$, $\abs{V_1}=\abs{U_2}=\sqrt{m}$, and $\abs{T}=\sqrt{m}-L$. We construct the edge set $E$ as follows: each node in $X$ has an edge to $x$; $x$ has an edge to every node in $U_1$; each node in $U_1$ has an edge to each node in $V_1$, such that each node in $V_1$ has an in-degree of $L$; each node in $U_2$ has edges to each node in $V_2$ and each node in $T$, 
such that each node in $U_2$ has an out-degree of $\sqrt{m}$ and each node in $V_2$ has an in-degree of $\sqrt{m}$; each node in $V_2$ has an edge to node $t$; node $t$ has a self-loop. Let $E_1$ denote subset of edges from $U_1$ to $V_1$, and let $E_2$ denote the subset of edges from $U_2$ to $V_2$. See~\Cref{fig:sn-wc-j-d} for an illustration, which also includes a swap. 

We note that in $G$, $\abs{V} = \Theta(n)$, $\abs{E} = \Theta(m)$, 
and $\pi(t) = \Theta(L/n)$. If we perform a swap on any $e_1\in E_1$ and any $e_2 \in E_2$ as in the proof of~\Cref{sp-wc-j-s-a}, we get a modified graph $G'$, where $\pi(t)$ has increased by $\Theta(1/(L\sqrt{m}))$, i.e. by a constant fraction. So assuming $\epsilon$ is at most this constant, an algorithm must distinguish between $G$ and $G'$. Note that $\InOrd$ is no more useful than $\In$, as every in-neighbor of nodes in both $G$ and $G'$ has the same out-degree. We will ensure that $1\le L\le n$, $L\sqrt{m} \le m$, and $L/n=L\sqrt{m}$. As a result, we set $L=n^{1/2}/m^{1/4}$. 
Then we get a lower bound of $\Omega(L\sqrt{m})=\Omega(n^{1/2}m^{1/4})$. 
\end{proof}

In the average case, a story similar to that of the single pair problem turns up.
If we have $\Jump$ together with $\InOrd$ or $\Adj$, but not both, we get a lower bound matching~\Cref{thm:sn-upper-avg-jump}.

\begin{theorem}\label{sn-ac-j-s-xor-a}
    Consider the adjacency-list model with $\Jump$ and either $\InOrd$ or $\Adj$, but not both.
    For any $n$ and $m$ with $n \leq m \leq n^2$, there exists a graph $G = (V, E)$ with $\Theta(n)$ nodes and $\Theta(m)$ edges, such that for any algorithm solving the single-node problem, the expected running time on $G$, averaging over all targets $t \in V$, is $\Omega(m^{1/2})$.
\end{theorem}
\begin{proof}
    \begin{figure}[h]
        \centering
        \begin{tikzpicture}[
    style1/.style={circle, draw, inner sep=0, minimum size=12pt, line width=0.5pt},
    style2/.style={-latex, line width=0.5pt},
    >=latex
    ]
    \pgfmathsetmacro{\n}{9};
    \pgfmathsetmacro{\L}{3}; % Should be divisor of \n
    \pgfmathsetmacro{\D}{5}; % Prettiest if odd
    \pgfmathsetmacro{\d}{6};
    \pgfmathsetmacro{\labelOffset}{.3}
    \pgfmathsetmacro{\layerZero}{0};
    \pgfmathsetmacro{\layerOne}{\layerZero-1};
    \pgfmathsetmacro{\layerTwo}{\layerOne-0.75};
    \pgfmathsetmacro{\layerThree}{\layerTwo-1};
    \pgfmathsetmacro{\layerFive}{\layerThree-1};
    \pgfmathsetmacro{\layerSix}{\layerFive-2};
    \pgfmathsetmacro{\layerSeven}{\layerSix-0.75};
    \pgfmathsetmacro{\layerEight}{\layerSeven-0.75};
    % Layer 2 = U1
    % Layer 3 = V1
    % Layer 5 = U2
    % Layer 6 = V2
    % Layer 7 = x
    % Layer 8 = W2
    \foreach \i in {1,...,\n} {
        \pgfmathsetmacro{\xi}{\i-1-(\n-1) / 2}
        \node[style1] (w1\i) at (\xi, \layerZero) {};
    }
    \node[style1] (u) at (0, \layerOne) {$u$};

    \foreach \i in {1,...,\n} {
      \draw[style2] (w1\i) -- (u);
    }
    
    \foreach \i in {1,...,\L} {
        \pgfmathsetmacro{\xi}{\i-1-(\L-1) / 2}
        \node[style1] (u1\i) at (\xi, \layerTwo) {};
    }
    
    \foreach \i in {1,...,\L} {
      \draw[style2] (u) -- (u1\i);
    }
    
    \foreach \i in {1,...,\d} {
        \pgfmathsetmacro{\xi}{\i-1-(\d-1) / 2}
        \node[style1] (v1\i) at (\xi, \layerThree) {};
    }
    \foreach \i in {1,...,\n} {
        \pgfmathsetmacro{\xi}{\i-1-(\n-1) / 2}
        \node[style1] (u2\i) at (\xi, \layerFive) {};
        \node[style1] (v2\i) at (\xi, \layerSix) {};
        \node[style1] (w2\i) at (\xi, \layerEight) {};
        \draw[style2] (w2\i) edge [loop below] (w2\i);
    }
    \pgfmathsetmacro{\g}{int(\n / \L)} 
    \foreach \i in {1,...,\g} {
        \pgfmathsetmacro{\xi}{(\i-1/2)*\L - \n/2}
        \node[style1] (g\i) at (\xi, \layerSeven) {};
        \foreach \j in {1,...,\L} {
            \pgfmathsetmacro{\k}{int((\i-1)*\L+\j)}
            \draw[style2] (v2\k) -- (g\i);
            \draw[style2] (g\i) -- (w2\k);
        }
    }
    \foreach \i in {1,...,\L} {
        \foreach \j in {1,...,\d} {
            % \pgfmathsetmacro{\k}{int(mod(\i+\j-1+\L-\d/2, \n)+1)}
            \ifthenelse{\i = 1 \AND \j = 6}{
                \draw[style2, densely dashed, color=red] (u1\i) -- (v1\j)
            }{
                \draw[style2] (u1\i) -- (v1\j)
            };
        }
    }
    \foreach \i in {1,...,\n} {
        \foreach \j in {1,...,\D} {
            \pgfmathsetmacro{\k}{int(mod(\i+\j-1+\n-\D/2, \n)+1)}
            \ifthenelse{\i = 7 \AND \k = 5}{
                \draw[style2, densely dashed, color=red] (u2\i) -- (v2\k)
            }{
                \draw[style2] (u2\i) -- (v2\k)
            };
        }
    }
    \draw[style2, color=blue] (u11) -- (v25);
    \draw[style2, color=blue] (u27) -- (v16);

    \node (W1) at (\n/2+\labelOffset, \layerZero) {$W_{1}$};
    \node (V1) at (\n/2+\labelOffset, \layerTwo) {$U_{1}$};
    \node (W1) at (\n/2+\labelOffset, \layerThree) {$V_{1}$};
    \node (W1) at (\n/2+\labelOffset, \layerFive) {$U_{2}$};
    \node (V2) at (\n/2+\labelOffset, \layerSix) {$V_{2}$};
    \node (X) at (\n/2+\labelOffset, \layerSeven) {$X~$};
    \node (w2) at (\n/2+\labelOffset, \layerEight) {$W_{2}$};
\end{tikzpicture}
        \caption{Hard instance for the average-case single-node problem with $\InOrd$ and $\Adj$.}
        \label{fig:sn-ac-j-s-xor-a}
    \end{figure}

    Let us construct the graph $G = (V, E)$.
    First, we let the node set $V$ be the disjoint union of sets $W_1$, $\{u\}$, $U_1$, $V_1$, $U_2$, $V_2$, $X$, and $W_2$.
    We give these sets sizes $\abs{W_1}=\abs{U_2}=\abs{V_2}=\abs{W_2} = n$, $\abs{U_1}=L$, $\abs{V_1}=d$ and $\abs{X}=n/L$ where $L$ is a parameter to be set later.
    We form a family of subsets $\{\mathcal V_1,\ldots,\mathcal V_{n/L}\}$ (resp. $\{\mathcal W_1,\ldots,\mathcal W_{n/L}\}$) partitioning $V_2$ (resp. $W_2$) into subsets of size $L$, and enumerate the nodes of $X = \{x_1, \ldots, x_{n/L}\}$.
    We construct the edge set $E$ as follows: each node in $W_1$ has an edge to $u$; $u$ has an edge to every node in $U_1$; each node in $U_1$ has an edge to every node in $V_1$; each node in $U_2$ has $d$ edges to $V_2$ such that every node in $V_2$ has in-degree $d$; for each $i \in \{1,\ldots,n/L\}$, each node in $\mathcal V_i$ has a node to $x_i$ which has an edge to every node in $\mathcal W_i$; and each node in $W_2$ has a self-loop.
    See~\Cref{fig:sn-ac-j-s-xor-a} for an illustration, including also a swap.
    Note that $\abs{V} = \Theta(n)$ and $\abs{E} = \Theta(m)$.

    It suffices to prove the lower bound for a given $t \in \mathcal W_g$ for a given $g$.
    Note that $\pi(t) = \Theta(1/n)$ in $G$.
    Let $E_1$ be the subset of edges from $U_1$ to $V_1$, and let $E_2$ be the subset of edges from $U_2$ to $\mathcal V_g$.
    If we perform a swap on an $e_1 \in E_1$ and $e_2 \in E_2$ as in the proof of~\Cref{sp-ac-j-s-xor-a}, we get a modified graph $G'$, where $\pi(t)$ increases by $\Omega((1/(L^2d))$, i.e. by a constant factor if we set $L = (n/d)^{1/2}$.
    So an algorithm must distinguish between $G$ and $G'$.
    Note that $\InOrd$ is no more useful than $\In$ in this construction, so analogously to previous proofs, we get a lower bound of $\Omega(Ld)=\Omega(m^{1/2})$ if we don't allow $\Adj$.

    Let us now handle the case where $\Adj$ is present and $\InOrd$ is absent.
    Here, we change the sizes of $U_1$ and $V_1$, just as in~\Cref{sp-ac-j-s-xor-a}, to $\abs{U_1} = d$ and $\abs{V_1} = L$.
    Analogously to the proof of~\Cref{sp-ac-j-s-xor-a} we again get a lower bound of $\Omega(Ld)=\Omega(m^{1/2})$.
\end{proof}

If we have $\Jump$ together with not only one of $\InOrd$ and $\Adj$ but both, we get a lower bound matching~\Cref{upper-sn-ac-j-s-a}.

\begin{theorem}\label{sn-ac-j-s-a}
    Consider the adjacency-list model with $\Jump$, $\InOrd$, and $\Adj$.
    For any $n$ and $m$ with $n \leq m \leq n^2$, there exists a graph $G = (V, E)$ with $\Theta(n)$ nodes and $\Theta(m)$ edges, such that for any algorithm solving the single-node problem, the expected running time on $G$, averaging over all targets $t \in V$, is $\Omega(\min\{m^{1/2}, n^{2/3}\})$
\end{theorem}
\begin{proof}
    Reuse the construction from~\Cref{sn-ac-j-s-xor-a}, but replacing the degree $d$ by a parameter $D$.
    Analogously to the proof of~\Cref{sp-ac-j-s-a} we get a lower bound of $\Omega(\min\{LD, L^2\}) = \Omega(\min\{m^{1/2}, n^{2/3}\})$ for $L=n^{1/3}$ and $D=m^{1/2}n^{-1/3}$.
\end{proof}

\subsubsection{Other Related Work}\label{subsec:other-subsection}
For the single-node PageRank problem, there is a lot of work, e.g., involving different graph parameters.
Let $\Deltain$ and $\Deltaout$ be the maximum in- and out-degree of the graph, respectively. In this paper, we have not considered these extra graph parameters. Our bounds are only in terms of $n$ and $m$,
but there is a line of research that parameterizes the complexity in terms of $\Deltain$ and $\Deltaout$, which we briefly review below.

The upper bound established by \cite{BPP18} is actually $\tilde{O}(\min\{n^{3/4}\Deltaout^{1/4}, n^{5/7}m^{1/7}\})$. The bound is later improved to $O(n^{2/3} \min\{\Deltaout^{1/3}, m^{1/6}\})$ by \cite{BPP23}, and $O(n^{1/2} \min\{\Deltain^{1/2}, \Deltaout^{1/2}, m^{1/4}\})$ by \cite{stoc/WangWW024}. 
A recent work~\cite{soda_Thorup0W026} shows that, by exploiting knowledge of $\Deltain$, the upper bound can be further improved to $\tilde{\Theta}\left(n^{1/2}\min\left\{ \Deltain^{1/2} \big/ n^{\smallexpo},\Deltaout^{1/2} \big/ n^{\smallexpo},m^{1/4}\right\}\right)$, where $\smallexpo = \frac{1}{2} \left(2\max\left\{\log_{1/(1-\alpha)}\Deltain,1\right\}-1\right)^{-1}$. When $\Deltain=n^{o(1)}$, this result is polynomially better than the previous bounds.  

Additionally, a very recent work~\cite{PageRank_TW} studies the instance-optimality of $\BiPPR$. It considers the adjacency-list model with $\Jump$, and shows that the instance-wise complexity of $\BiPPR$ is $\tilde{\Theta}(\max_{r\in [0, 1]}\left\{\min\left\{T_r, r/\vpi(t)\right\}\right\})$, where $V_r=\{v\in V \mid \vpi(v,t) \ge r\}$ and $ T_r=\sum_{v\in V_r}\left(1+\din(v)\right)$. When considering the worst-case graph instances, 
this bound matches our worst-case complexity result when parameterized only by $n$ and $m$.

On the lower bound side, the lower bound established by~\cite{BPP18, BPP23} is actually $\Omega(n^{1/2}\Deltaout^{1/2}, n^{1/3}m^{1/3})$, which is later improved to $\Omega(n^{1/2} \min\{\Deltaout^{1/2}, m^{1/4}\})$~\cite{stoc/WangWW024}. \cite{soda_Thorup0W026} further improves this bound to  $\Omega\left(n^{1/2}\min\left\{ \Deltain^{1/2} \big/ n^{\smallexpo},\Deltaout^{1/2} \big/ n^{\smallexpo},m^{1/4}\right\}\right)$, matching their upper bound. However, this improvement relies on the prior knowledge of $\Deltain$.

\bibliographystyle{alphaurl}
\bibliography{references}

\newpage

\appendix
\addtocontents{toc}{\protect\setcounter{tocdepth}{1}}
\section{Table of Notations} \label{sec:table_notations}

\begin{table*}[!h]
    \centering
    \renewcommand{\arraystretch}{1.3}
    \caption{Table of notations.} \label{tbl:def-notation}
    \begin{tabular}{ll}
    \toprule
    \textbf{Notation} & \textbf{Description} \\
    \midrule
    $G=(V,E)$ & underlying directed graph with node set $V$ and edge set $E$ \\
    $n, m$ & number of nodes and edges in $G$  \\
    $\din(v),\dout(v)$ & in-degree and out-degree of $v$  \\
    $\Nin(v),\Nout(v)$ & set of in-neighbors and out-neighbors of $v$  \\
    $d=m/n$ & average degree of $G$  \\
    $\vpi(s,t), \vpi(t)$ & PPR score of $t$ w.r.t. $s$, PageRank score of $t$ (\Cref{sec:intro}) \\
    $\alpha$ & decay factor in defining PageRank and PPR, $\alpha\in(0,1)$ (\Cref{sec:intro}) \\
    $\delta$ & threshold parameter in estimating $\vpi(s,t)$ (Section~\ref{sec:sp_upper}, \ref{sec:sp_lower}, \ref{sec:single-source}, \ref{sec:single-target}) \\
    $\rela$ & constant relative error parameter (Section~\ref{sec:sp_upper}, \ref{sec:sp_lower}, \ref{sec:single-source}, \ref{sec:single-target}, \ref{sec:single-node}) \\
    $p_f$ & constant failure probability parameter (Section~\ref{sec:sp_upper}, \ref{sec:sp_lower}, \ref{sec:single-source}, \ref{sec:single-target}, \ref{sec:single-node}) \\
    $\reserve(),\residue()$ & reserves and residues in \Cref{alg:push}: $\Push(v)$ (\Cref{subsubsec:backpush}) \\
    $\pushthreshold$ & push threshold (\Cref{subsubsec:backpush}) \\
    \midrule
    $L$ & maximum number of push levels (\Cref{sec:new-pair-alg})\\
    & ($L$ has a different meaning in the lower-bound proofs; defined where used)\\
    $\ph_i(v), \rh_i(v)$ & randomized reserves and residues at level $i\in [0,L]$ (\Cref{sec:new-pair-alg})\\
    $\rh'_i(v)$ & independent copy of $\rh_i(v)$ (\Cref{sec:new-pair-alg})\\
    $\ph(v)$ & randomized reserves, $\ph(v)=\sum_{i=0}^{L}\ph_i(v)$ (\Cref{sec:new-pair-alg}) \\
    $\rh(v)$ & randomized residues, $\rh(v)=\sum_{i=0}^{L}\rh_i(v)$ (\Cref{sec:new-pair-alg}) \\
    $q(s,t)$ & bidirectional estimator (\Cref{sec:new-pair-alg}) \\
    $n_r$ & number of random walk simulations (\Cref{sec:new-pair-alg}) \\
    $\rmax_i$ & push threshold at level $i\in [0,L]$ (\Cref{sec:new-pair-alg}) \\
    $\rmax$ & push threshold, functionally analogous to $\pushthreshold$, $\rmax=\sum_{i=0}^{L}\rmax_i$ (\Cref{sec:new-pair-alg})\\
    $\dr_{i+1}(u,v)$ & (expected) increment to $\rh_{i+1}(u)$ in \Cref{alg:random-push}: Push $\rh_i(v)$ (\Cref{sec:new-pair-alg})\\
    $\gap_{i}$ & fine-grained push threshold at level $i$ (\Cref{sec:new-pair-alg})\\
    $\er(u), \er_i(u)$ & “derandomized” version $\rh(u)$ and $\rh_i(u)$ (\Cref{sec:new-pair-alg})\\
    $\pth$ & threshold parameter in \Cref{alg:estimate-R}: Compute $\erh(u_k)$ (\Cref{sec:new-pair-alg})\\
    $\Vp$ & set of all nodes $v$ with $\ph(v)>\pth$ (\Cref{sec:new-pair-alg})\\
    $\rmaxx$ & lower bound on $\gap_i\rmax_i$ for all $i$: $\gap_i\rmax_i\ge\rmaxx$ for $\forall i\in [0,L]$ (\Cref{sec:new-pair-alg})\\
    \bottomrule
    \end{tabular}
\end{table*}

\section{Deferred details in \cref{sec:new-pair-alg}}

\subsection{Pseudocodes}
\label{subsec:app_notation_code}

\begin{algorithm}[H]
\label{fullcode}
\caption{$\spppr (s, t, L, \nr, \rmax_i, \gap_{i})$}
    \begin{algorithmic}[1]
        \State $\rh_0(t) \gets 1$, $\rhh_0(t) \gets 1$.
        \For{$i=0,1,2,\dots,L-1$}
            \For{each $v\in V$ with $\rhh_i(v)>\rmax_i$}
                \For{each $u\in \Nin(v)$}
                    \State $\dr_i(u,v)\gets \frac{(1-\alpha)\rh_i(v)}{\dout(u)}$.
                    \If{$\dr_i(u,v)\ge\gap_i\rmax_i$}
                        \State $\rh_{i+1}(u)\gets \rh_{i+1}(u)+\dr_i(u,v)$.
                        \State $\rhh_{i+1}(u)\gets \rhh_{i+1}(u)+\dr_i(u,v)$.
                    \Else
                        \State $\rh_{i+1}(u)\gets \rh_{i+1}(u)+\gap_i\rmax_i$ with probability $\frac{\dr_i(u,v)}{\gap_i\rmax_i}$.
                        \State $\rhh_{i+1}(u)\gets \rhh_{i+1}(u)+\gap_i\rmax_i$ with probability $\frac{\dr_i(u,v)}{\gap_i\rmax_i}$.
                    \EndIf
                \EndFor
                \State $\ph(v)\gets \ph(v)+\alpha \rh_i(v)$.
                \State $\rh_i(v)\gets 0$.
            \EndFor
        \EndFor

        \State $\pitt(s,t)\gets \ph(s)$.
        \For{$k=1,2,\dots,\nr$}
            \State Generate a random walk from $s$, stopping at $u_k$.
            \State $\erh(u_k) \gets 0$.
            \For{each $v\in \Vp$} $\quad$ \textcolor{gray}{\small // The set $\Vp$ contains all nodes $v$ in $G$ with $\ph(v)>\pth$.}
                \If{$(u_k,v)\in E$}
                    \State $\erh(u_k)\gets \erh(u_k)+\sum_i \nopush_i(u_k)\dr_i(u_k,v)$.
                \EndIf
            \EndFor
        
            \For{$j=1,2,\dots,\ns$}
                \State $v_j\gets$ a uniformly random vertex in $\Nout(u_k)\setminus \Vp$.
                \State $\erh(u_k)\gets \erh(u_k)+\frac{|\Nout(u_k)\setminus V_P|}{\ns} \sum_i \nopush_i(u_k)\dr_i(u_k,v_j)$.
            \EndFor
            \State $\pitt(s,t)\gets \pitt(s,t)+\frac{1}{\nr}\erh(u_k)$
        \EndFor

        \State \Return $\pitt(s,t)$.
    \end{algorithmic}
\end{algorithm}

\begin{algorithm}[H]
\caption{$\pw (s, t, L$)}
\label{alg:push_by_level}
\textbf{Input:} source node $s$, target node $t$, maximum round $L$\\
\textbf{Output:} estimate for $\pi(s,t)$
    \begin{algorithmic}[1]
        \State $\residue_0()\gets 0$, $\residue_0(t)\gets 1$
        \State $\reserve_0()\gets 0$
        \For{$i$ from $1$ to $L$}
            \State $\residue_i()\gets 0$
            \State $\reserve_i()\gets 0$
            \For{each $v\in V$}
                \State $\reserve_{i-1}(v)\gets \reserve_{i-1}(v)+\alpha \residue_{i-1}(v)$
                \For{each $u\in \Nin(v)$}
                
                \State $\residue_{i}(u) \gets \residue_i(u) + (1-\alpha)\residue_i(v) / \dout(u)$
                \EndFor
                \State $\residue_{i-1}(v)\gets 0$
            \EndFor
        \EndFor
    \State \Return $\sum_{i=0}^{L}p_i(s)$
    \end{algorithmic}
\end{algorithm}

\subsection{Proof of \cref{lem:pushback-time}}
\label{app:pushback-time}

Consider the invariant in \cref{lem:exp-inv}. Summing over all $w\in V$, we have:
$$\E{\sum_{w\in V}\ph(w)+\sum_{w\in V}\sum_{u\in V}\pi(w,u)\rh(u)}=\sum_{w\in V}\pi(w,t)=n\pi(t).$$
Notice that all $\pi(w,u)$ and $\rh(u)$ are non-negative, and $\pi(w,w)\ge\alpha$ for all $w\in V$. So:
$$\E{\sum_{w\in V}\p{\ph(w)+\alpha\rh(w)}}\le n\pi(t).$$
It is straightforward to check that, for all $w\in V$,
$$\E{\ph(w)+\alpha\rh(w)}=\alpha\sum_{u\in\Nout(w)}\dr(w,u).$$
Let $N$ denote the total number of calls to \cref{alg:random-push}. When $\gap_i\rmax_i\ge\rmaxx$ for all $i$, by \cref{lem:random-push-time}, the total time complexity for the push-back process is
$$O\p{\frac{\sum_{(u,v)\in E} \dr(u,v)}{\rmaxx}+N} = O\p{\frac{n\pi(t)}{\alpha\rmaxx}+N}.$$
Note that $O\p{\frac{n\pi(t)}{\alpha\rmaxx}}$ here is the upper bound of the number of times we push some residue along an edge.
On the other hand, we only need to call \cref{alg:random-push} to push $\rh_i(w)$ when $\rh_i(w)>\rmax_i$, which means it must receive some residue from its out-neighbors. So we have $N=O\p{\frac{n\pi(t)}{\alpha\rmaxx}}$, finishing the proof.

\subsection{Proof of \cref{lem:gap-1}}
\label{app:gap-1}
    
Consider any level $i$ and any vertex $u$. Given any $\{\rh_{i-1}(v),\rhh_{i-1}(v)\}_{v\in V}$, both $\rh_i(u)$ and $\rhh_i(u)$ are the sum of independent random variables in $[0,\gap_{i}\rmax_{i}]$\footnote{When some $\dr_{i-1}(u,v)>\gap_{i}\rmax_{i}$, since we will push it deterministically, we can split it into several deterministic variables in $[0,\gap_{i}\rmax_{i}]$.} with total expectation $\er_i(u)$. Then, by the Chernoff bound, we have
$$\P{\rhh_i(u)>\rmax_i \wedge \er_i(u)\le\rmax_i/2} \le \P{\rhh_i(u)>\rmax_i \mid \er_i(u)\le\rmax_i/2} \le e^{-\Theta(1/\gap_i)}$$
and
$$\P{|\rh_i(u)-\er_i(u)|>\eps\er_i(u) \mid \er_i(u)>\rmax_i/2,\rhh_i(u)>\rmax_i} \le e^{-\Theta(\eps^2/\gap_i)}.$$
Then
\begin{align*}
    \P{\rhh_i(u)>\rmax_i \wedge |\rh_i(u)-\er_i(u)|>\eps\er_i(u)}
    \le & \ \P{\rhh_i(u)>\rmax_i \wedge \er_i(u)\le\rmax_i/2} \\
    & + \P{|\rh_i(u)-\er_i(u)|>\eps\er_i(u) \mid \er_i(u)>\rmax_i/2,\rhh_i(u)>\rmax_i} \\
    \le & \ e^{-\Theta(\eps^2/\gap_i)}.
\end{align*}
Finally, the lemma follows by a union bound on all levels and all vertices.

\subsection{Proof of \cref{lem:err-1}}
\label{app:err-1}

Let
$$X=\ph(s)+\sum_{u\in V}\pi(s,u)\er(u).$$
We investigate the changes in $X$ across different levels.
For any previously defined variable (e.g., $\rh, \er, \dr, \nopush$), we use the superscript $(j)$ to indicate its value at the beginning of the randomized push at level $j$. That is, the point at which all $\rh_{j-1}(u)$ values have been pushed, where $j \in [0, L]$.

Recall that $X^{(0)}=\pi(s,t)$ and we want to show that with high probability, $$|X^{(L)}-\pi(s,t)|\le \eps\pi(s,t).$$
By a union bound, it suffices to show that with high probability, for all $j\in[0,L)$ we have $$|X^{(j+1)}-X^{(j)}|\le \eps'X^{(j)}$$ for some $\eps'=\Theta(\eps/L)$. 
The following claim computes the value of $X^{(j+1)} - X^{(j)}$. Before presenting the detailed proof, we first provide an intuitive explanation.
Consider each vertex $u$. If we push it in round $j$, we will subtract $\er_j(u)$ from $\er(u)$, but use $\rh_j(u)$ to compute how much we need to push. Note that the push from $\rh_j(u)$ to $\er_{j+1}(\cdot)$ is deterministic, so the error only comes from the difference between $\rh_j(u)$ and $\er_j(u)$.

\begin{claim}
\label{clm:Xj}
    $$X^{(j+1)}-X^{(j)}=\sum_{u\in V}\pi(s,u)\p{1-\nopush_j^{(j+1)}(u)}\p{\rh_j^{(j)}(u)-\er_j^{(j)}(u)}.$$
\end{claim}

\begin{proof}
    In round $j$, the only thing we do is to push residues from level $j$ to level $j+1$. It is straightforward to see:
    \begin{itemize}
        \item $\nopush_i^{(j+1)}(u)=\nopush_i^{(j)}(u)$ for all $i\neq j$ and $u\in V$.
        \item $\er_i^{(j+1)}(u)=\er_i^{(j)}(u)$ for all $i\neq j+1$ and $u\in V$.
        \item $\ph^{(j+1)}(u)-\ph^{(j)}(u)=\p{1-\nopush_j^{(j+1)}(u)}\alpha\rh_j^{(j)}(u)$ for all $u\in V$.
    \end{itemize}
    On the other hand, at the beginning of round $j$, we have not tried to push from levels $i\ge j$, so we further have:
    \begin{itemize}
        \item $\nopush_{j+1}^{(j+1)}(u)=\nopush_j^{(j)}(u)=1$ for all $u\in V$.
        \item $R_{j+1}^{(j)}(u)=0$ for all $u\in V$.
    \end{itemize}
    Also notice that: 
    \begin{itemize}
        \item $\dr_{j+1}^{(j+1)}(u,v)=\frac{\p{1-\nopush_j^{(j+1)}(v)}(1-\alpha)\rh_j^{(j)}(v)}{\dout(u)}$ for all $(u,v)\in E$.
    \end{itemize}
    So we have:
    \begin{align*}
        & X^{(j+1)}-X^{(j)} \\
        & = \p{\ph^{(j+1)}(s)-\ph^{(j)}(s)} + \sum_{u\in V}\pi(s,u)\p{\er^{(j+1)}(u)-\er^{(j)}(u)} \\
        & = \p{1-\nopush_j^{(j+1)}(s)}\alpha\rh_j^{(j)}(s) + \sum_{u\in V}\pi(s,u)\p{\er_{j+1}^{(j+1)}(u)+\p{\nopush_{j}^{(j+1)}(u)-1}\er_{j}^{(j)}(u)} \\
        & = \sum_{u\in V}\pi(s,u)\sum_{v\in\Nout(u)}\dr_{j+1}^{(j+1)}(u,v) + \sum_{v\in V}\pi(s,v)\p{1-\nopush_j^{(j+1)}(v)}\p{-\er_j^{(j)}(v)+\Ind{v=s}\alpha\rh_j^{(j)}(v)} \\
        & = \sum_{v\in V}\sum_{u\in\Nin(v)}\pi(s,u)\dr_{j+1}^{(j+1)}(u,v) + \sum_{v\in V}\pi(s,v)\p{1-\nopush_j^{(j+1)}(v)}\p{-\er_j^{(j)}(v)+\Ind{v=s}\alpha\rh_j^{(j)}(v)} \\
        & = \sum_{v\in V}\p{1-\nopush_j^{(j+1)}(v)}\rh_j^{(j)}(v)\p{\Ind{v=s}\alpha+\sum_{u\in\Nin(v)}\frac{\pi(s,u)(1-\alpha)}{\dout(u)}} + \sum_{v\in V}\pi(s,v)\p{1-\nopush_j^{(j+1)}(v)}\p{-\er_j^{(j)}(v)} \\
        & = \sum_{v\in V}\p{1-\nopush_j^{(j+1)}(v)}\rh_j^{(j)}(v)\pi(s,v) + \sum_{v\in V}\pi(s,v)\p{1-\nopush_j^{(j+1)}(v)}\p{-\er_j^{(j)}(v)} \tag{equation~\eqref{eqn:onestep_walkproperty}}\\
        & = \sum_{v\in V}\pi(s,v)\p{1-\nopush_j^{(j+1)}(v)}\p{\rh_j^{(j)}(v)-\er_j^{(j)}(v)}.
    \end{align*}
\end{proof}

By \cref{clm:Xj}, we have
$$|X^{(j+1)}-X^{(j)}| \le \sum_{u\in V}\pi(s,u)|\rh_j^{(j)}(u)-\er_j^{(j)}(u)|.$$
On the other hand, by \cref{lem:gap-1}, with high probability, we have
$$|\rh_j^{(j)}(u)-\er_j^{(j)}(u)| \le \eps'\er_j^{(j)}(u)$$ for all $j$ and $u$, which means
$$|X^{(j+1)}-X^{(j)}| \le \eps'\sum_{u\in V}\pi(s,u)\er_j^{(j)}(u) \le \eps' X^{(j)}.$$

\subsection{Proof of \cref{lem:gap-2}}
\label{app:gap-2}
Consider any level $i$ and any vertex $u$. Given any $\{\rh_{i-1}(v),\rhh_{i-1}(v)\}_{v\in V}$, $\rhh_i(u)$ is the sum of independent random variables in $[0,\gap_{i}\rmax_{i}]$ with total expectation $\er_i(u)$. Then, by the Chernoff bound, we have
$$\P{\rhh_i(u)\le\rmax_i \wedge \er_i(u)>2\rmax_i} \le \P{\rhh_i(u)\le\rmax_i \mid \er_i(u)>2\rmax_i} \le e^{-\Theta(1/\gap_i)}.$$
The lemma then follows by a union bound on all levels and all vertices.

\subsection{Proof of \cref{lem:last-level}}
\label{app:last-level}

Recall that $\rh_j^{(j)}(u)$ and $\er_j^{(j)}(u)$ denote the value of $\rh_j(u)$ and $\er_j(u)$ at the beginning of round $j$ (when they are fully computed and have not been cleared).
By \cref{lem:gap-1}, with high probability, we have
$$\rh_j^{(j)}(u) \le \p{1+\frac{1}{L}}\er_j^{(j)}(u)$$ for all $j$ and $u$. When this holds, we will show that
$$R_j^{(j)}(u) \le \p{1+\frac{1}{L}}^j\p{1-\alpha}^j$$ for all $j$ and $u$  by induction on $j$. It clearly holds for $j=0$. For $j>0$, we have
\begin{align*}
    R_j^{(j)}(u)
    & \le \sum_{v\in\Nout(u)}\frac{(1-\alpha)\rh_{j-1}^{(j-1)}(v)}{\dout(u)} \\
    & \le \sum_{v\in\Nout(u)}\frac{(1-\alpha)\p{1+\frac{1}{L}}\er_{j-1}^{(j-1)}(v)}{\dout(u)} \\
    & \le \sum_{v\in\Nout(u)}\frac{(1-\alpha)^j\p{1+\frac{1}{L}}^j}{\dout(u)} \\
    & = \p{1+\frac{1}{L}}^j\p{1-\alpha}^j.
\end{align*}

So for all $u\in V$, we have
$$R_L(u) \le \p{1+\frac{1}{L}}^L\p{1-\alpha}^L \le \rmax_L.$$

\subsection{Proof of \cref{lem:err-2}}
\label{app:err-2}
    
By \cref{lem:gap-2,lem:last-level}, with high probability, $\er(u)\le2\rmax$ for all $u\in V$. Fix any final state of the backward exploration process that satisfies the above condition. In the remaining part of the proof, all probabilities and expectations are conditioned on this final state.
Each $q(s,t)-\ph(s)$ is a random variable in $[0,2\rmax]$, so $\pit(s,t)$ is the sum of independent variables that are in $[0,2\rmax/\nr]$.
Consider the following two cases:
\begin{enumerate}
    \item $\E{\pit(s,t)}>\delta$. By the Chernoff bound, we have
    \begin{align*}
        \P{|\pit(s,t)-\E{\pit(s,t)}| > \eps\E{\pit(s,t)}}
        & \le e^{-\Omega(\eps\E{\pit(s,t)} / (2\rmax/\nr))} \\
        & \le e^{-\Omega(\eps\delta\nr/\rmax)} \\
        & \le \pf
    \end{align*}
    \item $\E{\pit(s,t)}\le \delta$. By the Chernoff bound, we have
    \begin{align*}
        \P{|\pit(s,t)-\E{\pit(s,t)}| > \eps\delta}
        & \le e^{-\Omega(\eps\delta / (2\rmax/\nr))} \\
        & \le e^{-\Omega(\eps\delta\nr/\rmax)} \\
        & \le \pf
    \end{align*}
\end{enumerate}
By \cref{lem:pit-exp}, $\E{\pit(s,t)}=\ph(s)+\sum_{u\in V}\pi(s,u)\er(u)$, then the lemma follows.

\subsection{Proof of \cref{lem:err-3}}
\label{app:err-3}
We now fix any final state of the backward exploration process and $\{u_k\}_{k\in[1,\nr]}$, and in the remaining part of the proof, all probabilities and expectations are conditioned on this state. It is straightforward to check that each $\erh(u_k)$ is an unbiased estimator of $\er(u_k)$, so $\E{\pitt(s,t)}=\pit(s,t)$. 
For any $v\in\Nout(u_k)$, let 
$$X(v)=\sum_{i=0}^L \nopush_i(u_k)\dr_i(u_k,v)$$
denote $v$'s contribution to $\er(u_k)$. Notice that
$$X(v) \le \sum_{i=0}^L \dr_i(u_k,v) = \frac{(1-\alpha)\ph(v)}{\alpha\dout(u_k)}.$$
So, for any $v\in\Nout(u_k)\setminus\Vp$, we have
$$X(v)=O\p{\frac{\pth}{\alpha\dout(u_k)}},$$
which means $\erh(u_k)$ is the sum of independent random variables in $\square{0,O\p{\frac{\pth}{\alpha\ns}}}$. Then, $\pitt(s,t)$ is the sum of independent random variables in $\square{0,O\p{\frac{\pth}{\alpha\ns\nr}}}$. The lemma then follows from the same case analysis as \cref{app:err-2}.

\end{document}